# Updated Inflight Calibration of Hayabusa2's Optical Navigation Camera (ONC) for Scientific Observations during the Cruise Phase

Eri Tatsumi[1]

Toru Kouyama[2]

Hidehiko Suzuki[3]

Manabu Yamada[4]

Naoya Sakatani[5]

Shingo Kameda[6]

Yasuhiro Yokota[5,7]

Rie Honda[7]

Tomokatsu Morota[8]

Keiichi Moroi[6]

Naoya Tanabe[1]

Hiroaki Kamiyoshihara[1]

Marika Ishida[6]

Kazuo Yoshioka[9]

Hiroyuki Sato[5]

Chikatoshi Honda[10]

Masahiko Hayakawa[5]

Kohei Kitazato[10]

Hirotaka Sawada[5]

Seiji Sugita[1,11]

1 Department of Earth and Planetary Science, The University of Tokyo, Tokyo, Japan

2 National Institute of Advanced Industrial Science and Technology, Ibaraki, Japan

3 Meiji University, Kanagawa, Japan

4 Planetary Exploration Research Center, Chiba Institute of Technology, Chiba, Japan

5 Institute of Space and Astronautical Science, Japan Aerospace Exploration Agency, Kanagawa, Japan

6 Rikkyo University, Tokyo, Japan

7 Kochi University, Kochi, Japan

8 Nagoya University, Aichi, Japan

9 Department of Complexity Science and Engineering, The University of Tokyo, Chiba, Japan

10 The University of Aizu, Fukushima, Japan

11 Research Center of the Early Universe, The University of Tokyo, Tokyo, Japan

## Abstract

The Optical Navigation Camera (ONC-T, ONC-W1, ONC-W2) onboard Hayabusa2 are also being used for scientific observations of the mission target, C-complex asteroid 162173 Ryugu. Science observations and analyses require rigorous instrument calibration. In order to meet this requirement, we have conducted extensive inflight observations during the 3.5 years of cruise after the launch of Hayabusa2 on 3 December 2014. In addition to the first inflight calibrations by Suzuki et al. (2018), we conducted an additional series of calibrations, including read-out smear, electronic-interference noise, bias, dark current, hot pixels, sensitivity, linearity, flat-field, and stray light measurements for the ONC. Moreover, the calibrations, especially flat-fields and sensitivities, of ONC-W1 and -W2 are updated for the analysis of the low-altitude (i.e., high-resolution) observations, such as the gravity measurement, touchdowns, and the descents for MASCOT and MINERVA-II payload releases. The radiometric calibration for ONC-T is also updated in this study based on star and Moon observations. Our updated inflight sensitivity measurements suggest the accuracy of the absolute radiometric calibration contains less than 1.8% error for the ul-, b-, v-, Na-, w-, and x-bands based on star calibration observations and ~5% for the p-band based on lunar calibration observations. The radiance spectra of the Moon, Jupiter, and Saturn from the ONC-T show good agreement with the spacecraft-based observations of the Moon from SP/SELENE and WAC/LROC and with ground-based telescopic observations for Jupiter and Saturn. Our calibration results suggest that the 0.7-μm absorption band typically observed on Ch and Cgh asteroids at the ~3-4% level can be detected with the ONC's signal-to-noise ratio (SNR) of ~2. We also demonstrate a decrease in SNR due to CCD temperature increases caused by radiant heat when the spacecraft is close to the surface, as the SNR is measured to be 150 at a CCD temperature of 20 °C (the worst case scenario). Since Ryugu may possess a significant amount of internal volatiles, a sodium atmosphere around Ryugu is considered to be highly plausible. We evaluated the upper limit of detectability of a sodium atmosphere around Jupiter using the Na-filter as 100 R with 100 images. This implies that the ONC-T can detect a sodium atmosphere of several 10s kR based on a single image set of v- and Na-bands and of several 100s R based on 100 image sets. Finally, we report the first inflight observation of Ryugu by ONC-T from $1.3 \times 10^6$ km away on 26 February 2018. The ONC-T v-band observation displays consistency with ground-based observation, which confirms the capability of ONC-T.

## 1. Introduction

The Hayabusa2 spacecraft was launched by the Japan Aerospace eXploration Agency (JAXA) in December 2014 and arrived at its target, the C-complex asteroid 162173 Ryugu (formerly 1999JU$_3$) in 27 June 2018. Its mission is to rendezvous with Ryugu and to return samples from the asteroid's surface. In addition to the sampling system, Hayabusa2 also carries remote-sensing instruments, such as the Optical Navigation Camera (ONC), the 3μm Near-Infrared Spectrometer (NIRS3), the Thermal Infrared Imager (TIR), and the Light Detection and Ranging laser (LIDAR). Before sampling, the asteroid surface will be mapped precisely and accurately with the remote-sensing instrument payload, for example 0.5-2 m/pix for the whole asteroid surface by ONC-T, and characterized from both a mineralogical and morphological point of view. This paper describes the inflight calibrations of the charge-couple device (CCD) for the telescopic camera (ONC-T) and the two wide-angle cameras (ONC-W1 and ONC-W2) onboard the Hayabusa2 spacecraft. ONC's designs are shown in **Table 1.1**. Their exposure time settings of three cameras are different because ONC-T engages optical navigations by observing stars and Ryugu with long exposure time, ONC-W1 and -W2 are mainly used for observing during close encounter to the Ryugu surface with short exposure time. Especially, ONC-W1 which needs to capture target markers with flash light, are set to have shortest exposure times. More detailed specifications such as a performance of CCD and hardware weight and size are shown in **Appendix A**.

A number of missions are planned to visit spectrally different types of asteroids, such as Hayabusa2 to a C-type asteroid, OSIRIS-REx to a B-type asteroid (Lauretta et al., 2011), Psyche to a M-type asteroid (Elkins-Tanton, 2018), and Lucy to a set of Jupiter Trojan asteroids (Levison et al., 2017). The comparison of the absolute reflectance from the remote sensing observations across these missions is important for understanding the difference between the various target asteroids, including differences in composition, aqueous alteration, the degree of space weathering, and thermal metamorphism. Furthermore, the compositional constraints and connections to meteorite analogs are important for understanding the properties of primitive materials in the early Solar System. Absolute reflectance spectra, sometimes referred to as spectral albedo, is one of the most powerful indices for the mineralogical classification (e.g., Johnson and Fanale, 1973; Gaffy, 1976) of solar system objects. Also, reflectance variations across an object's surface can be used to determine levels of aqueous alteration (Fornasier et al., 2014), thermal metamorphism (Hiroi et al., 1996), grain-size variation (Cloutis et al., 2013), and space weathering (e.g., Matsuoka et al., 2015; Lantz et al., 2017).

Another important feature of the Hayabusa2 mission is the Mobile Asteroid Surface Scout (MASCOT) lander, which has an instrument payload with overlapping wavelength coverage with the main spacecraft's payload, both of which will be observing the asteroid's surface (e.g., Ho et al., 2017) at different spatial resolutions. The counterpart of the ONC is the MASCOT Camera (MasCam), which has a 1024x1024 pixel complementary metal oxide semiconductor (CMOS) image sensor sensitive in the 400-1000 nm wavelength range with four color LEDs. In order to compare and connect the spectra from MasCam and ONC-T, a robust spectral radiometric calibration is necessary.

The main objective of this study is to describe the inflight calibration of three cameras, ONC-T, ONC-W1, and ONC-W2, during the 3.5 year cruise phase from December 2014 to May 2017, and to describe the calibration pipeline which will be generating products for JAXA's Data Archives and Transmission System (DARTS; https://www.darts.isas.jaxa.jp/planet/project/hayabusa2/) and NASA's Planetary Data System (PDS). The results of preflight calibrations of ONC-T and the results of inflight geometric calibrations were described in detail by Kameda et al. (2017) and Suzuki et al. (2018), respectively. However, differences in temperature and environment between the laboratory and inflight calibration measurements requires a validation and update to the preflight radiometric calibration. Thus, we conduct re-calibration of the CCD sensitivity based on inflight star observations obtained at the reference temperature (~-30°C). We also measured the sensitivity temperature dependence based on inflight Jupiter observations under varying hardware temperatures. Furthermore, we summarize the sensitivities of ONC-W1 and -W2 based on both preflight and inflight measurements. The goal of this paper is to provide reliable sensitivity measurements for the ONC system and evaluate its use in mineralogical mapping.

First we will describe the calibration flow of ONC images in **Sec. 2**, and each detailed analysis in the calibration processes is shown in **Sec. 3** for ONC-T and **Sec. 5** for ONC-W1 and -W2. Detectability of sodium atmosphere will be discussed in **Sec. 4**. ONC system alignment with the spacecraft and NIRS3 will be analyzed in **Sec. 6** and **7**, respectively. Finally, applications in scientific analyses will be discussed in **Sec. 8**, including the report of the first inflight observation of Ryugu in **Sec. 8.4**.

**Table 1.1.** Designs of ONC.

| | ONC-T | ONC-W1 | ONC-W2 |
|---|---|---|---|
| Effective lens aperture | 15.1 mm | 1.08 mm | 1.08 mm |
| Focal length | 120.50±0.01 mm for wide-filter* (Suzuki et al., 2018) | 10.22 mm | 10.38 mm |
| Field of view | 6.27° (Nadir view) | 69.71° (Nadir view) | 68.89° (Slanted ~30° from nadir) |
| F number (F#) | 9.05 | 9.6 | 9.6 |
| Color filters | 7 color bandpass filters (ul: 0.40 μm, b: 0.48 μm, v: 0.55 μm, Na: 0.59 μm, w: 0.70 μm, x: 0.86 μm, | Clear filter | Clear filter |
| | p: 0.95 μm) and 1 clear filter (wide). | | |
| Exposure time | 5.44 ms, 8.20 ms, 10.9 ms, 16.4 ms, 21.8 ms, 32.8 ms, 43.5 ms, 65.6 ms, 87.0 ms, 131 ms, 174 ms, 262ms, 348 ms, 525 ms, 696 ms, 1.05 s, 1.39 s, 2.10 s, 2.79 s, 4.20 s, 5.57 s, 8.40 s, 11.1 s, 16.8 s, 22.3 s, 33.6 s, 44.6 s, 67.2 s, 89.1 s, 134 s, 178 s, 0 s (for smear) | 170 μs, 256 μs, 340 μs, 513 μm, 680 μs, 1.03 ms, 1.36 ms, 2.05 ms, 2.72 ms, 4.1 ms, 5.44 ms, 8.20 ms, 10.9 ms, 16.4 ms, 21.8 ms, 32.8 ms, 43.5 ms, 65.6 ms, 87.0 ms, 131 ms, 174 ms, 262 ms, 348 ms, 525 ms, 696 ms, 1.05 s, 1.39 s, 2.10 s, 2.79 s, 4.20 s, 5.57 s, 0 s (for smear) | 1.36 ms, 2.05 ms, 2.72 ms, 4.1 ms, 5.44 ms, 8.20 ms, 10.9 ms, 16.4 ms, 21.8 ms, 32.8 ms, 43.5 ms, 65.6 ms, 87.0 ms, 131 ms, 174 ms, 262 ms, 348 ms, 525 ms, 696 ms, 1.05 s, 1.39 s, 2.10 s, 2.79 s, 4.20 s, 5.57 s, 8.40 s, 11.1 s, 16.8 s, 22.3 s, 33.6 s, 44.6 s, 0 s (for smear) |

*Focal lengths for v- and Na-filters of ONC-T are in Appendix. A.

## 2. Image Calibration Flow

Raw images acquired by the ONC system need to be processed in a sequence of steps to scientifically calibrate the image data. In this section, we summarize the calibration flow of the ONC images from raw data (Level 0) to higher-level, radiometrically and geometrically calibrated data (Level 2). **Figure 2.1** displays the data calibration flowchart for the ONC and shows the product levels to be archived in both DARTS and PDS (L0: raw data, L1: raw data with a header, L2a: spacecraft information added data, L2b: flat-field corrected digital unit data, L2c: radiance, L2d: radiance factor (I/F), L2e: reflectance factor at (incidence angle, emission angle, phase angle)=(30°, 0°, 30°)).

The digital number for each pixel in a raw image is given as a 8-, 10-, or 12-bit binary number. For higher-level products, we use the FITS (Flexible Image Transport System) formatted files. A FITS formatted file for the ONC system includes a primary header data unit and an image extension. Headers include hardware status, navigation, and other ancillary information, such as spacecraft position and attitude calculated from the SPICE kernels. As described in **Fig. 2.1**, the calibration flow requires input parameters for each step, and those parameters are included in the headers. The bias and smear subtraction is conducted inflight during home position (at alt. ~20 km) observations. However, for images acquired during the cruise phase and the proximity operations during the rendezvous phase, the bias and smear has to be corrected onground through the pipeline process. One more thing to note is that some steps for ONC-T have not yet been implemented in the calibration pipeline (gray boxes in **Fig. 2.1**). Dark current and stray light will be negligible in most asteroid observations, except for some proximity observations, such as observations during the very close approach to the surface or scan observations. When delicate analyses, such as photometric or spectral analyses close to the asteroid limb are conducted, point-spread-function (PSF) corrections need to be incorporated. However, because the same PSF correction method may not always be applicable, depending on where in the FOV the target falls or if it fills or only partially fills the FOV, we do not apply this step for first public product release. Nevertheless, the inner part of the target's disk will be little affected by the PSF; the value without the PSF correction is sufficiently accurate. The effect of the PSF will be discussed more in **Sec. 3.7.1**. This process could be implemented in a future update to the pipeline process.

Each calibration step is described in detail in the following sections. Calibration campaigns were conducted prefight and inflight during the cruise phase. The ONC-T includes 7 color bandpass filters and one "wide" clear filter, whereas the ONC-W1 and ONC-W2 include a single "wide" clear filter. We carefully test and validate the spectral reconstruction of the ONC-T using multiple spectrally well-known stars and planets, since the absolute sensitivity is critical for the spectral characterization of the Ryugu's surface.

*File Naming Rules*

ONC imaging data, Level 2a – 2d are produced in FITS file format. The file naming conventions for ONC images are

“hyb2_onc_(Date)_(Time)_t(Filter)(i)_(Level).fit” for ONC-T,

and

“hyb2_onc_(Date)_(Time)_(Cam)(i)_(Level).fit” for ONC-W1 and -W2,

where (Date) indicates the observation date in 8 digit number, YYYYMMDD, (Time) indicates the observation time in 6 digit number, hhmmss, (Filter) is the band name of filter (u: ul-filter, b: b-filter, v: v-filter, n: Na-filter, w: w-filter, x: x-filter, p: p-filter, and i: wide-filter), (i) is “f” for 1024x1024 pixels fear data and “b” for 32x1024 pixels dark-reference data (Optical Black) (see **Fig. 5** in Kameda et al. (2017)), (Cam) is “w1” or “w2” corresponding to the instruments ONC-W1 and W2 respectively, and (Level) indicates the product levels. Note that the higher-level products may be provided different naming rules and formats.

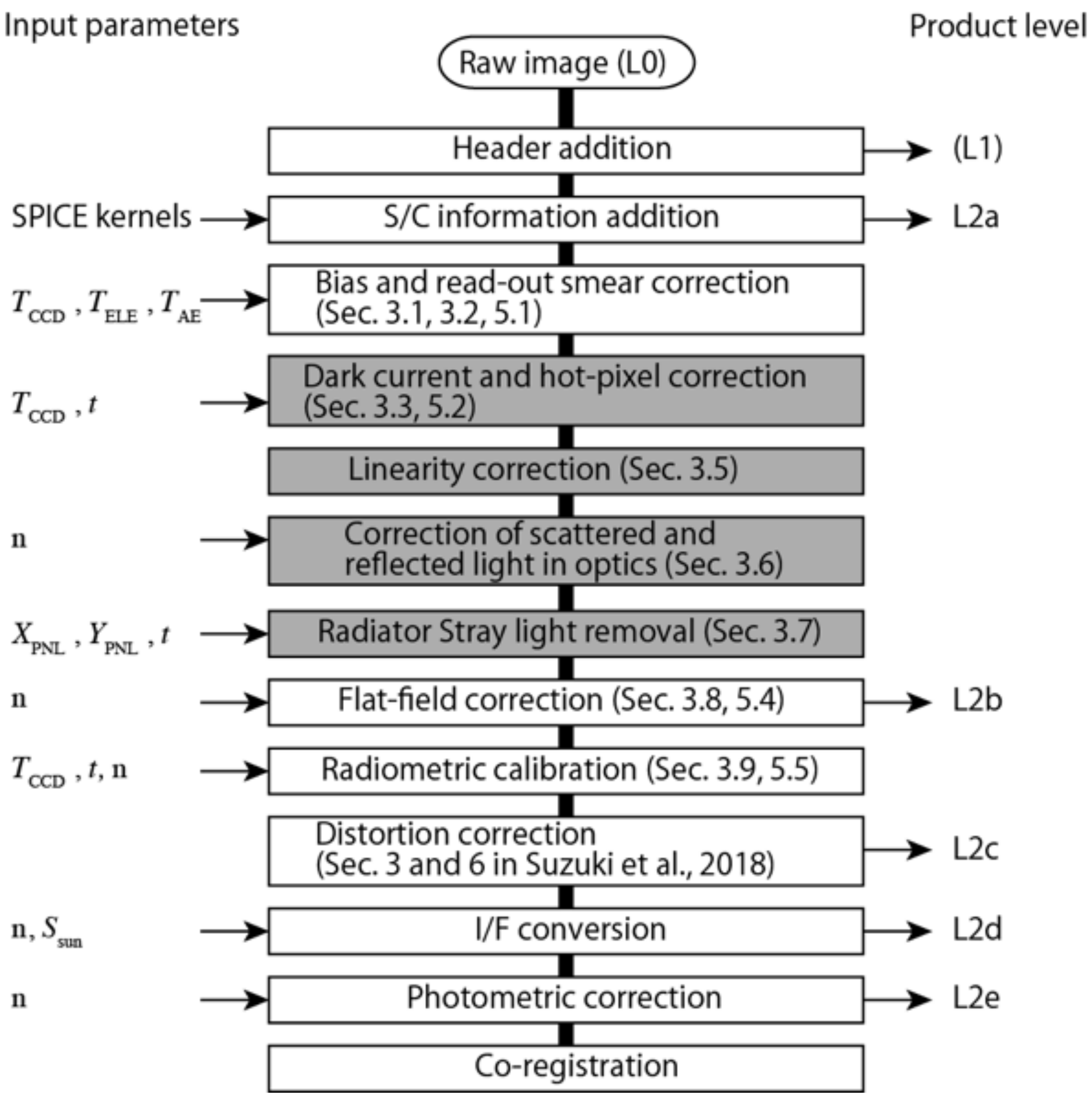

**Figure 2.1.** ONC data calibration flowchart, starting from raw images to higher-level products. Calibration processes input parameters are acquired from the FITS file headers. Note that gray boxes have not been implemented in the current version of the pipeline, but are expected in a future update. Products levels in brackets are intermediate products and not going to be archived. Some L2a products have been smear and bias corrected onboard. (n: filter, $t$: exposure time, $T_{AE}$: electronics package of ONC system (AE) temperature, $T_{CCD}$: CCD temperature of ONC-T, $T_{ELE}$: the electric circuit temperature of ONC-T, ($X_{PNL}$, $Y_{PNL}$): spacecraft attitude, $S_{sun}$: solar irradiance)

## 3. Radiometric calibration of ONC-T

For radiometric calibration, image digital number (DN) is converted to physical units of radiance (Wm$^{-2}$μm$^{-1}$sr$^{-1}$). The radiometric calibration equation has the form:

$$F = \frac{L^{-1}(I_{\mathrm{raw}} - I_{\mathrm{bias}}(T_{CCD,T}, T_{\mathrm{ELE},T}, T_{AE}) - I_{\mathrm{smear}}(I_{\mathrm{raw}}) - I_{\mathrm{dark}}(t, T_{CCD,T})) - I_{\mathrm{sl}}}{f_{T,n} * S_n(T_{CCD,T}) * t} \ [\mathrm{Wm^{-2}\mu m^{-1}sr^{-1}}], \quad (n = \mathrm{ul, b, v, Na, w, x, p}) \qquad (3.1)$$

where $F$ [Wm$^{-2}$μm$^{-1}$sr$^{-1}$] is the radiance from the surface, $I_{\mathrm{raw}}$ [DN] is a raw image, $I_{bias}(T_{CCD,T}, T_{ELE,T})$ [DN] is the bias level, $I_{\mathrm{dark}}(t, T_{CCD,T})$ [DN] is the dark current, $I_{\mathrm{smear}}(I_{\mathrm{raw}})$ [DN] is the smear image, $I_{\mathrm{sl}}$ [DN] is the stray light level. $L^{-1}(I)$ is the linearity correction function, $f_{T,n}$ is the flat-field image of ONC-T, $S_n(T_{CCD,T})$ [(DN/s)/(Wm$^{-2}$μm$^{-1}$sr$^{-1}$)] is the sensitivity, $t$ [s] is exposure time, $T_{CCD,T}$ [°C] is the CCD temperature of ONC-T, $T_{ELE,T}$ [°C] is the electric circuit (ELE) temperature of the ONC-T, $T_{AE,T}$ [°C] is the AE temperature of the ONC system, and n indicates filter name. Each term is described in detailed in following subsections. Calibrations conducted so far for both inflight and preflight are summarized in **Table 3.1**. Note that all the calibration images during the cruise phase were obtained in 12-bit format, but the rendezvous phase data will be 12-bit, 10-bit, or 8-bit formats for different purposes, such as 10-bit images for spectroscopic observations and 8-bit images for geometric observations. The bit depth information is included in the header of each image.

In Kameda et al. (2017) the effective wavelengths were calculated using only the transmission of band-pass filters. To obtain a more precise measurement, we recalculate the effective wavelengths using the system efficiency $\Phi_n(\lambda)$ (**Fig. 3.1**), including the transmittance of the band-pass filters, the neutral density (ND) filter, lenses, CCD cover glasses, the quantum efficiency of the CCD, and the spectrum of the light source. The effective wavelength for each filter, using the solar spectrum (ASTM E-490) and the white light as a light source, respectively, are tabulated in **Table 3.2**. **Table 3.2** also shows the bandwidth, the effective bandwidth, and the effective solar irradiance through each filter. The bandwidth is defined in the same way as Kameda et al. (2017), but here we used the system efficiencies instead of the transmissions themselves. The effective wavelength is thus defined as

$$\lambda_{\mathrm{eff},n} = \frac{\int \lambda J(\lambda)\Phi_n(\lambda)d\lambda}{\int J(\lambda)\Phi_n(\lambda)d\lambda}, \qquad (3.2)$$

where the system efficiency $\Phi_n(\lambda) = \tau_n(\lambda)\tau_{\mathrm{ND}}(\lambda)\tau_{\mathrm{L}}(\lambda)\tau_{CG}(\lambda)Q(\lambda, T_{CCD,T})$; $\tau_n(\lambda)$, $\tau_{ND}(\lambda)$, $\tau_L(\lambda)$, and $\tau_{CG}(\lambda)$ are the transmissions of the band-pass filters, the ND filter, the lenses, and the CCD cover glasses, respectively, $Q(\lambda, T_{CCD,T})$ is the quantum efficiency, and $J(\lambda)$ is the light source spectrum. The effective bandwidth $\Delta\lambda_{\mathrm{eff}}$ is defined as

$$\Delta\lambda_{\mathrm{eff},n} \cdot \Phi_n(\lambda_{\mathrm{eff},n}) = \int \Phi_n(\lambda)d\lambda. \qquad (3.3)$$

Hereafter we use the effective wavelength with respect to the solar spectrum. Note that the quantum efficiency used here is extrapolated to -30 °C of CCD temperature using the temperature dependence function provided by its manufacturer, which is defined for the temperature range from -20 °C to 40 °C and normalized at 20 °C. The quantum efficiency at 20 °C and other components of the system efficiency are shown in detail in **Appendix B**. Except for some proximity operations, the ONC will be operated at a CCD temperature around -30 °C (the reference temperature).

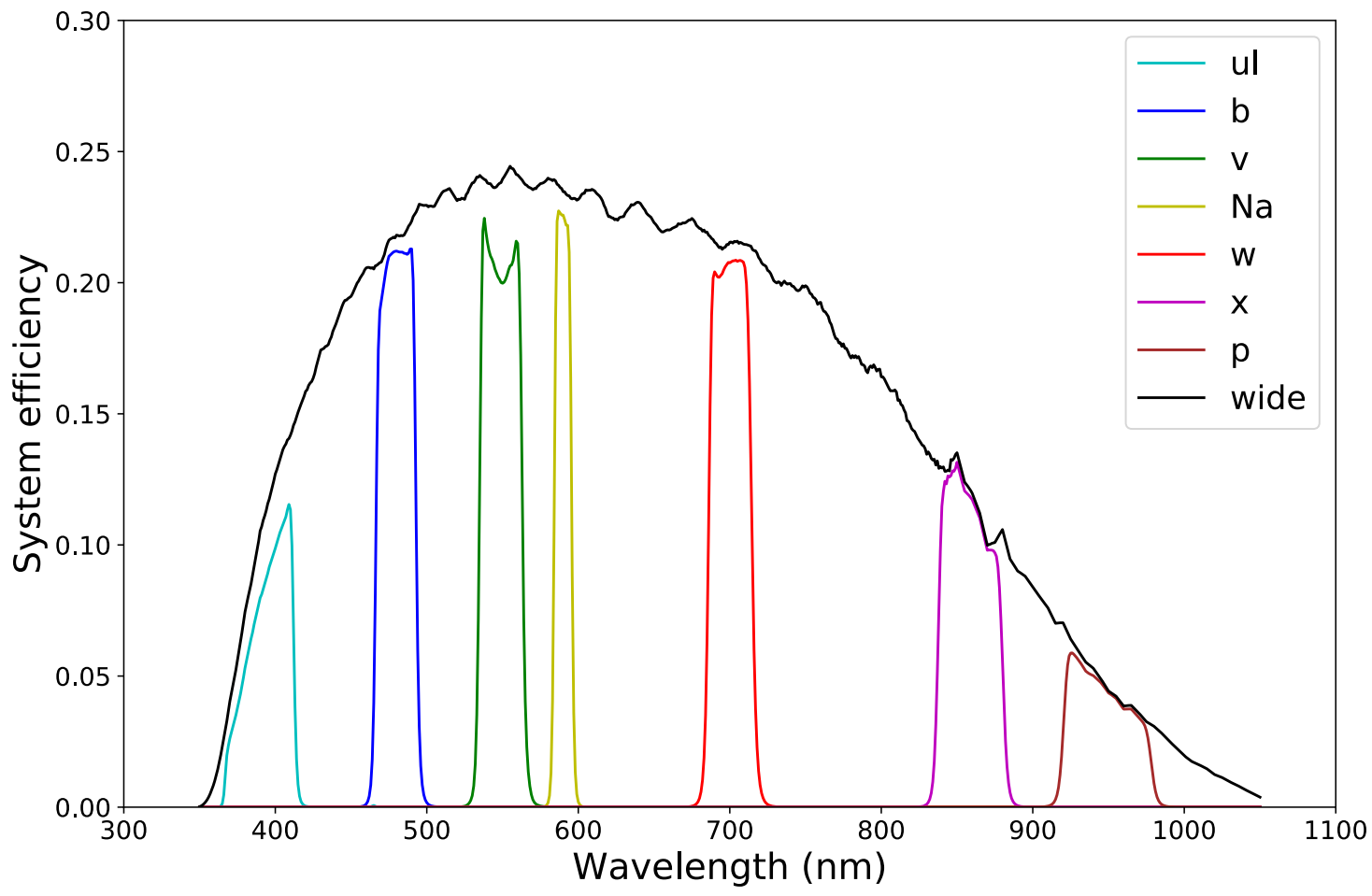

**Figure 3.1.** System efficiency of ONC-T bandpass photometric system.

**Table 3.1.** Summary of data used for calibration during the cruise phase.

| | Preflight measurement | Inflight measurement | Inflight Observation Date |
|---|---|---|---|
| Bias | 0-sec exposures (Kameda et al., 2017) | 0-sec exposures | Continuously during the cruise phase |
| EMI | N/A | 0-sec exposures | Continuously during the cruise phase |
| Dark current and Hot pixels | Thermal vacuum test (variation of CCD, ELE, and AE temperatures) | Deep sky observation with variation of CCD and ELE temperatures | 17 April 2017, 23 May 2017, 25 May 2017 |
| Sensitivity | Integrating sphere (Kameda et al., 2017), temperature dependence provided the manufacturer (Suzuki et al., 2018) | ESO standard and Alekseeva (1996) stars. Temperature dependence by Jupiter with variation of CCD temperature. The sensitivities derived from stars are tested by the Moon, Mars, Jupiter, and Saturn. | (Stars) 19 October 2016, 23 May 2017, 10 October 2017, 12 October 2017 (Moon) 5 December 2015 (Mars) 24 May 2016, 31 May 2016, 7 June, 2016 (Jupiter) 16 to 18 May 2017 (Saturn) 3 November 2017 |
| Sensitivity Time-dependence | N/A | Health check (FF lamp) | 11 December 2014, 16 April 2015, 10 September 2015, 8 July 2017, 16 October 2017, 5 December 2017 |
| Linearity | Integrating sphere | FF lamp with variation of exposure time | 2 December 2017 |
| Flat-Field | Integrating spheres (Kameda et al., 2017; Suzuki et al., 2018) | Star observation with variation of spacecraft attitude | 12 to 14 October 2017 |
| Geometric distortion | N/A | Stars (Suzuki et al., 2018) | 11 December 2017 |
| Alignment to the S/C | N/A | Stars | 11 December 2017 |
| Sharp PSFs | Pinhole (Kameda et al., 2017) | Stars (Suzuki et al., 2018) | 11 December 2017 |
| Broad PSFs | Small extended sources | Earth and Mars | 4 December 2015, 25 May 2016 |
| Stray lights | N/A | Radiator stray lights with variation of the spacecraft attitude. Ghost with Moon and Earth. | 11 December 2014, 23 June 2015, 12 to 17 October 2015, 9 November 2015, 14 November 2015, 22 December 2015, 9 February to 17 March 2016, 21 March 2016, 28 June to 23 July 2016, 13 to 20 June 2017, 17 to 21 October 2017 |

**Table 3.2.** Characteristics of the ONC-T photometric system.

| | ul | b | v | Na | w | x | p |
|---|---|---|---|---|---|---|---|
| $\lambda_{eff}^{1)}$ (nm) | 397.5 | 479.8 | 548.9 | 589.9 | 700.1 | 857.3 | 945.1 |
| $\lambda_{eff}^{2)}$ (nm) | 395.2 | 480.0 | 549.0 | 590.0 | 700.3 | 857.7 | 945.7 |
| $\lambda_{eff}$ (nm) Kameda et al. (2017) | 390.4 | 479.8 | 549.0 | 590.2 | 700.4 | 858.9 | 949.7 |
| Effective bandwidth $\Delta\lambda_{eff}$ (nm) | 36.0 | 26.6 | 30.6 | 11.8 | 29.2 | 41.7 | 56.0 |
| Bandwidth (nm) | 34.7 | 26.6 | 27.9 | 11.6 | 28.8 | 42.4 | 57.2 |
| Bandwidth (nm) Kameda et al. (2017) | 45 | 25 | 28 | 10 | 28 | 42 | 57 |
| Effective solar irradiance (W/m²/um) | 1343.7 | 1969.1 | 1859.7 | 1788.0 | 1414.4 | 985.8 | 834.9 |

1) with respect to solar spectrum. 2) with respect to white light.

### 3.1. Read-out smear removal

ONC is not equipped with a mechanical shutter, instead, each exposure is divided with an electronical mechanism. Because of the mechanical shutter-less design of the ONC, the CCD is irradiated during frame transfer along the vertical direction. So that, even 0-sec exposure images actually include read-out smear during this very short exposure time (<10 ms). For the global observations at the home position the ONC-T will take 0-sec exposures and proper exposures alternately, and the read-out smear will be removed onboard by subtracting the 0-sec exposures from the proper exposures. However, during the touchdown sequences the spatial coverage of the field-of-view (FOV) changes rapidly over the short time duration of these operations, and appropriate 0-sec exposures with the same spatial coverage as the proper exposure image cannot be acquired. In this operational case, the read-out smear removal process will be conducted after the images have been downlinked using the method described by Ishiguro et al. (2010):

$$I_{smear}(H) = \frac{t_{VCT}}{t_{VCT}+t}\sum_{H=0}^{N_V-1}(I - I_{bias} - I_{dark})/N_V. \quad (3.4)$$

The exposure duration during vertical charge transfer $t_{VCT}$ is $7.2\ \mu s \times 1024 = 7.373\ ms$ for ONC-W2, which has been derived from the Earth observation images acquired during the Earth-Moon flyby. Because the same manufactured CCDs are equipped on ONC-T, -W1 and -W2, this vertical transfer time similar between the three cameras. For example, **Fig. 3.2** shows an original image and the smear-removed images of an ONC-W2 (hyb2_onc_20151203_084458_w2f_l2a.fit) set of observations. The residual of the smear-removed image is less than 1% of the intensity of the Earth (~900 DN), for 95% of pixels. For the case of Ryugu, the effect of the smear will be less than in the Earth observations case due to the longer exposures needed for Ryugu due to its corresponding lower brightness. This bias removal method works sufficiently for the ONC-W2 image. We will obtain the vertical transfer times for ONC-T and W1 using Ryugu images during the rendezvous phase.

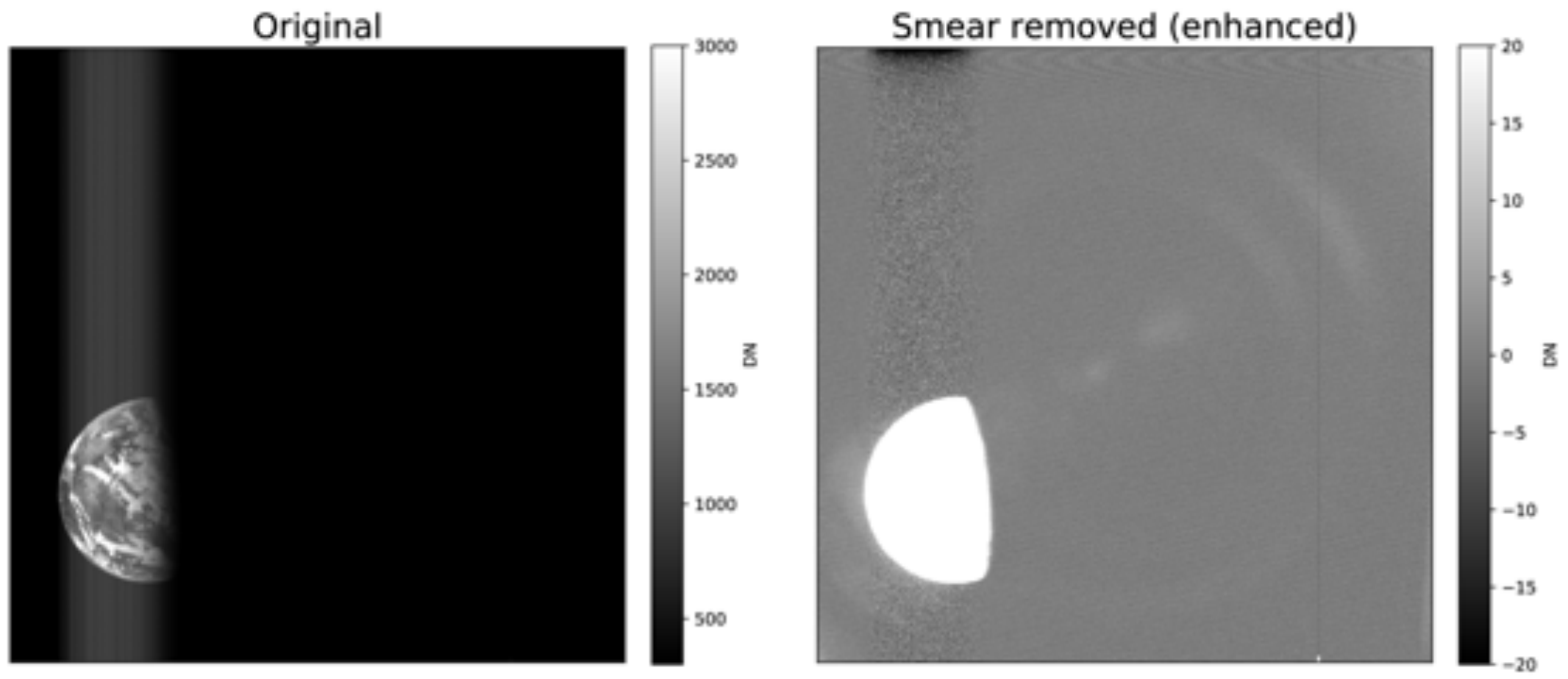

**Figure 3.2.** An example of smear removal of an ONC-W2 image (hyb2_onc_20151203_084458_w2f_l2a.fit) using **Eq. (3.4)**, showing the read-out smear was well removed <1% intensity of the object.

### 3.2. Bias Current Correction

The ONC system has an electronic bias current to offset its zero level. The temperature dependence of the preflight bias level was assessed by Kameda et al. (2017). Here, we also verify the bias level using inflight observations. We analyzed images acquired with a zero-length exposure and targeted to sky in order to avoid strong smear. As Kameda et al. (2017) reported, bias level changes depending on the CCD temperature ($T_{CCD,T}$ [°C]) and the AE temperature ($T_{AE,T}$ [°C]). Bias level dependencies on the CCD and the ELE temperatures are also observed in the inflight image data (**Fig. 3.3**). It should be noted that due to inflexibility in the AE control, we instead changed the temperature of ELE that resides just behind the CCD. During the preflight test, the AE and the ELE temperatures were controlled simultaneously to almost same temperature. Thus, because the results of the pre-flight test include the effect of both the ELE and AE temperatures, we cannot directly compare the preflight and inflight test measurements. We plotted the preflight AE and ELE temperature dependence test with the inflight ELE temperature dependence test results together in **Fig. 3.3**. The bias level is well described by fitting the inflight data with an empirical linear combination of the CCD and ELE temperatures as

$$I_{bias}(T_{CCD,T}, T_{ELE,T}) = 320.66 + 0.652T_{CCD,T} - 0.953T_{ELE,T} \; [DN]. \quad (3.5)$$

**Figure 3.4** displays how well the empirical function describes the bias level of the inflight data acquired from launch until December 2017. **Figure 3.5** shows the time-dependent change of the bias level, which suggests an increase in the bias level during the first half-year after launch. Although the empirical relationship derived from the inflight image data does not match the pre-flight experiments, this is ascribed to the effect of the AE temperature variation in the preflight data. The bias level has stayed nearly constant, within 1% change, in the 3 years of cruise. Moreover, the effect of the AE temperature can be estimated from the preflight measurements of Kameda et al. (2017). After correcting for the effects of CCD and ELE temperatures, the remaining effect is ascribed to variations due to AE temperature differences. The empirical correction factor $B(T_{AE})$ for the AE temperature, derived from the preflight measurements, is given by

$$B(T_{AE}) = 0.987 - 0.00251T_{AE}, \qquad \text{for -30°C} < T_{AE} < 59\text{°C} \tag{3.6}$$

When the AE temperature changes, the bias level can be estimated by $I_{bias}(T_{CCD,T}, T_{ELE,T})B(T_{AE})$. Note that this relationship based on preflight measurements may include errors of >4%, as the bias level appears to have increased by 4% just after the launch.

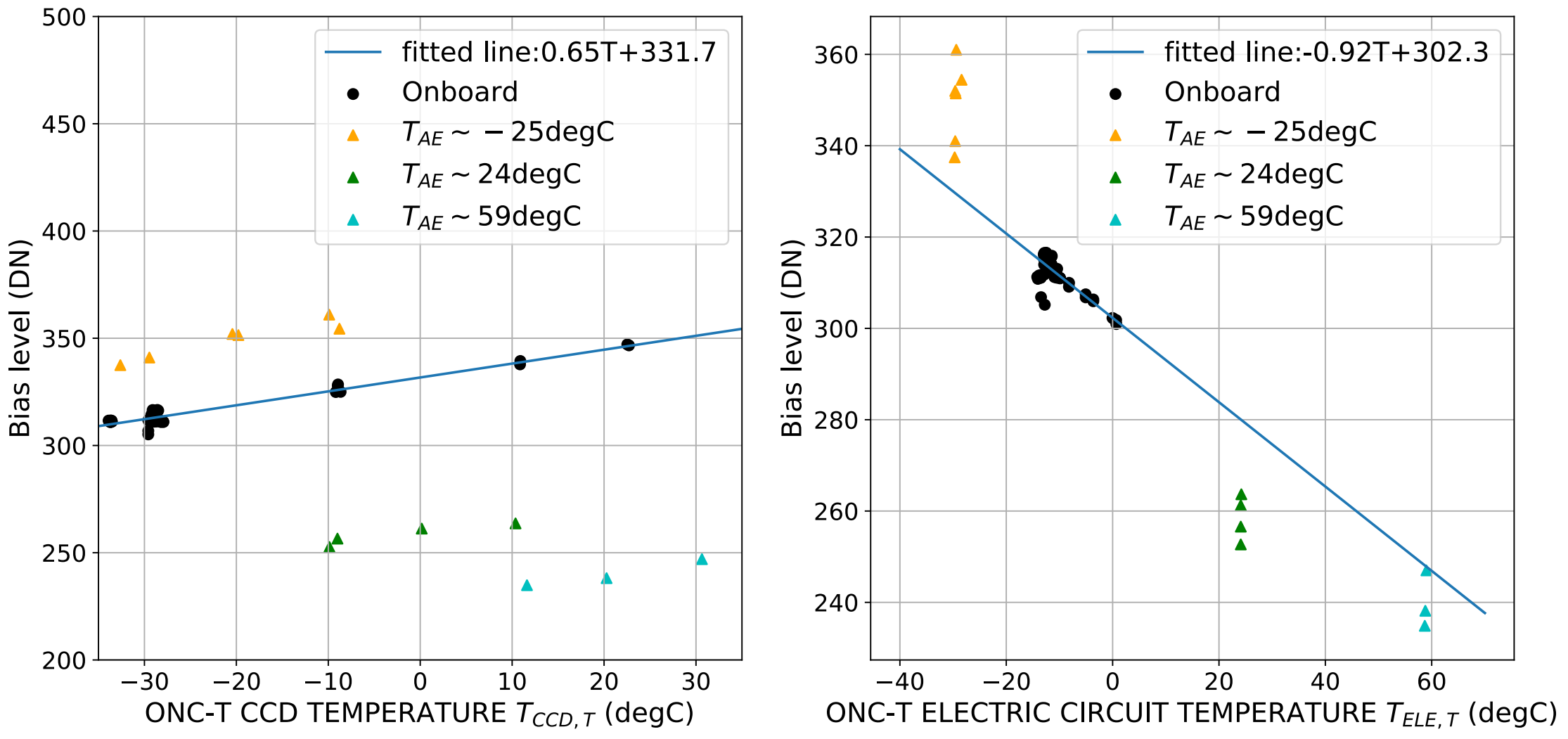

**Figure 3.3.** The bias level of inflight (dots) and preflight (triangles) 0-sec exposure images as a function of CCD and ELE temperatures. The AE temperature deference contributes the bias level significantly.

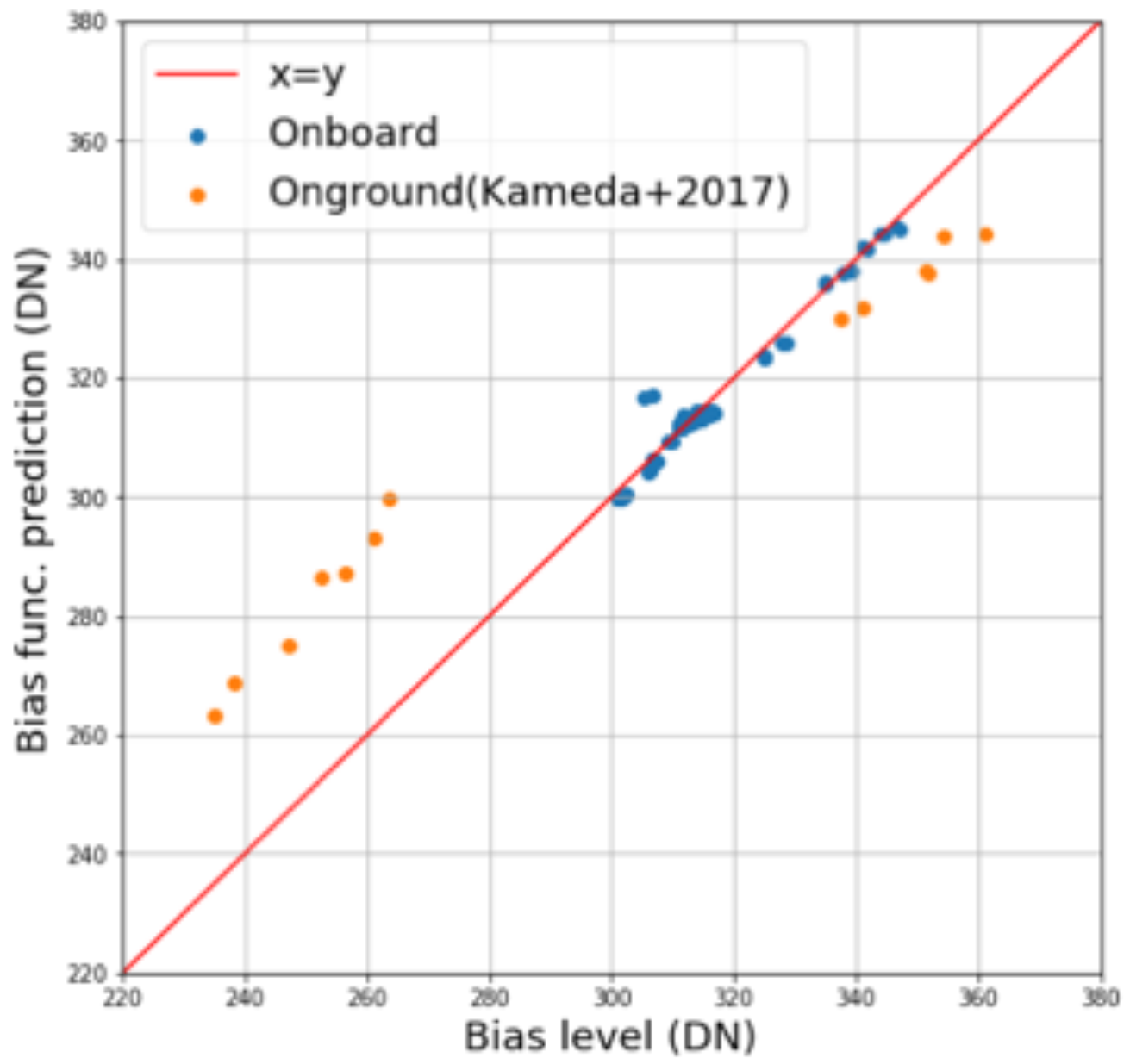

**Figure 3.4.** Bias level of the preflight data (Kameda et al., 2017) and the inflight data compared with the empirical function $I_{bias}(T_{CCD,T}, T_{ELE,T})$.

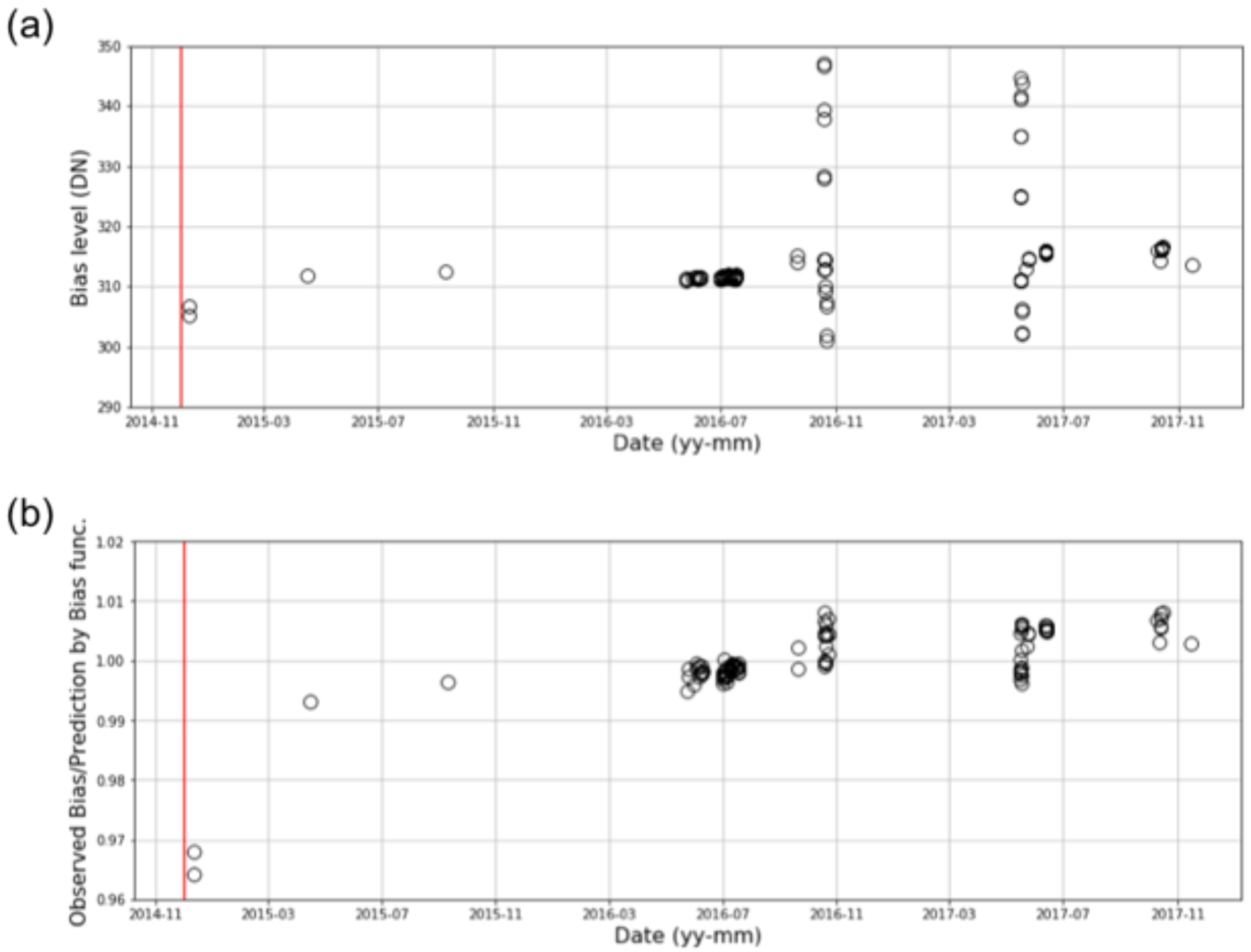

**Figure 3.5.** Bias level variation of ONC-T. Red lines indicate the launch date. (a) The observed bias level of 0-sec exposures during the cruise phase. (b) Comparisons of the observed bias and the bias model **Eq. (3.5)**. The bias model well predicts the inflight data within 1% change over the 3 years of cruise.

### 3.3. Dark Current and Hot Pixels Correction

The CCD constantly responses to thermal electrons even when the detector is not irradiated, which produces a dark current that should be subtracted from observed images to accurately reproduce reflectance levels. However, because the ONC does not have a physical shutter, we cannot directly assess the dark current at the time of each observation. Sky observations, however, can provide a measure of the signals from the dark current, hot pixels, EMI, stars, and stray light. As Suzuki et al. (2018) reported, the radiator stray light has a gradual structure over an image. The stray light cannot be completely separated from the dark current, which also has a contribution to the intensity over an image. Thus, we obtained upper limits of the dark current (<0.09 DN/s) contribution at the reference temperature, $T_{CCD,T} = -30$°C, based on the parts of images where the stray light contamination is minimized. This value is comparable to the preflight ground measurement (<0.05 DN/s at $T_{CCD,T} = -30$°C) by Kameda et al. (2017). Furthermore, the dark current level variation as a function of temperature can be measured using sky observations, assuming that the gradual structure in the images are due to radiator stray light as measured at the reference temperature. To obtain dark current levels at higher CCD temperatures (-10 °C, 10 °C, 20 °C, and 25 °C), we obtained sky images with a short exposure time (1.05 s) for all temperature conditions and long exposure times of 33.6 s at -10°C, 5.57 s at 10 °C, and 2.1 s at 20 °C and 25 °C. We took 3 images for each exposure condition, and removed the effects of cosmic rays by taking a median of these 3 images. After the cosmic ray correction, all images were stray light corrected based on images taken at $T_{CCD,T} = -30$ °C, where the stray light contribution was extracted using a fifth-order polynomial regression and neglecting outliers which are caused by stars and hot pixels. We obtained the dark current distribution except for star-containing regions within the image frame. **Figure 3.6 (a)** is a histogram of dark current levels for different CCD temperatures in the ONC-T. The averages and the standard deviations for the distributions are also shown in the **Fig. 3.6 (b)**. The dark current grows exponentially depending on the CCD temperature, which is consistent with the dark current being caused by thermal noise. The empirical relationship of the CCD temperature and the dark current level is given by

$$I_{dark} = t\ exp(0.10\ T_{CCD,T} + 0.52) \ \ [\text{DN}]. \tag{3.7}$$

Carefully monitoring the dark images at -30 °C, we observed some pixels exhibit anomaly high response levels (hot pixels). We defined hot pixels as a pixel with a response larger than 30 DN/s, or 0.3% the expected intensity of the asteroid. Hot pixel positions on 14 October 2017 are listed in Table 3.3 for the CCD temperature condition is cold (-30 °C). The number of hot pixels increases as the function of the CCD temperature (**Fig. 3.7**). This is especially important during the touchdown sequences, as the CCD temperature could increase as high as 20 °C. Thus, the absolute signal at the hot pixel locations should be analyzed carefully. Note that, however, the number of hot pixels can be increased with time due to cosmic ray irradiation. The temporal variation in hot pixels is shown in **Fig. 3.8**, indicating that the number of hot pixels gradually increases. Thus, we need to repeat this analysis during the approach and rendezvous phases to monitor the change in hot pixels.

**Table 3.3.** Hot pixel list at $T_{CCD,T} = -30$°C. H and V are the horizontal and vertical pixel coordinate in an image.

| H | V | Signal level [DN/s] |
|---|---|---|
| 24 | 653 | 47.0 |
| 52 | 276 | 89.3 |
| 296 | 939 | 55.8 |
| 407 | 239 | 37.1 |
| 505 | 29 | 31.4 |
| 582 | 204 | 75.1 |

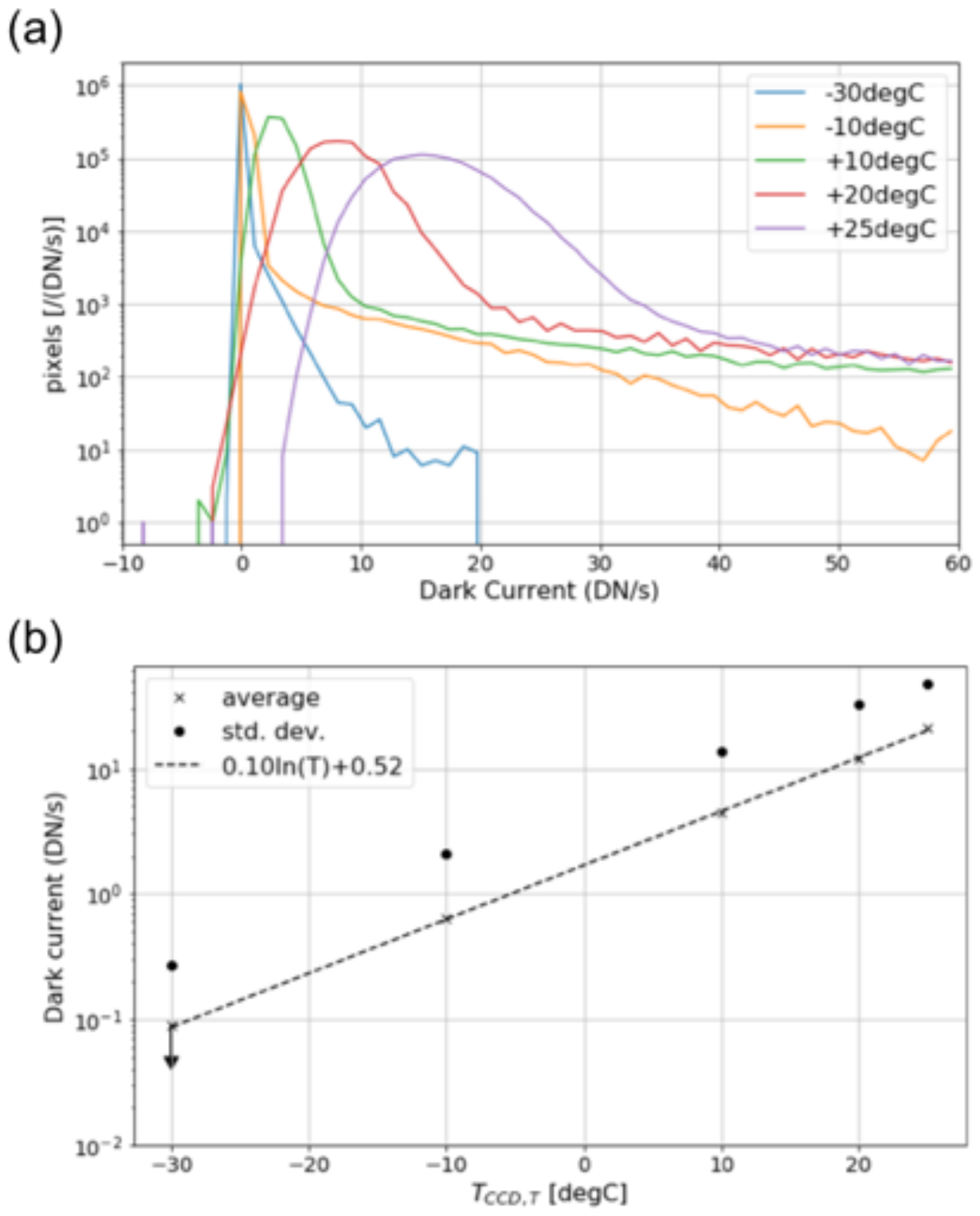

**Figure 3.6.** Dark current level dependence on the CCD temperature of ONC-T. (a) Dark current distribution. The deviations of dark current get larger as the CCD temperature increases. (b) Dark current level grows exponentially with CCD temperature, indicating dark current levels are well described by thermal noise.

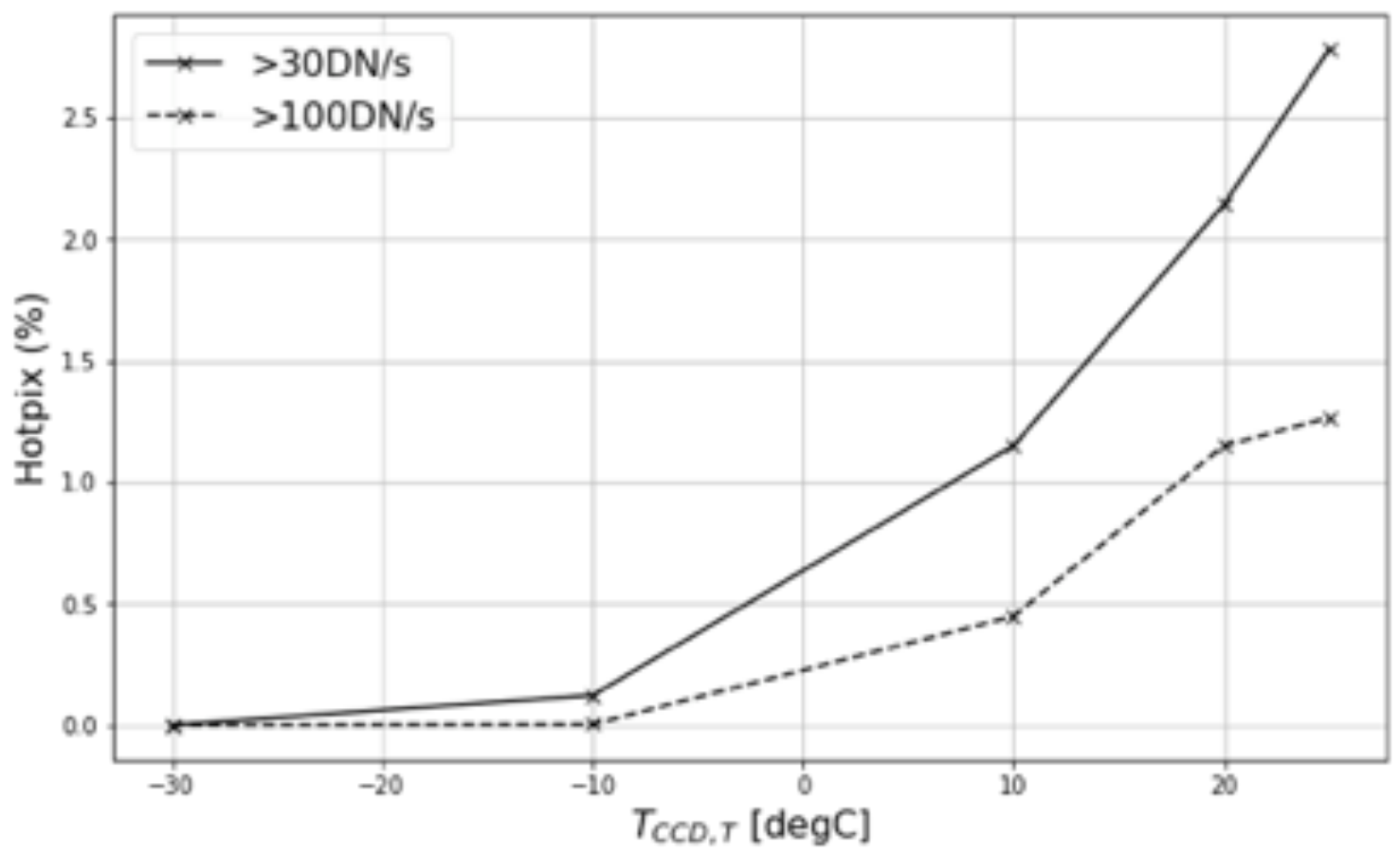

**Figure 3.7.** The amount of hot pixels in the ONC FOV. The CCD temperature could be as high as 20°C during the touchdowns.

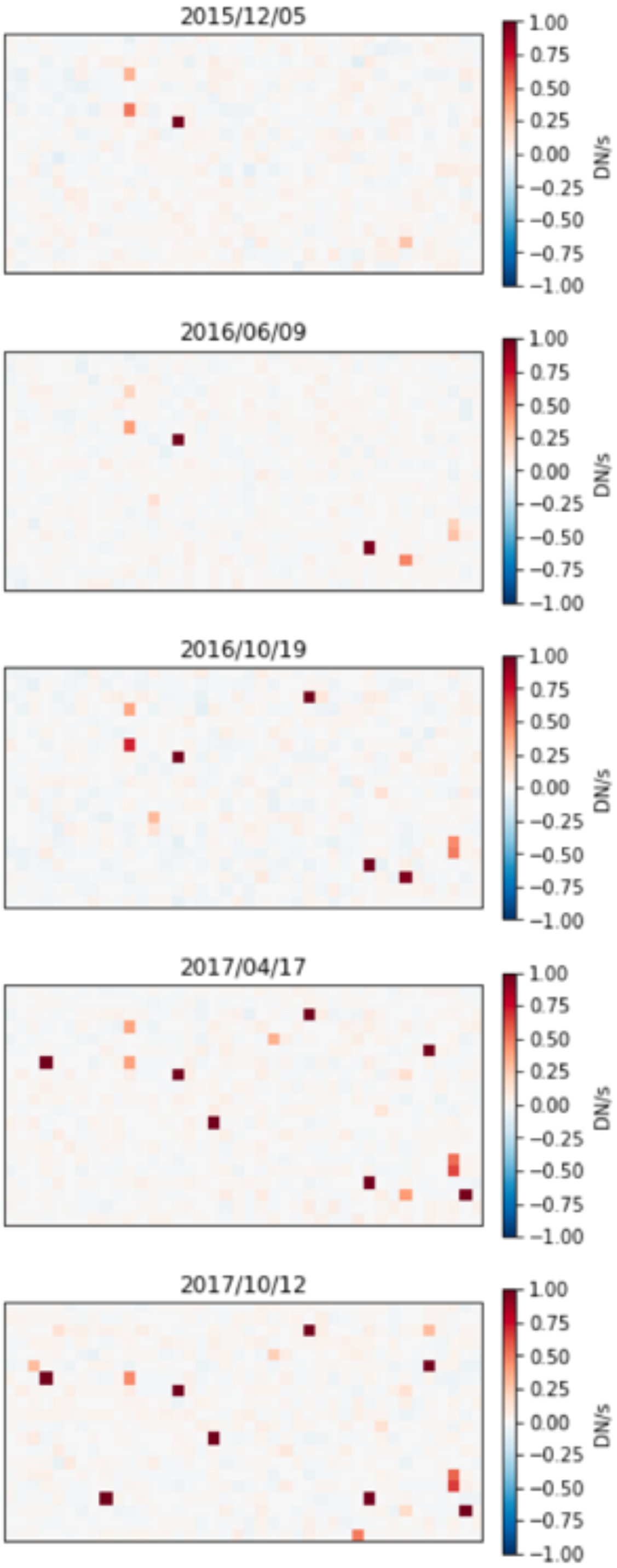

**Figure 3.8.** Increasing number of hot pixels (sensitive pixels) in the box {[492, 118];[532, 138]}.

### 3.4. Characterization of Electro-Magnetic Interference

**Figure 3.9** shows the presence of electro-magnetic interference (EMI) patterns embedded in the CCD noise (stripe patterns). This is caused by interference with the active circuits nearby the ONC system during the transfer of the electrons generated at the CCD to the digital electronics. Thus, the intensity of the EMI is independent of the exposure time. The EMI pattern changes with time depending on which systems with specific frequencies were active at that time of exposure. 0-sec exposure images were analyzed line-by-line based on Fourier Fast Transforms (FFTs) to extract the periodic patterns. **Figure 3.10 (a)** shows a power spectrum of 3 images taken at different times (16 April 2015, 24 May 2016, 16 October 2017), which indicates stronger peaks for some specific frequencies. We evaluated the EMI from inverted FFTs of the extracted data for frequencies > 0.001 Hz and amplitudes > 0.3 DN. **Figure 3.10 (b)** shows the EMI structure variation during the cruise phase. After the EMI is extracted from the images, the amplitude of the EMI is evaluated as a half of the gap between the maximum and minimum signals in the line (**Fig. 3.11**). This suggests the EMI intensities have not drastically changed during the cruise phase and the highest EMI intensity was ~5 DN, which is <0.2% of the expected asteroid disk intensity (~2500 DN). However, we continue to monitor the EMI contribution to accurately remove the EMI from images required for sensitive and advanced spectral analyses.

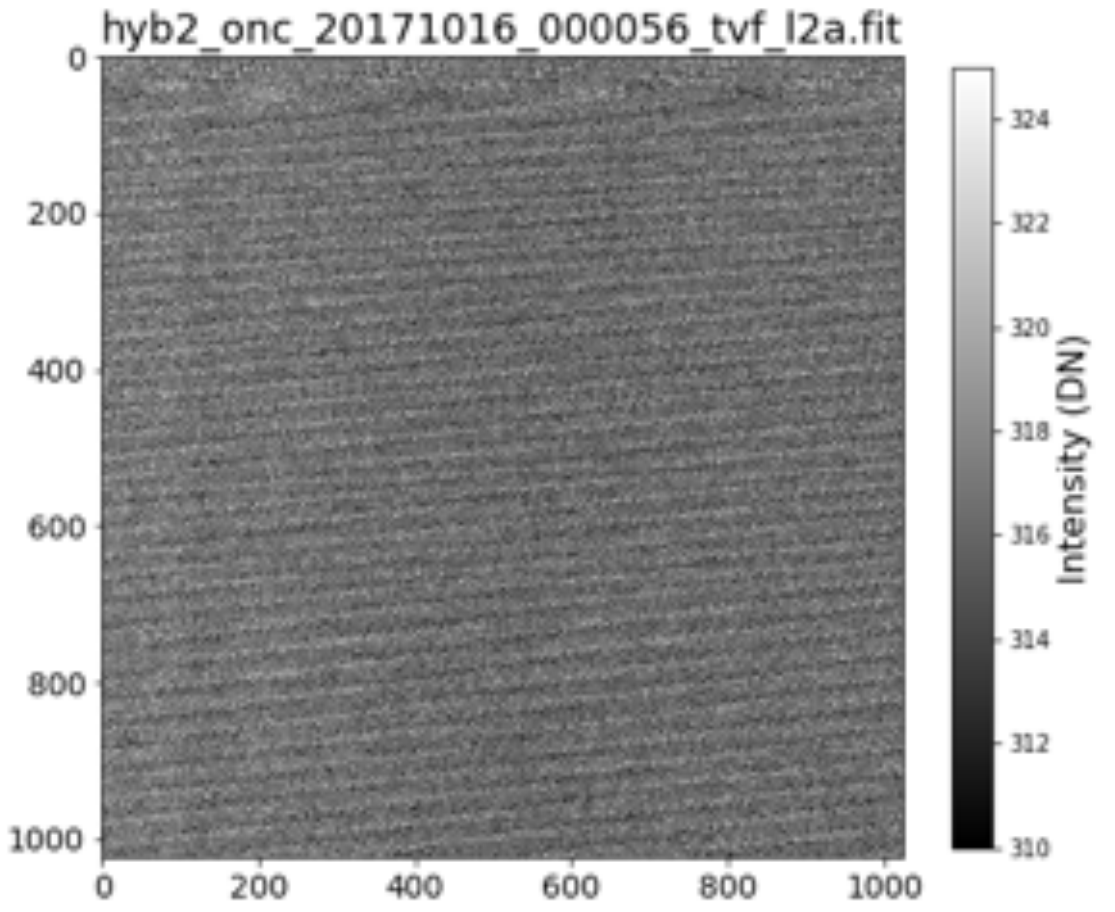

**Figure 3.9.** An example image of the EMI pattern (hyb2_onc_20171016_00056_tvf_l2a.fit). The presence of periodical stripe patterns over the image was observed.

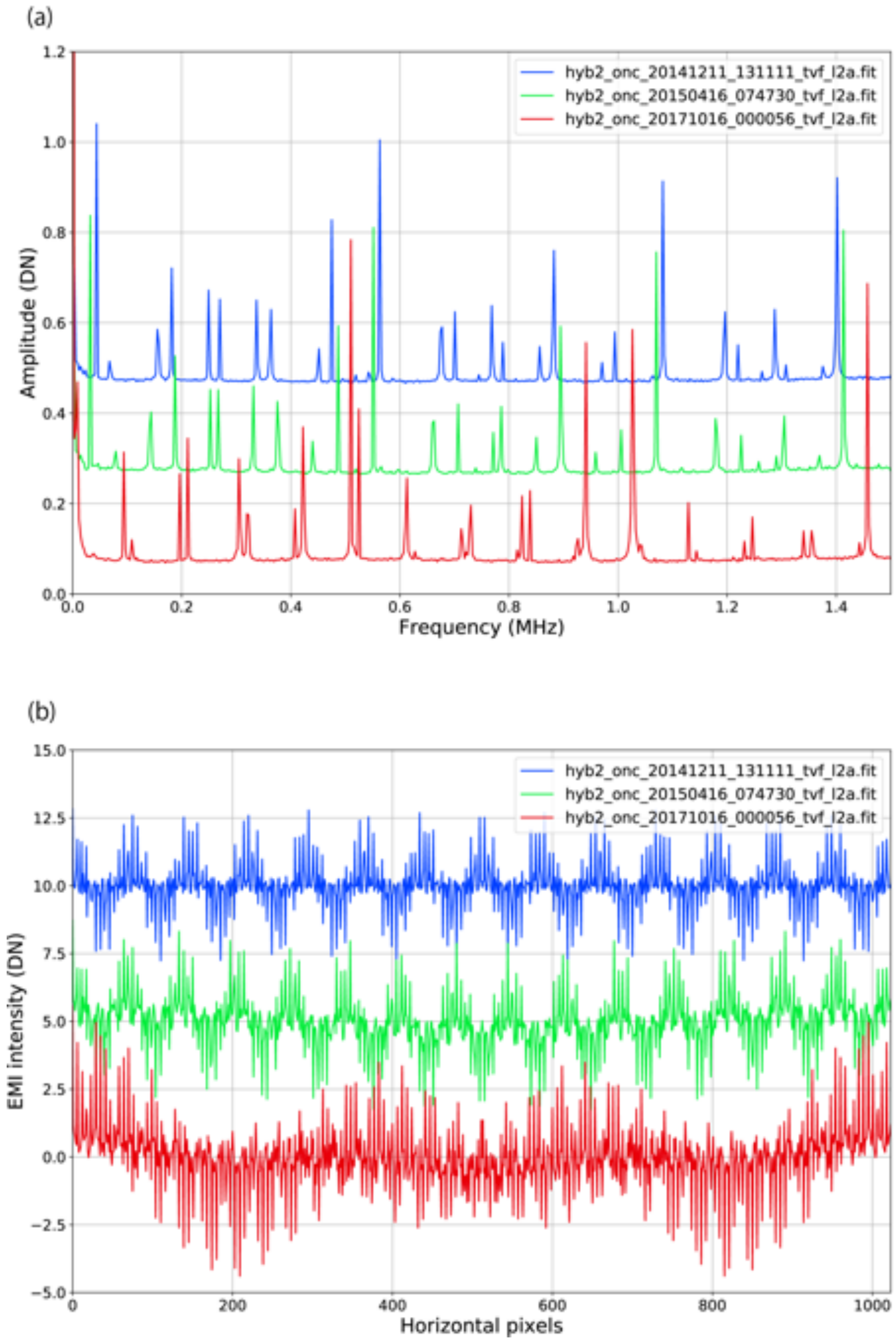

**Figure 3.10.** (a) The results of FFT analysis of 3 images (hyb2_onc_20150416_074730_tvf_l2a.fit, hyb2_onc_20160524_124508_tvf_l2a.fit, hyb2_onc_20171016_000056_tvf_l2a.fit). There are several strong peaks possibly associated with other system circuits. (b) The extracted EMI signals have different structures at different observations. Offsets are 5 DN from each other. Sampling rate is 3MHz.

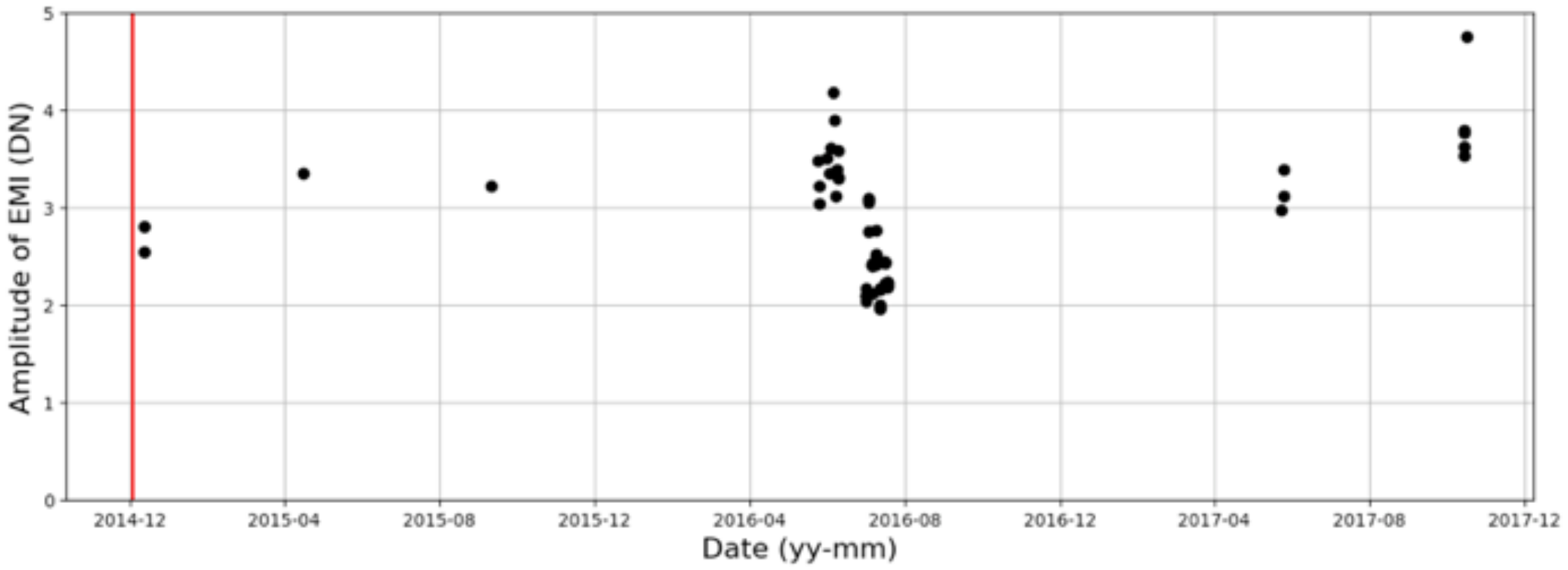

**Figure 3.11.** Variation in the amplitude of the EMI waves during the cruise phase. The amplitude is defined as a half of the gap between maximum and minimum signals of EMI waves. The red line indicates the launch date.

### 3.5. Linearity

In this section, we report on the linearity of the ONC-T CCD response based on inflight observations of Flat-Field (FF) lamp. The CCD response to the number of photons should ideally be linear. Preflight testing (**Fig. 3.12**) shows that the response of the ONC-T CCD was linear up to 3200 DN, within an error of 0.6%. This response was measured using an integrating sphere and taking images over a variety of exposure times at room temperature. The linearity was measured again inflight by observing the FF lamp through the v-band filter at exposure times 0.131, 0.348, 0.525 and 0.696 s. We took 3 shots every 2 seconds for each exposure time to measure the output stability of FF lamp. We also measured the stray light before and after these observations by turning off the FF lamp in order to examine stray light effects on this test. After being corrected for bias levels and radiator stray light, the bright {[384, 384] ; [415, 415]} and the dark {[384, 480] ; [415, 511]} boxes, shown in **Fig. 3.13**, in each image were averaged. Inflight-test results are shown in **Fig. 3.14**. The DN accumulation rate is sensitive to small departures in linearity. Though the DN counts up to 3100 DN appear linear with exposure time (**Fig 3.14**), small changes in the accumulation rate (shown by the variations along the horizontal lines in **Fig. 3.14**) indicate small deviations in the CCD linearity response. This test shows that strong linearity (less than 1% deviation) was achieved up to 3100 DN, whereas a 10-13% drop in linearity is seen at signal levels of 3600 DN. Thus, exposure times should be selected carefully in order not to exceed this quantitative limit of 3100DN. The decrease of accumulation rate can be fitted by a third-order polynomial function, providing, the empirical relationship between the observed intensity $I_{obs}$ and the ideal intensity $I_0$ of

$$I_{obs} = 1.0073 I_0 - 2.9285 \times 10^{-6} I_0^2 - 3.6434 \times 10^{-10} I_0^3. \qquad (I_{obs} < 3100\ \mathrm{DN}) \qquad (3.8)$$

Using this relationship, the linearity can be corrected within 0.1% error (red symbols in **Fig. 3.14**).

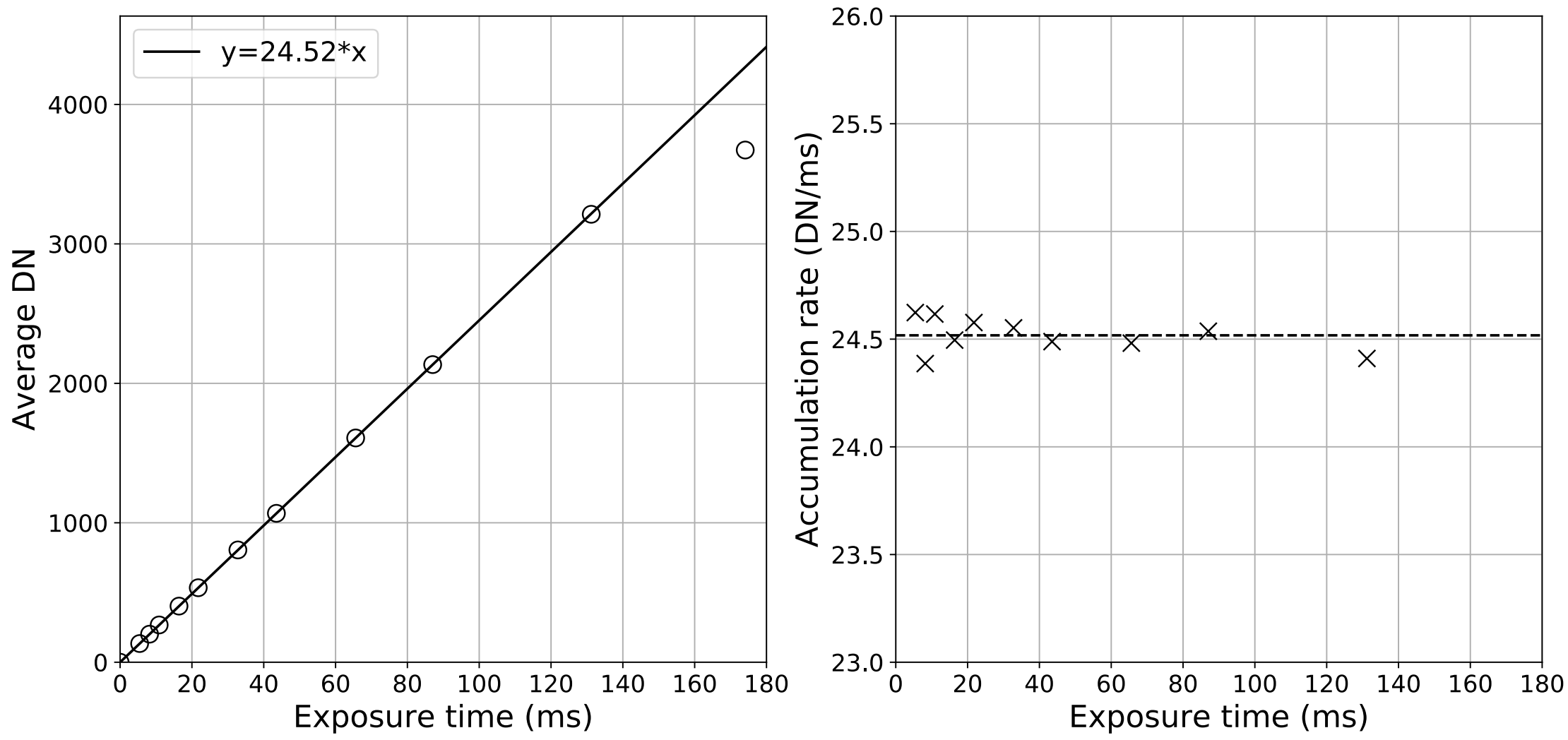

**Figure 3.12.** Preflight ground linearity test showing the linear relationship up to 3200 DN. (left) The averaged intensity from the integrating sphere images with different exposure times is shown by open circles. The solid line shows linear relationship of 24.52 DN/s. (right) The signal accumulation rate over each time duration is shown by crosses. The horizontal line indicates the ideal accumulation rate of 24.52 DN/s with perfect linearity. The deviation from the ideal accumulation rate was within 0.6%.

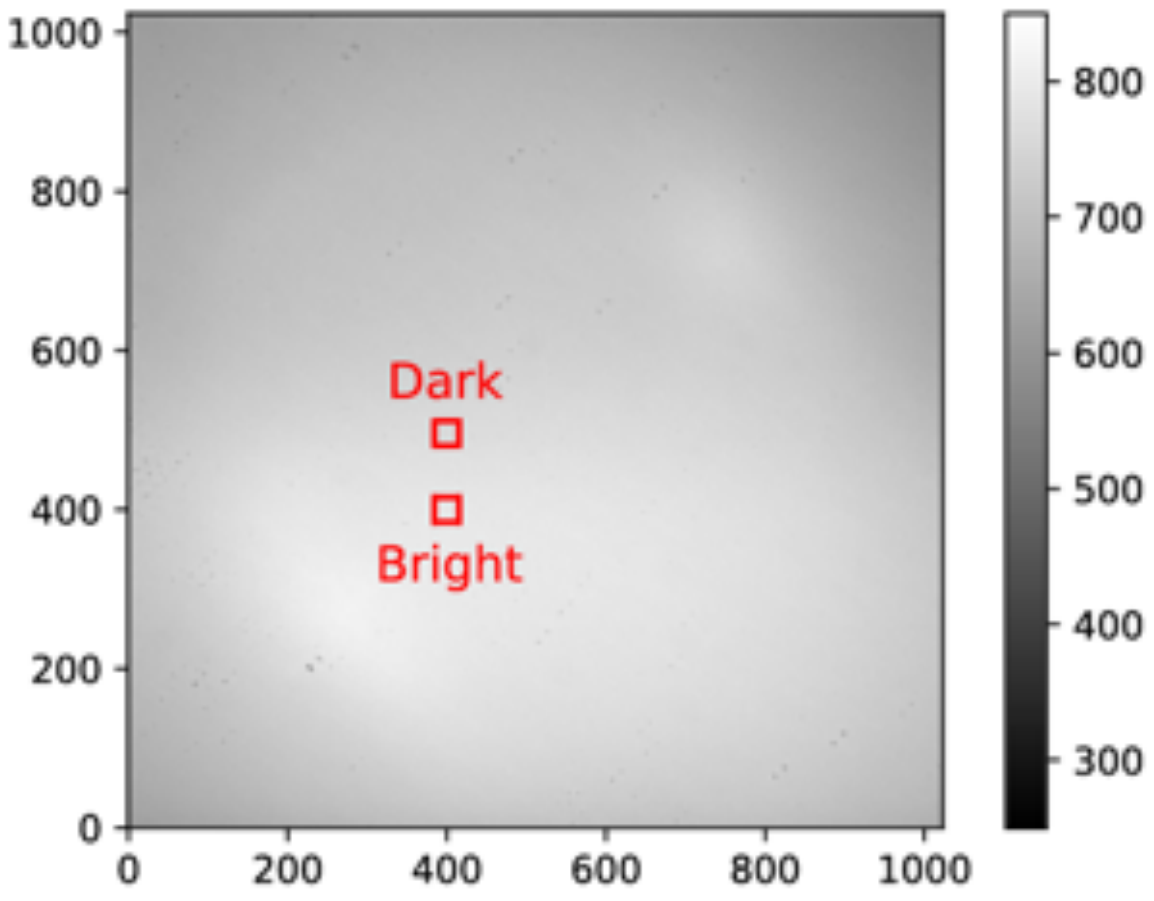

**Figure 3.13.** The red squares show the regions of interest in the inflight linearity test. Two different intensity regions, the bright and dark areas, of the FF lamp images are tested.

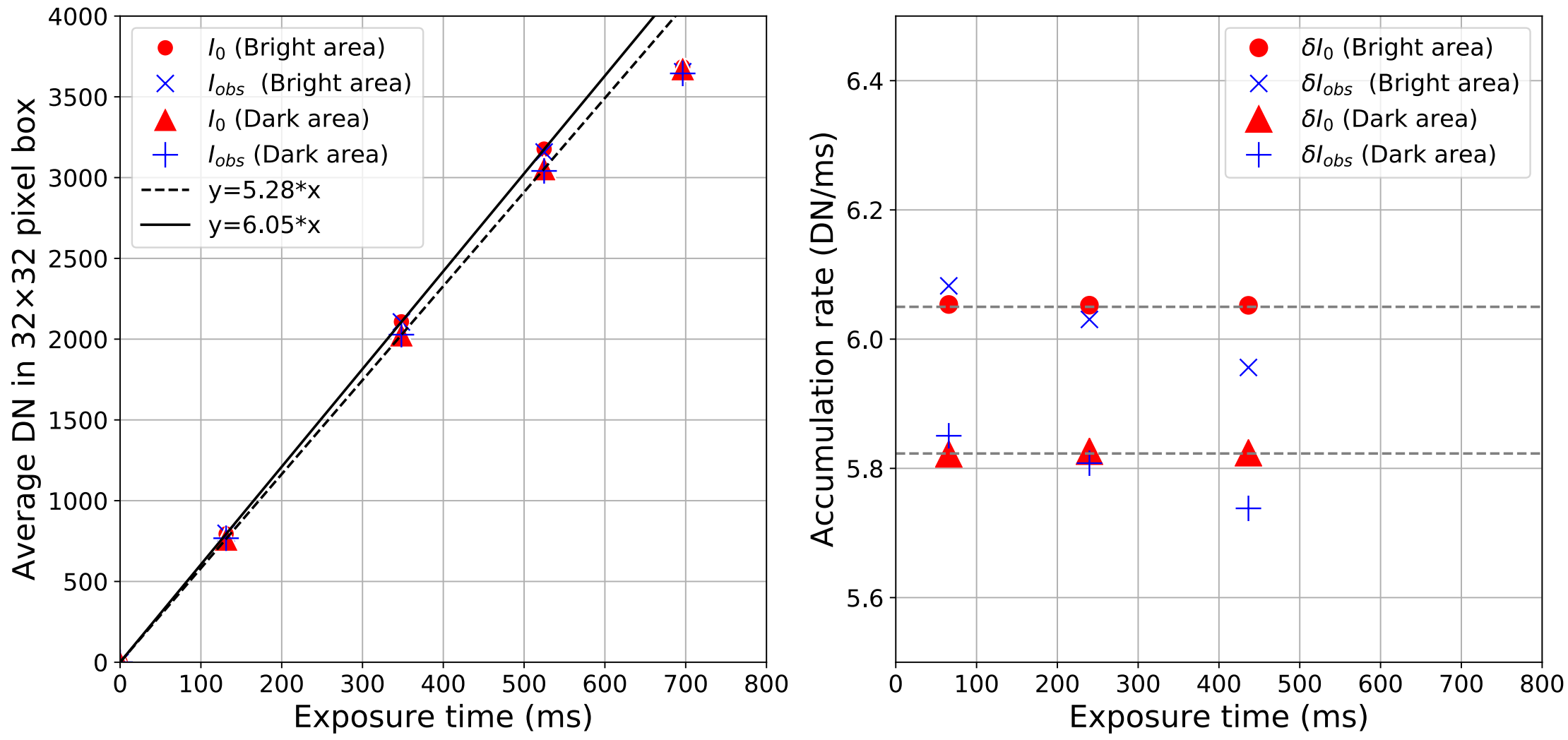

**Figure 3.14.** Inflight linearity test displaying the linear relationship up to 3100 DN. (left) Open circles and triangles are averaged intensity in the bright and dark boxes of the FF lamp images with different exposure times, respectively. (right) Crosses and plusses represent the accumulation rate over each time duration in the bright and dark boxes. Blues are the observed values and reds are the linearity corrected values. Before the linearity correction, both areas show similar trends with decay in accumulation rate for the longer exposure times. The horizontal lines indicate the ideal accumulation rates 6.05 DN/ms for bright area and 5.82 DN/ms for dark area with perfect linearity. The deviations from the ideal accumulation rates are < 1% for before and < 0.1% for after the linearity correction.

## 3.6. Correction of Scattered and Reflected Light in Optics

### 3.6.1. Scattered Light (Broad PSFs)

The presence of scattered light around objects (dim halo) was observed (**Fig. 3.15**) in all ONC filter bands. That is, the sky close to the imaged object was brighter than accountable for from the bias and dark current levels due to the scattering of light from adjacent pixels containing the object image. This is caused by the multiple scattering of photons within the optical system before detection by the CCD. This effect was also observed in AMICA (Asteroid Multiband Imaging Camera) images onboard Hayabusa (Ishiguro, 2014). This scattered light affect has severe effects on the spectral analysis of regions near the object limb since the scattering behavior has a wavelength dependency, which is reported and described in detail by Ishiguro (2014) for the AMICA instrument. They reported that for the AMICA p-band the intensity changed by more than 10% close to object limb in images acquired at the Home Position (altitude of ~7.5 km). Thus, the scattered light correction is especially critical for photometric and spectral analyses. Because the light is distributed isotopically about the object, we can evaluate the scattered light effect as a part of point spread function (PSF).

However, to distinguish this scattering effect from the sharp peak of a PSF, which is usually measured by observing around stars as infinitely small point light sources, we refer to this wide spread attenuation function as the "broad PSF". The sharp peak PSF have been already obtained as part of the preflight calibration by Kameda et al. (2017) and inflight calibration by Suzuki et al. (2018). Therefore, here we investigate the broad PSF based on the method described in Ishiguro (2014) and will show an application to an Earth image obtained during the Earth-Moon swing-by.

*Methodology:*

The total PSF $f_{PSF}$ is a summation of the sharp peak PSF, $f_s$, and the broad PSF, $f_b$:

$$f_{PSF} = f_s + f_b. \tag{3.9}$$

In principle, we don't obtain images without a PSF but what we obtain are blurred images.

$$I_{obs} = I_0 * f_{PSF} = I_0 * (f_s + f_b) \tag{3.10}$$

where $I_0$ is the image without blur and $I_{obs}$ is the observed image. Considering we have to start from blurred images $I_{obs}$, what we determine is the function $f_b{}'$, so as to satisfy the following equation in the sky region where counts from $f_s$ are negligible:

$$I_{obs} * f'_b = I_0 * f_b \tag{3.11}$$

Then we can evaluate blurred counts due to a PSF with $I_{obs}$ and $f_b{}'$. As is described in Ishiguro (2014), we also assumed that the $f_b{}'$ can be expressed as the summation of Gaussian functions:

$$f'_b(r) = \sum_{i=1}^{N} \frac{A_i}{\sqrt{2\pi}\,\sigma_i} \exp\left(-\frac{r^2}{2\sigma_i^2}\right) \tag{3.12}$$

where $r$ [pix] is the distance from a point source, $A_i$ and $\sigma_i$ are constants (N=1 to 6), where $\sigma_i = 2^{i+2}$ ($N \leq 4$), $\sigma_5 = 110$, and $\sigma_6 = 710$. The effect of scattered light can then be corrected as

$$I_0 * f_s = I_{obs} - I_{obs} * f'_b \tag{3.13}$$

Assuming that the $f_s$ is sharp enough to satisfy $f_s = C_s \delta(0,0)$, where $C_s$ is the contribution due to the sharp peak PSF, and

$$I_0 = (I_{obs} - I_{obs} * f'_b)/C_s \text{ where} \tag{3.14}$$

$$C_s = \int 2\pi r f_s \, dr.$$

Preflight disk object images were obtained through all filters with different sized disks placed in the FOV (**Fig. 3.16**). We determined the $A_i$ Gaussian coefficient values sequentially, using a bisection method from the broader components (beginning with $A_6$) to narrower component (ending with $A_1$). We performed this analysis to derive the coefficients $A_i$ for each filter to reduce the residual signal in sky region farther than 8 pixels from edge of the object to <1% of the object's signal by using **Eq. (3.14)**. Small disk objects in the FOV were useful in deriving $A_i$ for smaller values of $i$ and large disk objects for larger values of $i$. The coefficients we obtained are listed in **Table 3.4** and **Fig. 3.17** shows the broad PSF on AMICA and ONC-T. The contribution of the broad PSF on ONC-T is smaller than that observed within the AMICA image data sets.

*Evaluation:*

The validity of the broad PSF correction is assessed with applications to the Earth images obtained during the Earth-Moon gravity swing-by. The broad PSF corrected images are shown in **Fig. 3.18** with brightness variations across horizontal profiles displayed to show the reduction in scattered light. Comparing the sky close to the object limb, the residual is less than 1% and now the ghost can be recognized more clearly than the scattered light.

The broad PSF correction is also verified by using a saturated Mars image observed on 25 May 2016. In this test, the strategy is that if a point source object is observed with a long exposure time, the broad PSF can be observed more clearly. Mars was imaged almost as the point source (~1.4 pixel in diameter in the FOV). We obtained the images using the p-band filter with two different exposure times, a proper exposure time of 0.044 s and a long exposure time of 178.26 s. The dark current and the radiator stray light corrections were applied before the intensity analysis for the broad PSF. The proper exposure images are used to measure the radiation flux from the object by fitting the observed intensity with a Gaussian function $\mathrm{G}_{mars}$, which is equivalent to the Mars signal convolved by only the sharp peak PSF. Here, the Full-Maximum-Half-Width (FMHW) of the Gaussian function is 1.53, which is consistent with the sharp peak PSF obtained in Kameda et al. (2017) and Suzuki et al. (2018). The DN count of the long exposure image was estimated by multiplying by the ratio of two exposure times:

$$G_{long} = \left(\frac{t_{long}}{t_{prop}}\right) G_{prop} \qquad (3.15).$$

Thus, the scattered light around Mars is estimated by $G_{long} * f_b{}'$. **Figure 3.19** compares the observed scattered light with the long exposure and the estimation of our broad PSF. Thus, our broad PSF obtained based on preflight images are consistent with the observed scattered light from a point source.

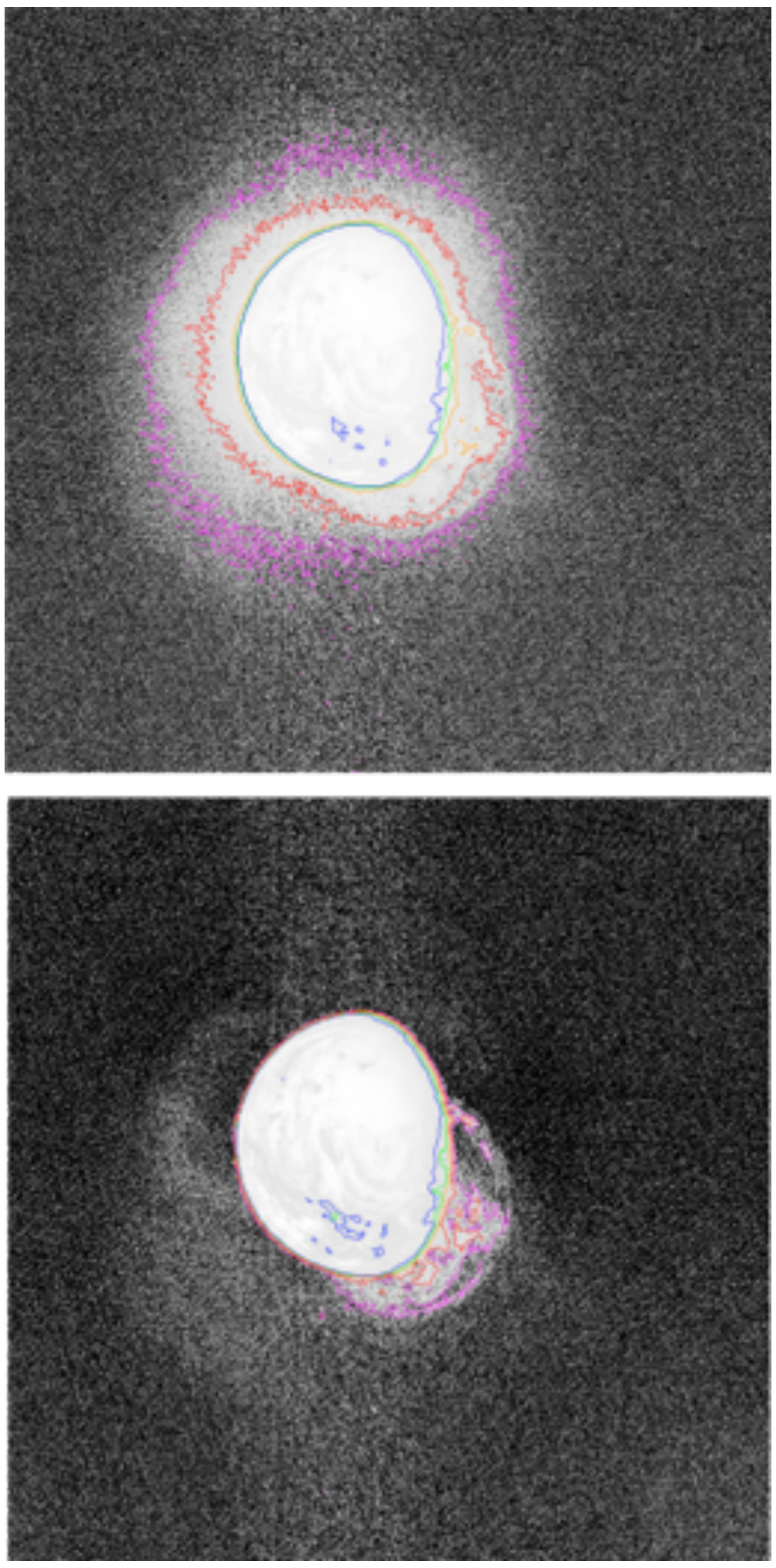

**Figure 3.15.** An Earth image (hyb2_onc_20151204_040959_tpf_l2a.fit) before (top) and after (bottom) the scattered light correction shown by histogram equalization, which enhances contrast. The contour lines correspond to 0.5, 1, 2, 4, 8% of the disk intensity (magenta: 0.5%, red: 1%, orange: 2%, green: 4%, blue: 8%).

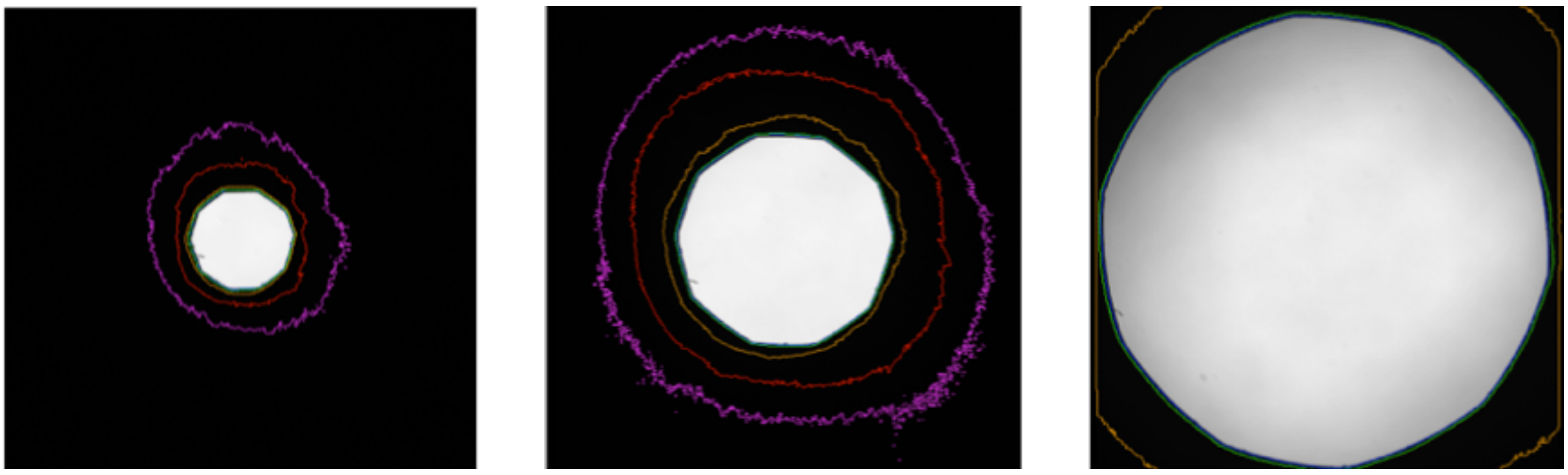

**Figure 3.16.** ONC-T preflight images (p-band) used for fitting the broad PSF. The size of objects in the FOV were 225, 468, and 979 pixels in diameter from left to right. The contour lines correspond to 0.5, 1, 2, 4, and 8% of the disk intensity (magenta: 0.5%, red: 1%, orange: 2%, green: 4%, blue: 8%).

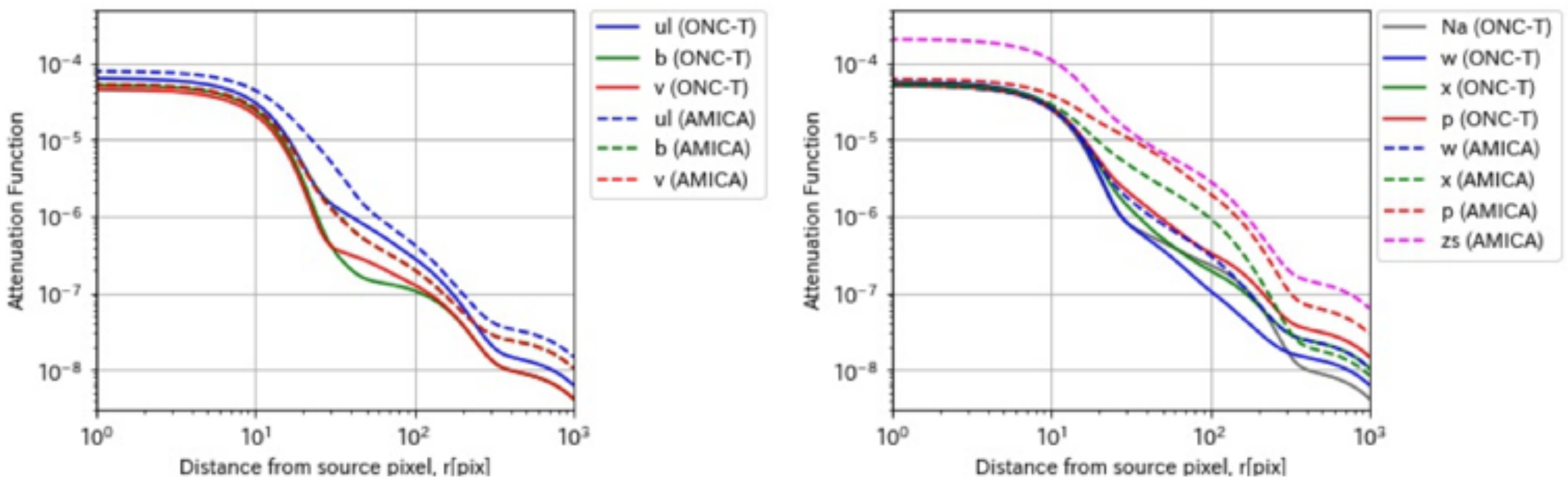

**Figure 3.17.** Comparison of the broad PSF on ONC-T (solid lines) and AMICA (dotted lines).

**Table 3.4.** The broad PSF coefficients $A_i$ ($10^{-4}$) and the broad PSF contribution $\int 2\pi r f'_b\, dr$ on ONC-T and AMICA by Ishiguro (2014). Note that the contribution of broad PSF on AMICA is evaluated by $f'_b = [\int 2\pi r f_s\, dr \,/\, \int 2\pi r f_{PSF}\, dr] \times f_b$.

| ONC-T | $A_1$ | $A_2$ | $A_3$ | $A_4$ | $A_5$ | $A_6$ | $\int 2\pi r f'_b\, dr$ |
|---|---|---|---|---|---|---|---|
| *ul* (0.4 µm) | 12.0 | 0.7 | 0.7 | 0.5 | 0.7 | 0.3 | 0.11 |
| *b* (0.48 µm) | 10.0 | 0.5 | 0.0 | 0.0 | 0.4 | 0.2 | 0.07 |
| *v* (0.55 µm) | 9.0 | 0.0 | 0.2 | 0.1 | 0.4 | 0.2 | 0.07 |
| *Na* (0.59 µm) | 10.0 | 0.3 | 0.4 | 0.0 | 0.3 | 0.2 | 0.08 |
| *w* (0.70 µm) | 11.0 | 0.4 | 0.5 | 0.2 | 0.2 | 0.3 | 0.09 |
| *x* (0.86 µm) | 10.0 | 1.3 | 0.6 | 0.1 | 0.6 | 0.5 | 0.14 |
| *p* (0.95 µm) | 9.0 | 1.5 | 1.2 | 0.1 | 1.1 | 0.7 | 0.19 |

| AMICA | $A_1$ | $A_2$ | $A_3$ | $A_4$ | $A_5$ | $A_6$ | Contribution of $f_b$ |
|---|---|---|---|---|---|---|---|
| *ul* (0.38 µm) | 11.4 | 7.6 | 1.1 | 1.0 | 0.8 | 0.7 | 0.23 |
| *b* (0.43 µm) | 9.7 | 1.5 | 0.3 | 0.4 | 0.4 | 0.5 | 0.13 |
| *v* (0.55 µm) | 9.6 | 1.3 | 0.3 | 0.4 | 0.4 | 0.5 | 0.13 |
| *w* (0.70 µm) | 9.2 | 1.4 | 0.6 | 0.7 | 0.6 | 0.5 | 0.15 |
| *x* (0.86 µm) | 7.9 | 3.1 | 1.8 | 2.4 | 1.9 | 0.4 | 0.20 |
| *p* (0.96 µm) | 7.9 | 4.0 | 6.6 | 3.2 | 5.1 | 1.4 | 0.53 |
| *zs* (1.01 µm) | 33.5 | 10.7 | 4.0 | 6.0 | 6.4 | 3.0 | 0.95 |

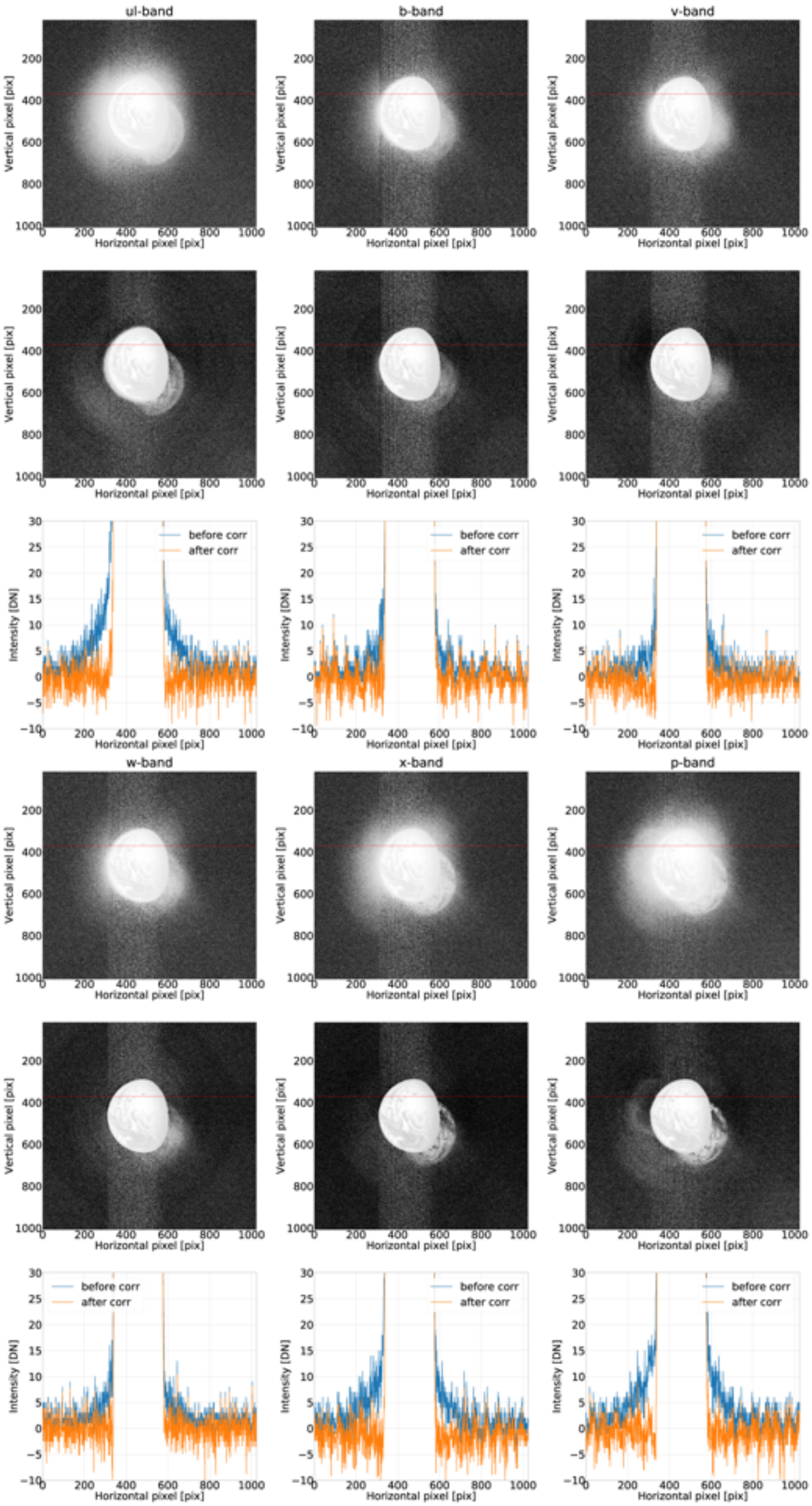

**Figure 3.18.** Enhanced Earth images before (the first row images) and after (the second row images) the broad PSF scattered light correction. Horizontal profiles along red lines of them are shown in the third row.

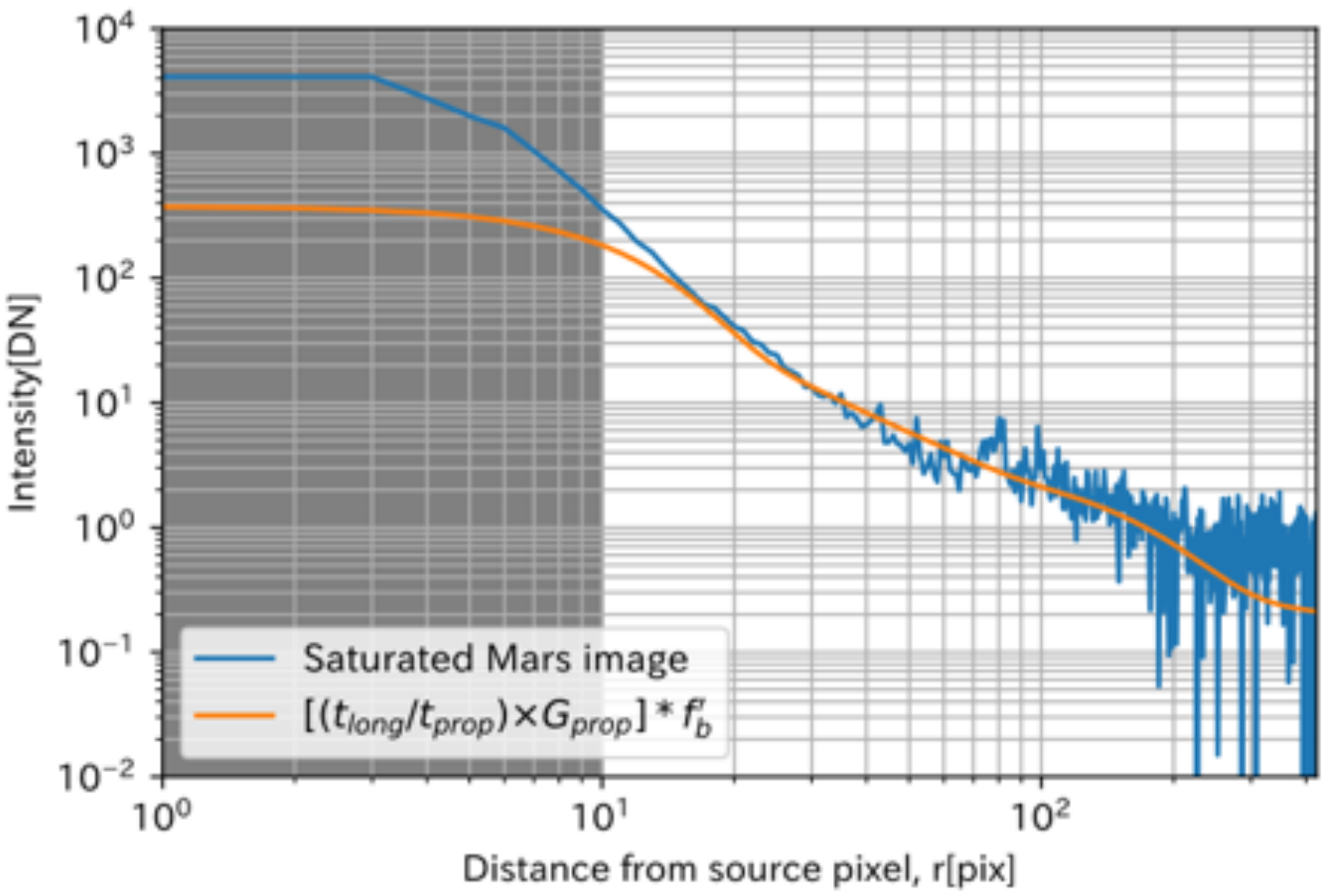

**Figure 3.19.** Observed scattered light as a function of distance from Mars, comparing with our broad PSF obtained preflight. Note that close part from Mars is affected by the sharp-peak PSF (gray hatch).

### 3.6.2. Ghost Effect

There is noise within the ONC system due to reflections between optics, such as filters, lens, and the CCD, that create an effect captured within the image hereafter referred to as the "ghost effect". Because the ONC optics are axi-symmetric, a ghost image usually appears at a symmetrical location about the optical axis. Here, we investigated the intensity of the ghost effect and the position of the ghost image within the image frame. **Figure 3.20** displays examples of ghosts in the images obtained during the Earth-Moon swing-by. The degree or magnitude of the ghost effect is measured to be weaker than 0.1 % of the intensity of the observed object for all filters, however, there is a wavelength dependency to the magnitude of the effect. For example, v-, Na-, and w-filters do not have ghost effects larger than other CCD noise sources, such as MEI noise, dark noise, and shot noise, while the ul-, x-, p-, and wide-filters all suffer from a small (<0.1%) but detectable ghost effect. From the Earth-Moon swing-by images, we determine the reflection coefficients (the intensity ratio between the ghost and object) for the filters as 0.0033, 0.0012, <0.0001, <0.0001, <0.0001, 0.0045, 0.0065, and 0.0028 from short to long wavelength and the wide filter, respectively. The light reflects symmetrically about the sub-pixel position at ~ (496.5, 501). Using these values, examples of ghost removal are shown in **Fig. 3.21**. It should be noted that when the object is far from the center of FOV, this correction method is insufficient (**Fig. 3.22**). The ghost effect on images that include the object at a corner is much weaker than that on images which include the object at the center. This may indicate that the intensity of the ghost effect is a function of the distance from the optical axis. Thus, the intensity assessment is valid for the center region of FOV, where Ryugu will be located on home position observations. Moreover, when the object fills the entire field of view, such as close approaches to the asteroid <10 km, this ghost removal method may work only in the area < 500 pix radius from the center.

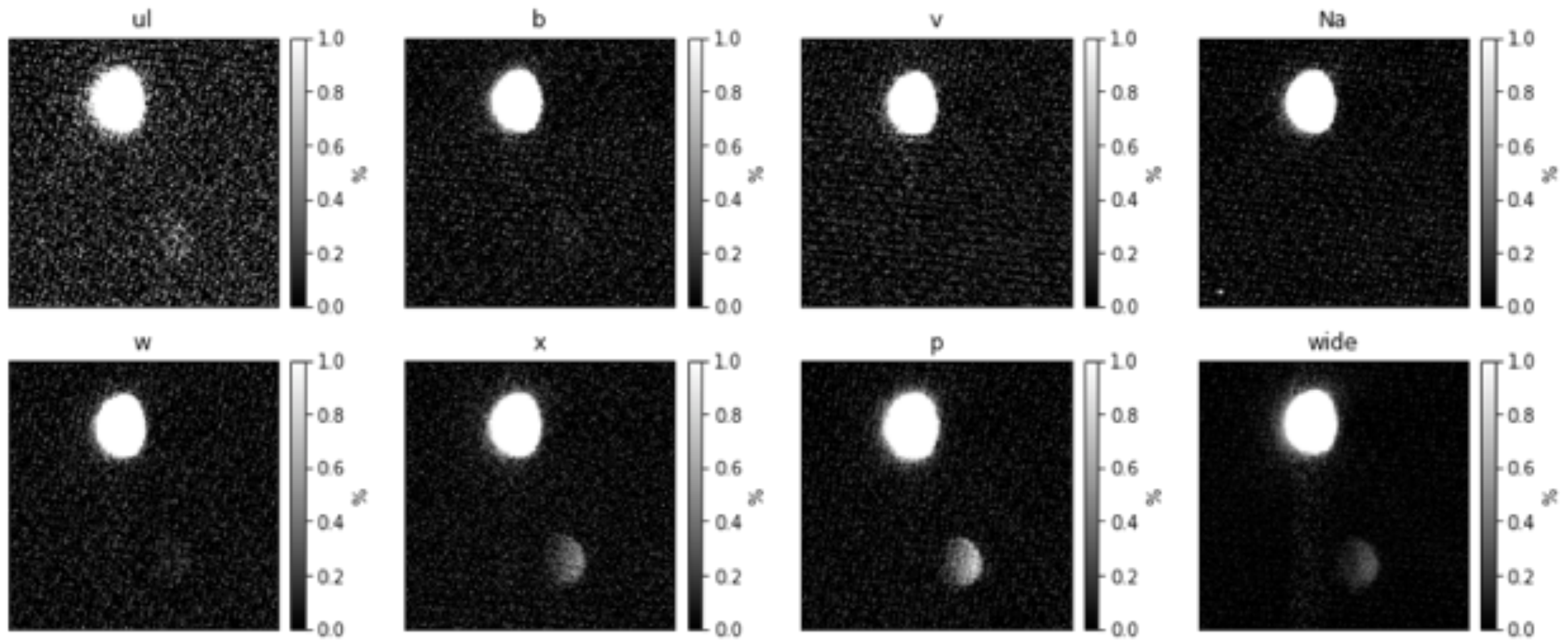

**Figure 3.20.** Ghost effect captured in images of the Moon with multiband filters (hyb2_onc_20151205_115044_tuf_l2a.fit, hyb2_onc_20151205_115024_tbf_l2a.fit, hyb2_onc_20151205_114908_tvf_l2a.fit, hyb2_onc_20151205_114952_tnf_l2a.fit, hyb2_onc_20151205_114920_twf_l2a.fit, hyb2_onc_20151205_114940_txf_l2a.fit, hyb2_onc_20151205_115012_tpf_l2a.fit, hyb2_onc_20151205_115056_tif_l2a.fit). The centers of images, 200x200 pix, are shown. Ghosts of the Moon appear point symmetrically at the optical axis (at lower left in this case). Color bars indicate the signal normalized by the disk average signal of the Moon.

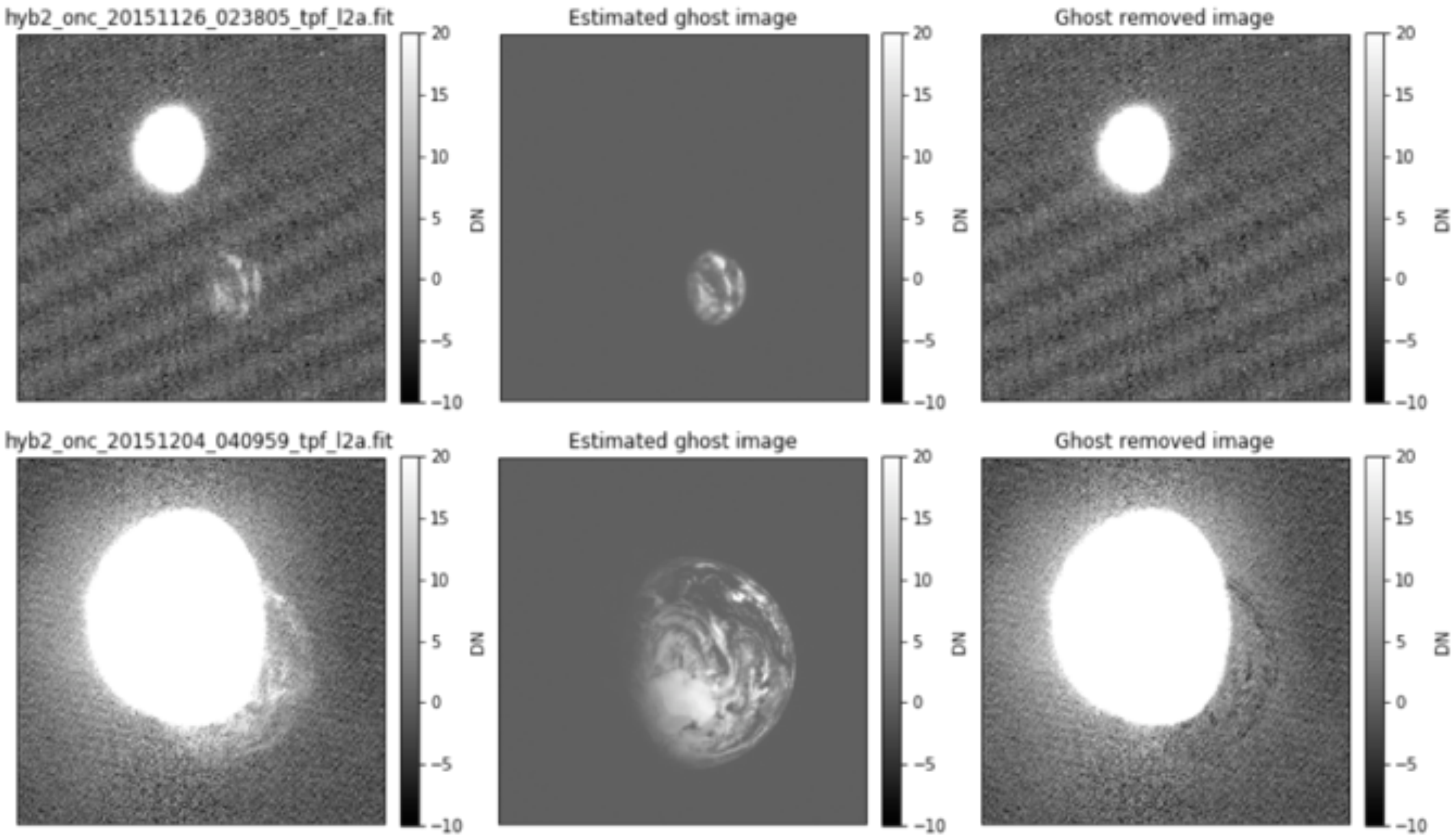

**Figure 3.21.** Two examples of ghost removal for Earth images taken through the p-filter (hyb2_onc_20151126_023805_tpf_l2a.fit, hyb2_onc_20151204_040959_tpf_l2a.fit). The centers of the FOV, 200x200 pix and 600x600 pix respectively, are shown.

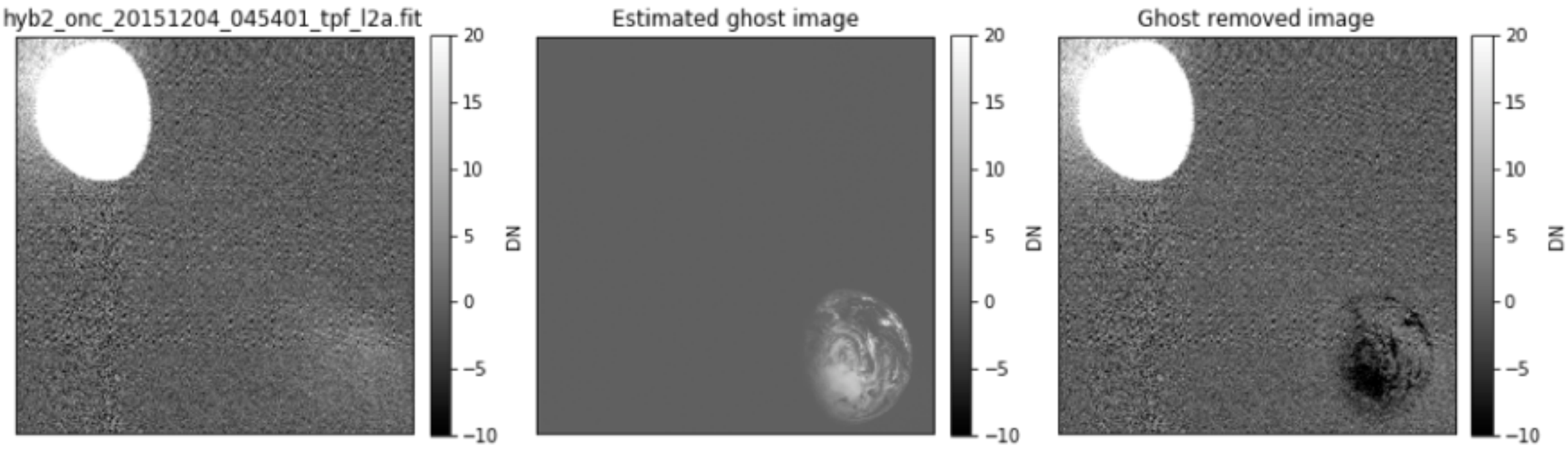

**Figure 3.22.** Over-correction of an image with the object observed within the corner of the image frame (hyb2_onc_20151204_045401_tpf_l2p.fit). The ghost effect is much weaker for the object image at a corner.

### 3.7. Radiator Stray Light Removal

The radiator stray light contamination is caused by the reflection of sunlight by a part of the ONC-T radiator which was reported in Suzuki et al. (2018). **Figure 3.23** summarizes the stray light intensity with respect to the spacecraft attitude, which was continuously monitored with onboard attitude sensors (i. e., star trackers). There are two ways to describe the spacecraft attitude with respect to the Sun, which are ($X_{PNL}$, $Y_{PNL}$) and ($\phi, \gamma$). The definition of attitudes and spacecraft coordinates are described in **Appendix C**. The intensity of the stray light was evaluated at the highest intensity part of the image frame, in the same way as Suzuki et al. (2018), and we normalized the intensity by the distance from the Sun. The definition of the spacecraft attitude was referred to Sec. 8.1 in Suzuki et al. (2018). The reference attitude of the spacecraft is that the antenna is pointing to the Earth. Thereby the value of $\phi$ corresponds to the phase angle, i.e., the Earth-Ryugu-Sun angle, and is automatically determined by the position of the Earth, Ryugu and the Sun at the observation time. We expect that $\phi$ will be change from -20° to -10° before the first touchdown. The trend of the stray light did not change during the 3.5 years of the cruise phase. That is, the strongest part is always in the (+H, +V) corner, and the stray light component gradually weakens towards the opposite (-H, -V) corner (see **Fig. 21** in Suzuki et al. (2018)). Because attitudes with $\gamma < -7°$ do not have strong stray light component, we are planning to obtain images of the target asteroid at those attitudes with negligible stray light, <3.8 DN/s at 1 AU, for global mappings. On the other hand, for the proximity observations, such as the lander releases, the Small Carry-on Impactor (SCI) crater scan observations, and touchdowns, when the observation attitudes allow stray light at the $\sim$1000 DN/s level may contaminate the images. Here, we model the stray light as a function of spacecraft attitude towards the Sun to provide a method for removing this component from the images.

Based on inflight observations, the spatial patterns of the radiator stray light does not change drastically with the spacecraft attitude. Moreover, the stray light component is reproducible for the same spacecraft attitude when the intensity is normalized by the square of the heliocentric distance of the spacecraft (**Fig. 3.23**). Thus, we separately modeled the stray light pattern with intensity variation (modeled at the center of FOV) and the spatial distribution.

Investigations of a strong stray light component for possible attitudes during the rendezvous phase was conducted on 17, 19, and 21 October 2017, especially with the spacecraft twisting angle of $\gamma < -5°$. First, we empirically modeled the intensity of the stray light at the center of FOV with a polynomial function:

$$I_{sl} = (-0.0879\phi^3 - 5.616\phi^2 - 119.4\phi - 603.8)(-0.0037\gamma^2 + 0.113\gamma + 0.956) \text{ [DN/s]}. \quad (3.16)$$

The comparison of the stray light intensity model and the observed intensity is shown in **Fig. 3.24**. By dividing the stray light images by the central intensity, the spatial distributions can be examined. We bin those normalized patterns into 8 × 8 (average of 64 × 64 pixels) images to smooth out the noise. After normalization and smoothing, we applied principal component analysis to those images and extracted the average pattern and the principal components of the stray light patterns. The principal component analysis extracts the orthogonal components with highest variance to lower variance. The component with highest variance, the first principal component, may explain the stray light pattern change most. The components with small variance could be noise. Thus, the components with high variance are enough to express the variation in stray light pattern. In our analysis, The first principal component (PC1) score contains 81% of variance and the second principal component (PC2) score contains 18 % of variance. Thus, 99% most of the spatial distribution variation can be explained by the PC1 and PC2 scores. The stray light patterns $M_{\mathrm{sl}}$ can be decomposed linearly by the principal components as

$$M_{\mathrm{sl}} = M_{\mathrm{ave}} + a_{\mathrm{PC1}} M_{\mathrm{PC1}} + a_{\mathrm{PC2}} M_{\mathrm{PC2}}, \tag{3.17}$$

Where $M_{\mathrm{ave}}, M_{\mathrm{PC1}},$ and $M_{\mathrm{PC2}}$ are the average spatial distribution image, the first principal component image, and the second principal component image which are shown in **Fig. 3,25**. $a_{\mathrm{PC1}}$ and $a_{\mathrm{PC2}}$ are the PC1 and PC2 scores, respectively. Moreover, we found that the PC1 score for stray light correlate with the value of $\phi$ (**Fig. 3.26**). Based on the intensity model and the spatial distribution model, we can estimate the stray light at an arbitral attitude in the range of $-30° \leq \phi \leq -10°$ and $-5° \leq \gamma \leq 10°$. Applying this removal method to the observed images, the residual of the stray light averaged over the FOV is smaller than 25 DN/s/pixel, corresponding to 0.25 % of the intensity of the asteroid disk. Note that, the edge of FOV tends to have more error than the center with this method.

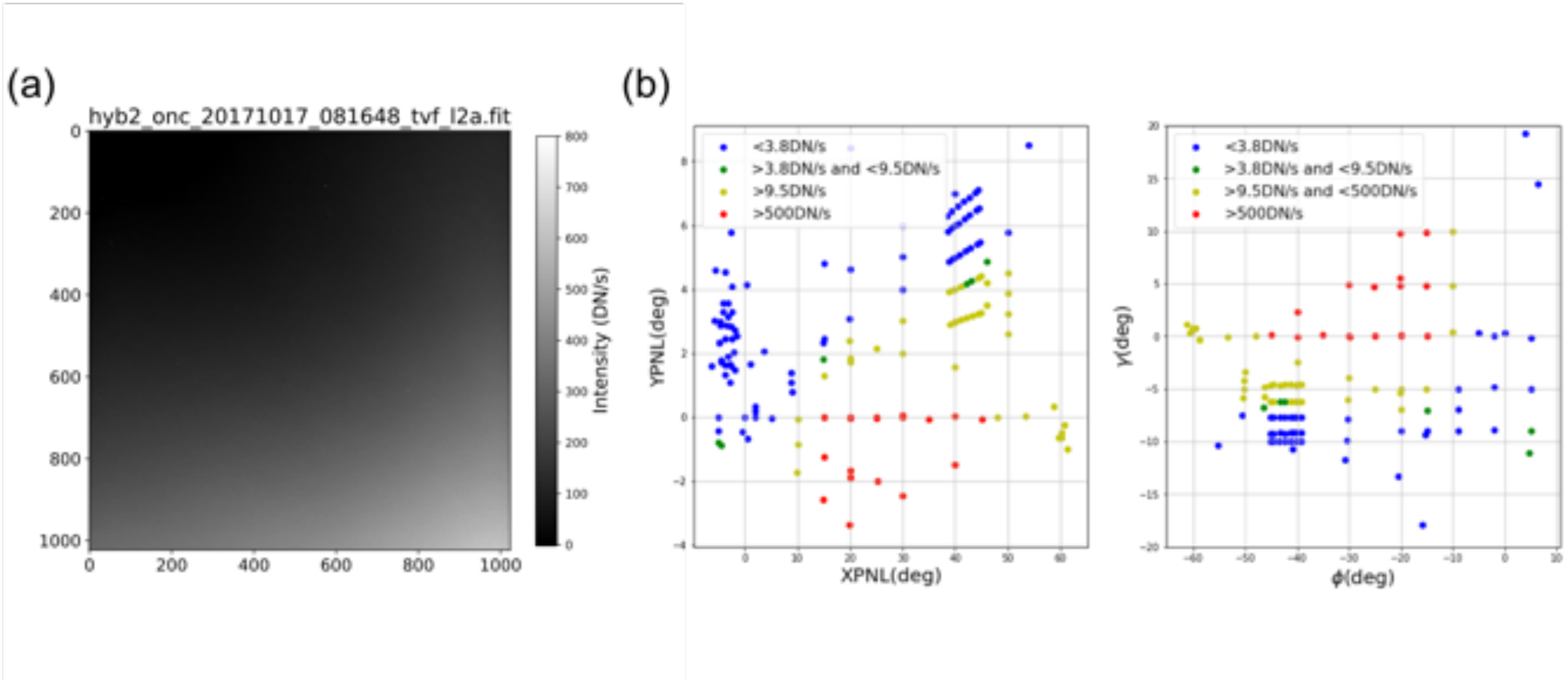

**Figure 3.23.** (a) Typical pattern of radiator stray light. (b) The intensity classification of radiator stray light. Radiator stray light occurs as a function of the angle of the sun to the $X_{PNL}$ and $Y_{PNL}$ (left), and the solar phase angle ($\phi$) and the twisting angle around $Z_{SC}$ axis ($\gamma$) (right). During the rendezvous phase, we are planning to twist the spacecraft to $\gamma < -7°$ as much as possible to avoid strong stray light contributions.

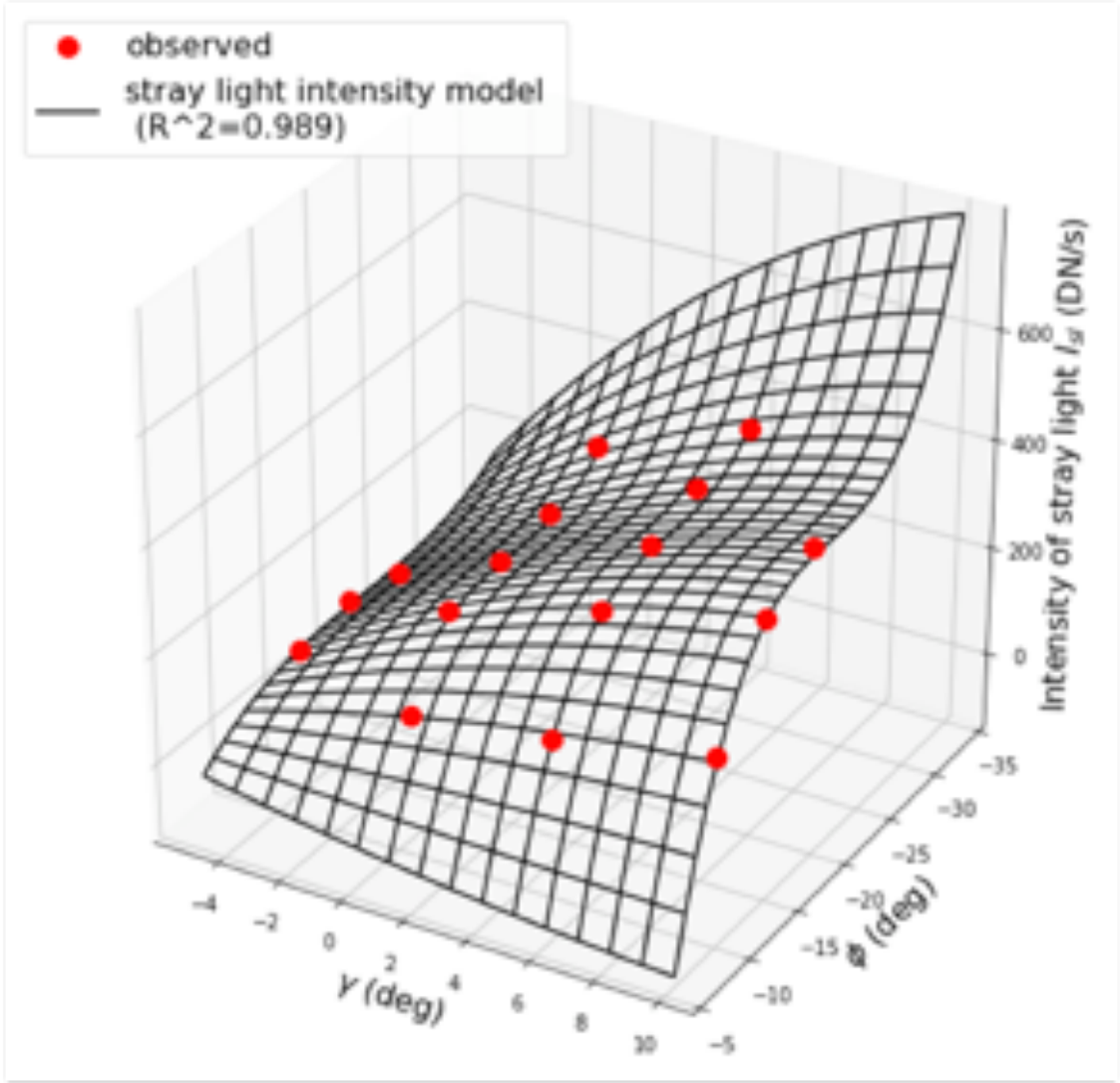

**Figure 3.24.** Intensity model (black mesh) of stray light as a function of $-30° \leq \phi \leq -10°$ and $-5° \leq \gamma \leq 10°$, comparing with observations (red).

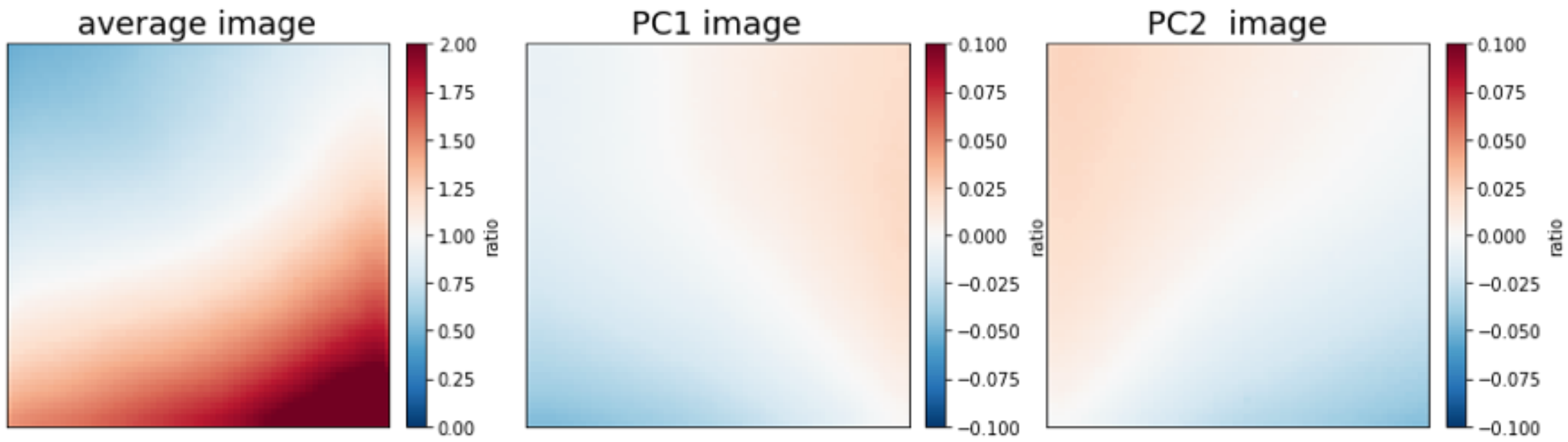

**Figure 3.25.** The average and the principal components of the stray-light spatial distributions.

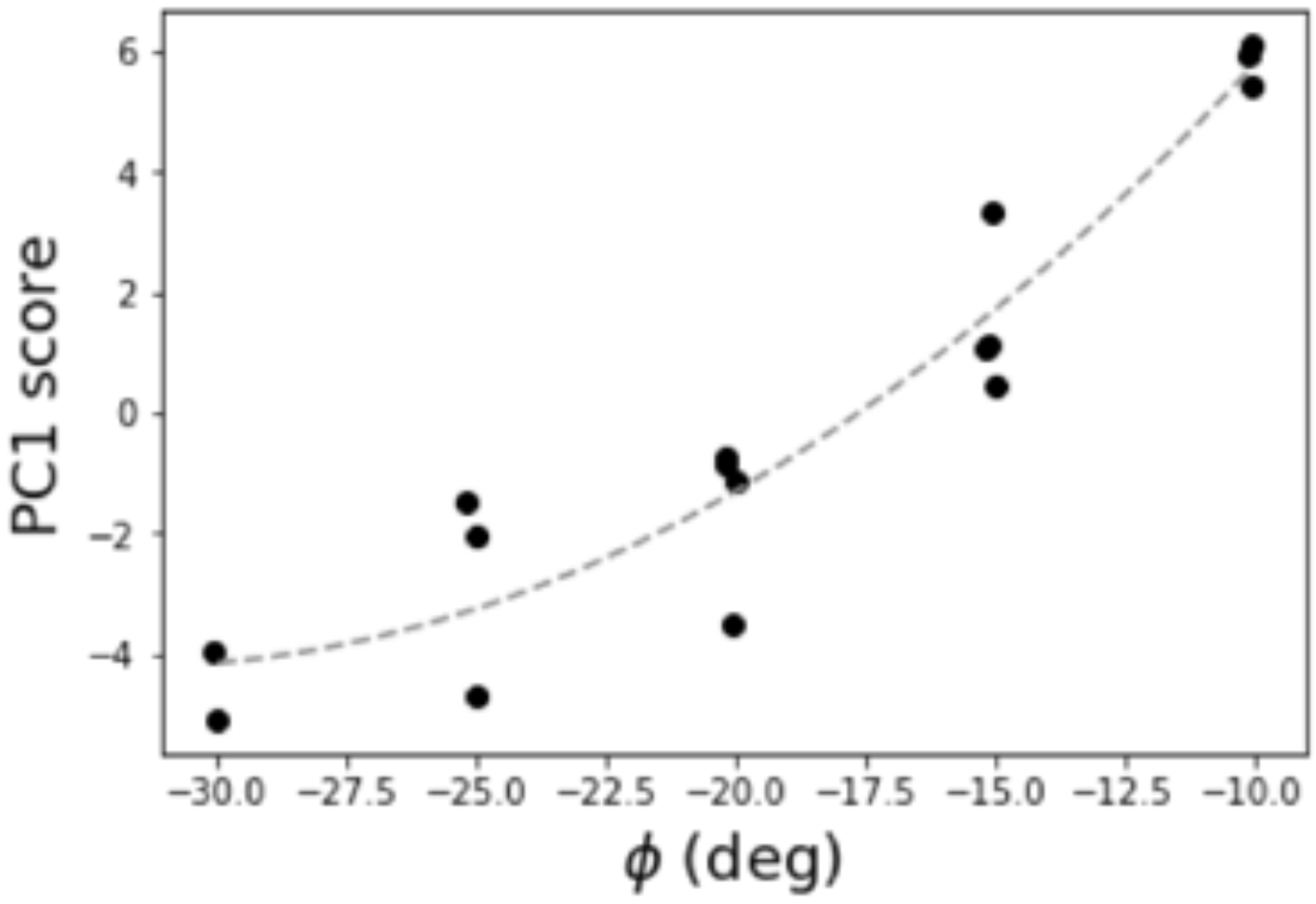

**Figure 3.26.** The PC1 scores and the phase angle $\phi$ of the stray-light spatial distributions.

### 3.8. Flat-Field Correction

Flat-field correction is one of the key calibrations needed for deriving reliable brightness properties at any position within an image frame. After deriving the preflight flat-field correction using the integrating sphere measurements, the front hood position of ONC-T was changed slightly during disassembly and assembly at the launching site. Thus, all ONC-T bands are expected to have different sensitivity distributions (flat-fields) inflight than those obtained preflight using the integrating sphere images (Kameda et al., 2017; Suzuki et al., 2018). In fact, ~10% differences in the lunar brightness were confirmed between images with the Moon located in the center of an image and those with the Moon located in the corner for all filter bands (Suzuki et al., 2018). The lunar images were acquired during Hayabusa2's Earth-Moon flyby on 3 December 2015, which provided the gravity assist to reach its target, 162173 Ryugu. The lunar images were corrected with an additional flat-field correction derived from measurements taken with a portable flat light source conducted after the hood position was slightly changed. By applying the additional flat-field correction, the discrepancy between lunar brightness values was reduced to 2% in all bands except for ul-band, while the relatively large discrepancies (~3%) still remain in the ul-band (Suzuki et al., 2018).

For evaluating the validity of the flat-field corrections and investigating the detailed structure of the current sensitivity distribution in ul-band, a star observation campaign was conducted from 12 to 14 October 2017. During this campaign four bright stars were observed under five different attitude conditions of the Hayabusa2 spacecraft to place the stars within different locations within the ONC-T FOV. The four bright stars observed were Phi, Sigma, Tau, and Zeta Sagittarii whose locations are close enough to each other to be observable at the same time within the FOV. Star locations in ONC-T image frame during the observation campaign cover almost the whole FOV of the image frame (**Fig 3.27(a)**). Star observations were conducted with exposure times of 16.7 and 22.8 seconds for v-band, and 16.7 and 67.2 seconds for ul-band, to obtain sufficient counts from the different star brightness to measure the CCD sensitivity.

The expected flux from a star $J_{\mathrm{star},n}$ through $n$-filter can be determined from

$$J_{\mathrm{star},n} = \frac{\int \lambda J_{\mathrm{star}}(\lambda)\Phi_n(\lambda)d\lambda}{\int \lambda \Phi_n(\lambda)d\lambda}, \tag{3.18}$$

where $J_{\mathrm{star}}$ [W/m$^2$/μm] is the irradiance from the star. This value should be proportional to the total digital count rate of the observed star obtained by ONC-T:

$$\frac{(\sum_{i,j}\frac{I_{i,j}}{f_{i,j}})/t}{\int \lambda J_{\mathrm{star}}(\lambda)\Phi_n(\lambda)d\lambda} = C\ (constant), \tag{3.19}$$

where $I_{i,j}$ is the DN counts and $f_{i,j}$ is a coefficient of flat field correction at (i, j) pixel. This indicates that left hand side of **Eq. (3.19)** should have the same value for any star and any exposure time, if flat field correction is perfect at all locations. In other words, if a stellar observation has a smaller value for the right hand side of **Eq. (3.19)** than other stars, the flat field correction at the position is insufficient, and thus the corrected sensor sensitivity is smaller than expected.

To evaluate sensor sensitivities at the star positions, we performed flat field correction with those proposed in Suzuki et al (2018) after performing dark signal subtraction and hot-pixel reduction. Then we measured the total DN counts for each star by integrating the DN counts in a 20-pixel radius around the brightness center of the star. The only exception to this was for Tau Sagittarii located at (H, V)=(86.7, 619.7) with an exposure time of 67.2 sec. In this case we used a 15-pixel radius for integrating the DN counts to avoid including huge counts from a cosmic ray hit located nearby the star in the FOV. In these calculations, we used the star flux from Alekseeva et al. (1996), which is available in the database VisieR (http://vizier.u-strasbg.fr/viz-bin/VizieR).

**Table 3.5** summarizes the values of normalized $C$ which are calculated from as $C$ values calculated from **Eq. (3.19)** in each observation are normalized to the mean $C$ value for each filter. Examination of the standard deviations of $C$ values shows that the v-band image frame achieves uniform sensitivity with a deviation < 2% with no significant outliers. Considering that the standard deviation includes errors in star irradiance and in noise reduction processes, the 2% deviation can be considered as an upper limit of the sensor's uniformity in sensitivity. In addition, gradual sensitivity variations across the image can be observed in the v-band FOV (**Fig. 3.27(b)**), indicating that additional flat field corrections could be performed, though it is hard to derive a high-accuracy flat-field correction only from this star observation campaign due to the sparseness of the star positions across the FOV.

The ul-band image frame has a larger sensitivity deviation (3.9%) across the image scene compared to the sensitivity deviation in the v-band. A gradual sensitivity variation across the FOV (low sensitivity around a lower-left image corner and high sensitivity in upper region) can also be observed in ul-band (**Fig. 3.27(c)**). Especially at positions of (H, V) = (85.8, 903.8) and (38.6, 958.1), $C$ values (denoted by underlines) are 6-9 % smaller than a mean value, indicating a decline in sensor sensitivity uniformity near the image edges. From these results, the brightness inconsistency seen in the lunar ul-band observation, reported by Suzuki et al. (2018), could be explained by the worse uniformity near the image corners. This could affect mapping and photometric analysis especially in the peripheral regions in the FOV. This decreased uniformity in the ul-band flat-field may be fundamentally caused by the preflight measurements. During the preflight integrating sphere measurements stray light contributions from the gap between the filter wheel and baffle were detected (see "FW roundabout" stray light description in Suzuki et al., 2018). The FW roundabout stray light has a stronger contribution at the shorter wavelengths, thus preferentially contaminating the ul-band images. It should be noted that if we focus only on the stars observed inside of the radius of 400 pix from the image center, where the standard deviation of $C$ values is 2.2% in the ul-band, indicating we may have sufficient sensitivity uniformity when we use only the center of FOV.

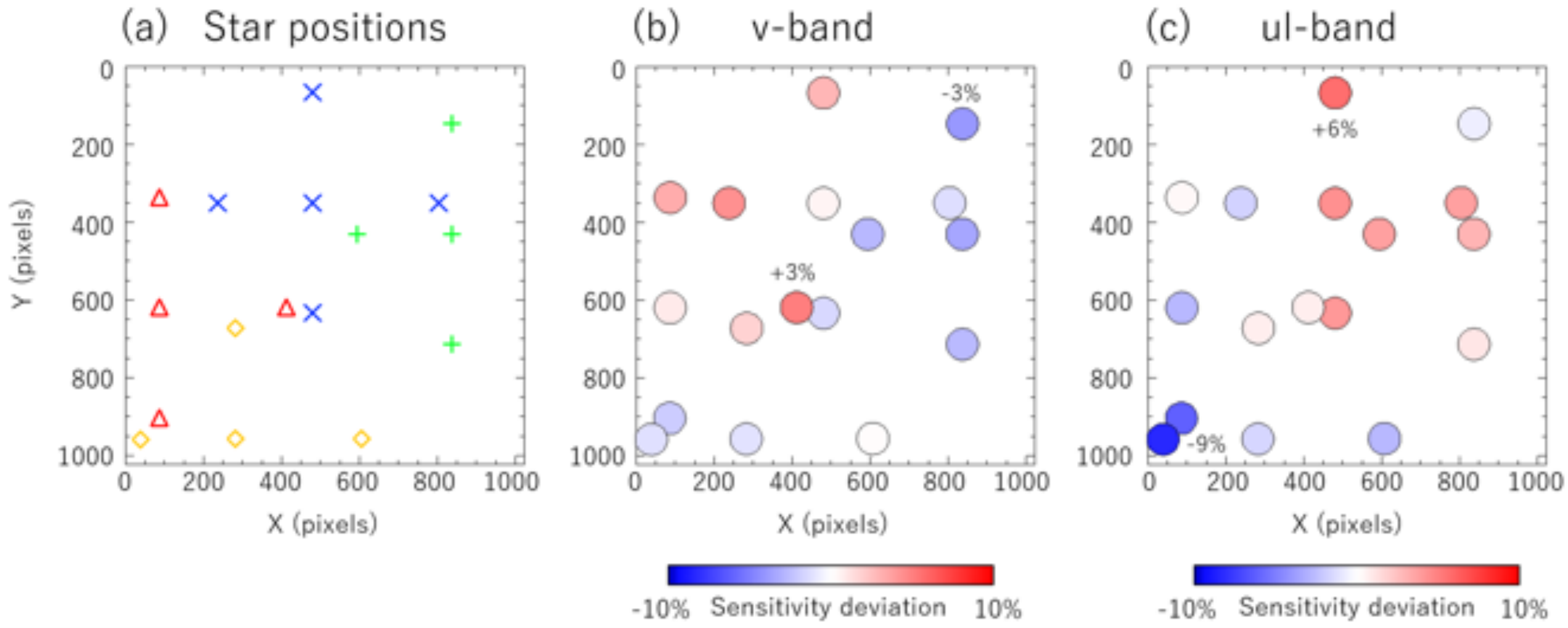

**Figure 3.27.** (a) Star positions obtained during the star observation campaign from 12 to 14 October 2017. Blue crosses, green crosses, red triangles, and orange diamonds indicate observed positions of Sigma, Phi, Tau, and Zeta Sagittarii, respectively. (b) Deviations of the measured sensor sensitivities at star positions from their mean value for v-band after performing the flat-field correction proposed in Suzuki et al (2018). (c) Same for ul-band.

**Table 3.5.** Obtained $C$ values from **Eq. (3.19)** and their standard deviations from star observations for ul- and v-bands. Each $C$ value is normalized by an averaged $C$ value of all stars together for each band.

| Star name | Position (pixel) | | Normalized $C$ (= normalized sensitivity) | | | |
|---|---|---|---|---|---|---|
| | H | V | v-band | | ul-band | |
| | | | 16.8 sec | 22.3 sec | 16.7 sec | 67.2 sec |
| Phi | 836.7 | 430.3 | 0.968 | 0.997 | 1.043 | 1.019 |
| | 837.6 | 146.3 | 0.995 | 0.965 | 0.985 | 1.003 |
| | 837.1 | 713.5 | 0.974 | 1.000 | 1.019 | 1.003 |
| | 594.3 | 430.2 | 0.989 | 0.984 | 1.046 | 1.031 |
| Sigma | 480.3 | 350.8 | 1.004 | 1.002 | 1.045 | (saturated) |
| | 480.5 | 66.8 | 1.016 | 1.014 | 1.058 | (saturated) |
| | 480.5 | 633.4 | 0.991 | 0.995 | 1.042 | (saturated) |
| | 238.0 | 350.4 | 1.025 | 1.020 | 0.983 | (saturated) |
| | 804.5 | 350.2 | 0.994 | 0.994 | 1.039 | (saturated) |
| Tau | 86.7 | 619.7 | 0.990 | 1.020 | (too low) | 0.973 |
| | 87.0 | 336.4 | 1.023 | 1.010 | (too low) | 1.004 |
| | 85.8 | 903.8 | 0.995 | 0.985 | (too low) | 0.938 |
| | 411.8 | 618.9 | 1.022 | 1.030 | (too low) | 1.008 |
| Zeta | 282.4 | 957.0 | 0.985 | 1.005 | 0.999 | 0.971 |
| | 283.3 | 672.4 | 1.003 | 1.016 | 1.015 | 1.002 |
| | 38.6 | 958.1 | 0.993 | 0.996 | 0.921 | 0.911 |
| | 607.4 | 956.2 | 0.993 | 1.009 | 0.972 | 0.974 |
| Standard deviation (%) | | | 1.6 | | 3.9 | |

### 3.9. Radiometric Calibration

In this section, the conversion factor from count per seconds (DN/s) to radiance (W/m²/μm/sr), referred to as "radiometric calibration", is discussed. First, the temporal variation in sensitivity is discussed in **Sec. 3.9.1**. Suzuki et al. (2018) used the sensitivity corrected by the typical temperature dependence provided by the manufacturer (E2V). However, the temperature dependence had not been evaluated inflight at that time. Thus, we second measured the effect of CCD temperature on the sensitivity based on inflight flux measurements taken of Jupiter (**Sec. 3.9.2**). The objective in this section is to update the sensitivity conversion based on inflight measurements. We evaluate the sensitivity conversion in two ways; 1) based on the hardware specifications of components, such as CCD, filters, and lens of the ONC-T (**Sec. 3.9.3**), and 2) based on star observations (**Sec. 3.9.4**). Finally, the updated conversion to radiance is validated based on ONC observations of the Moon, Jupiter and Saturn as discussed in **Sec. 3.9.5**.

#### 3.9.1. Time-Dependent Sensitivity Change

The FF lamp is designed to monitor the sensitivity (absolute value and spatial pattern) of the CCD during the mission. **Table 3.6** summarizes the FF lamp observations taken during the cruise phase. FF lamp observations are included in the default routine of the health-checkout sequence of operations. Health-checkouts have been conducted four times over the cruise phase after the initial post-launch checkout. An extra observation on 5 December 2017 was conducted to monitor the linearity properties of the CCD (see **Sec. 3.5**). Only v-band was used in this linearity measurement, and only a section (128×128 pixels) of the FOV was downlinked for examination.

**Figure 3.28** shows an example the ONC-T images taken of the FF lamp output on 16 October 2017. The brightness gradient in the vertical direction of the ul- and b-band images are characteristic of the FF lamp properties, due to the FW roundabout stray light. **Figure 3.29** examines the brightness gradient in the vertical direction in the FOV for all filters by displaying the ratio of top quarter of the image (#1) and the bottom quarter of the image (#4). The reason for the strong gradation at the short wavelengths can be explained as follows. The FF lamp radiance is close to a 2000 K black body radiator that has peak emission at long wavelengths. For example, the energy flux from the FF lamp at the p-band wavelength range is about 500 times larger than the flux at the ul-band wavelength. Therefore, the FW roundabout stray light is a negligible percentage of the transmitted light through the longer wavelength filters but is a measurable percentage of the transmitted light through the short wavelength (e.g., ul- and b-band) filters. The difference between segments #1 and #4 are due to the spatial distribution bias of the FW roundabout stray light. This FW roundabout stray light is not expected to affect the observations of Ryugu, since the reflectance spectrum of Ryugu is relatively flat and the solar radiance spectrum has a peak at the v-band wavelength range of only ~ 2 times larger than ul-band wavelength range.

Average signal values of the FF lamp images are used to monitor the time variations in the CCD response. **Figure 3.30** shows the time variation of the count flux of the FF lamp images obtained during the cruise phase. The FF lamp has two modes of voltage settings which is selected depending on the sensitivity of filters used to image the lamp. **Figure 3.30(a)** shows the "High voltage mode" data (used for the shorter wavelength filters: ul- to Na-bands), and **Fig. 3.30(b)** shows the "Low voltage mode" data (used for the longer wavelength filters: w- to p-, and wide-bands). To show relative sensitivity changes, the measured output of the FF lamp is normalized to the output measured on 16 April 2015. Note that the data acquired on 8 July 2017 are excluded from the plot lines due to possible contamination by radiator stray light affects (**Sec. 3.7**).

**Figure 3.30(a)** indicates that the sensitivity of the shorter wavelength filters (ul- to Na-bands) decreased after the initial checkout. **Figure 3.30 (b)** also indicates that the sensitivity of the longer wavelength filters (w- through p- and wide-bands) decreased after the initial checkout. However, we cannot conclude the degradation is real due to the unstable emission from the "Low voltage mode" of the FF lamp used for the preflight long wavelength measurements. Since the "High voltage mode" was more stable during the pre-flight measurements, we are limiting our discussion to the ul-to-Na bands which were characterized using the FF lamp "High voltage mode". There are two possible explanations for the decreasing trend in the "High voltage mode" data; 1) degradation of the FF lamp, e.g., low electric current and 2) degradation in the transmittance of the filters due to exposure to the space environment. During the initial checkout, the instrument temperatures were slightly higher than those of subsequent observations (**Table 3.6**). Due to these temperature differences (FF lamp and CCD), we inter-compare the observations acquired after 16 April 2015, which were obtained under similar temperature conditions. After 16 April 2015, the FF lamp flux is more stable, decreasing ~1% for the b-to-Na bands and decreasing ~2.5% for the ul-band. This demonstrates that the ul-band has a steeper decrease in sensitivity than the other three bands. To investigate the reason for the decrease in sensitivity, we compare the local measured output change. **Figure 3.29** shows the measured output change for segments #1 and #4. Although segment #1 is more contaminated by the FW roundabout stray light, the decreasing rate in the measured output is almost same. Thus, this decrease in sensitivity is more likely caused by degradation of the FF lamp.

Since the main purpose of this 5 December 2017 observation was to examine the linearity response, only the ONC-T v-filter was used. Additionally, to reduce the data downlink size, only a portion (128x128 pixels) of the FOV was downlinked, using the region of interest (ROI) function form the on-board processing algorithms. Time variation in the FF lamp output in this area of the CCD are plotted in **Fig. 3.31**. Although the temperature environment (**Table 3.6**) during this observation is very similar to the environment of 16 October 2017 observation, there is a small increase (~0.4%) from the 16 October 2017 observation. It is suspected that this random trend may be due to slight stabilities in the FF lamp output.

In summary, we conclude that the count flux from the FF lamp have a long-term decreasing trend (~1–2.5%) between 16 April 2015 to 5 December 2017. There are two plausible candidates for the decreased count flux: (1) degradation of the FF lamp, and (2) degradation of the filters under the space environment. To resolve whether the sensitivity degradation in the CCD is real or not, additional observations are needed; especially, repeat calibration observations using stars that have already been observed by the ONC.

**Table 3.6.** Summary of the FF lamp observations held after launch.

| Date | Observation purpose | Note | Temperature [°C] | | | |
|---|---|---|---|---|---|---|
| | | | ONC-AE | | ONC-T CCD | |
| | | | Start | End | Start | End |
| 11 December 2014 | Initial checkout | 7 band + wide band | -4.49 | -4.10 | -14.79 | -14.79 |
| 16 April 2015 | Health checkout | 7 band + wide band | -6.46 | -6.20 | -29.56 | -29.54 |
| 10 September 2015 | Health checkout | 7 band + wide band | -6.95 | -6.46 | -29.46 | -29.43 |
| 8 July 2017 | Health checkout | 7 band + wide band. Solar Stray light was contaminated | -7.05 | -7.02 | -28.77 | -28.77 |
| 16 October 2017 | Health checkout | ONC-T (7 band + wide) | -5.77 | -5.77 | -28.59 | -28.91 |
| 5 December 2017 | Linearity observation | ONC-T v band. A small part of the FOV is cutout. | -5.77 | -5.77 | -28.72 | -28.72 |

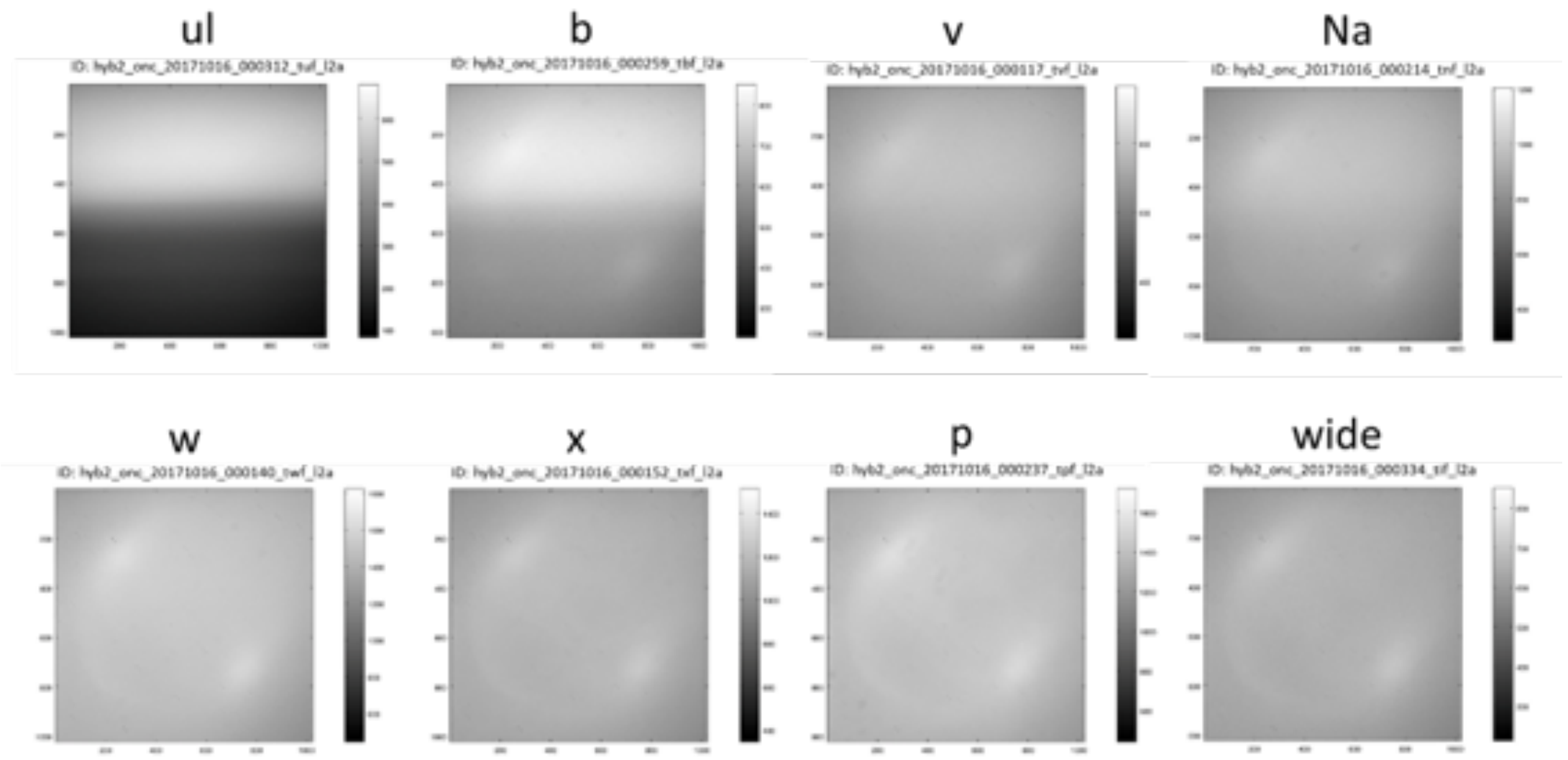

**Figure 3.28.** Examples of the FF lamp images. These observations were taken on 16 October 2017.

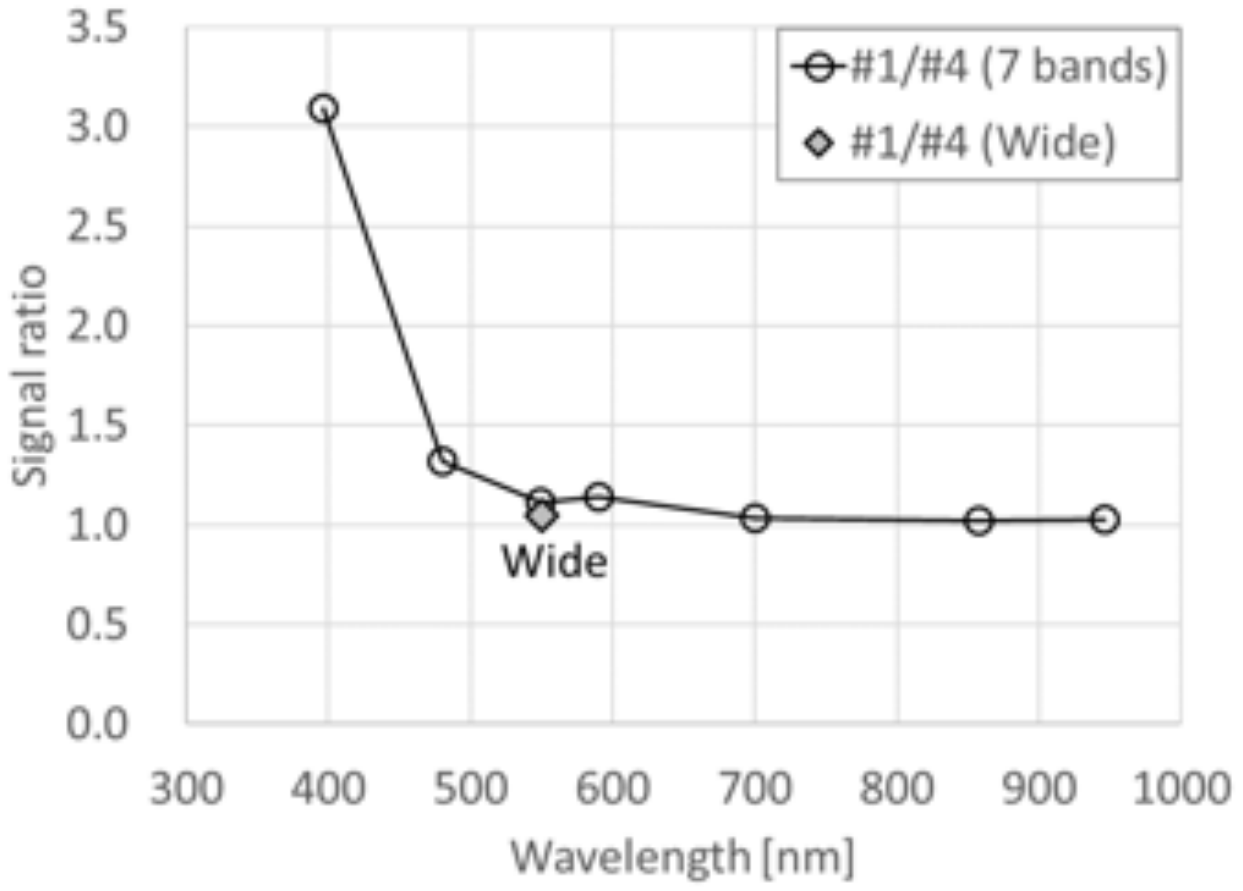

**Figure 3.29.** Vertical brightness gradients in an FF lamp image. The observation was taken on 16 October 2017. The image was divided into 4 segments in the vertical direction, and the average signal ratio between segment #1 (top quarter of an images) and segment #4 (bottom quarter) as a function of wavelength is shown. The ONC-T wide band data is plotted at 550 nm for reference.

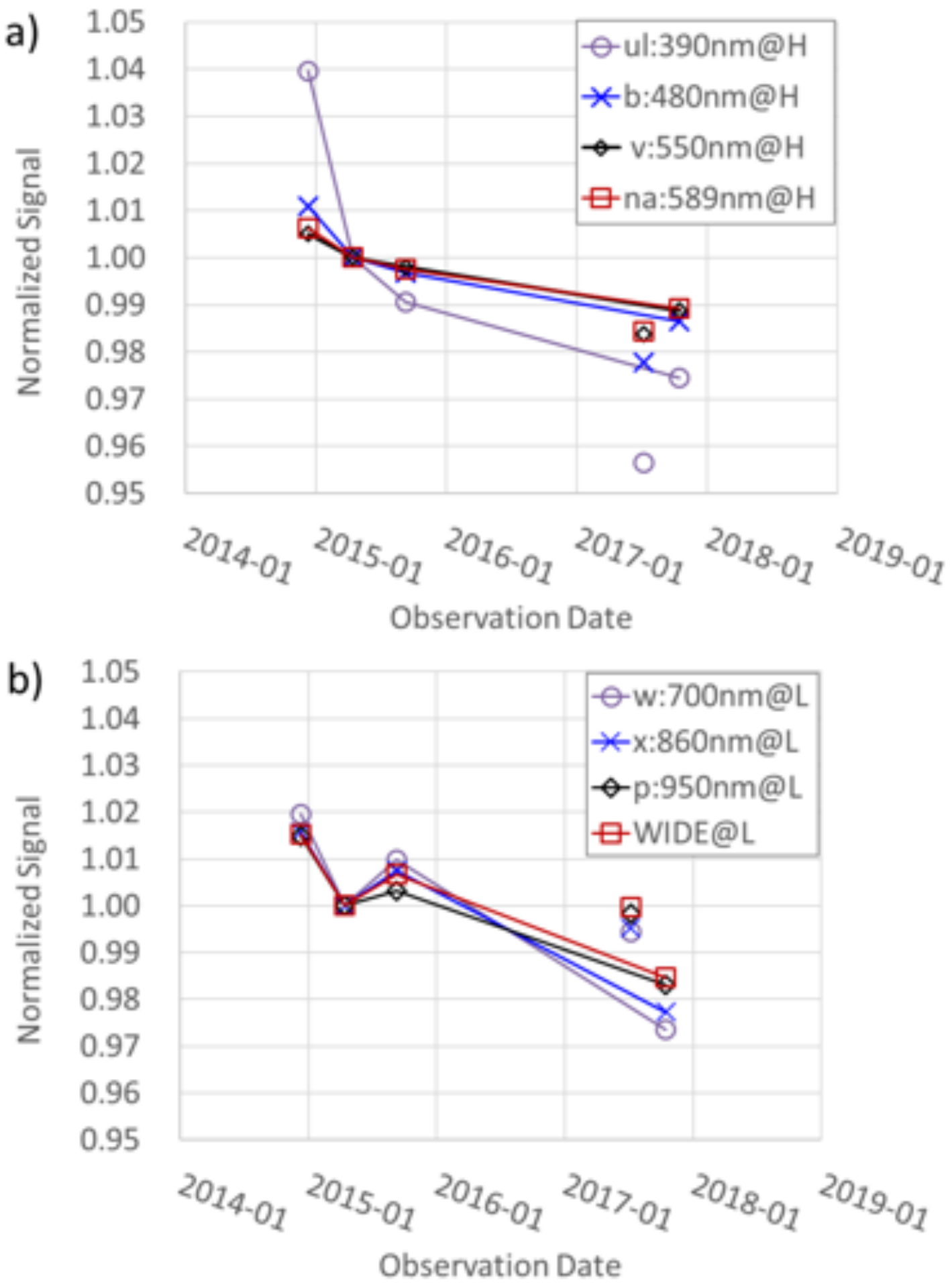

**Figure 3.30.** The time variation in the FF lamp count flux during the cruise phase. (a) ul- to Na-band data were obtained in the "High voltage mode". (b) w- to p-, and wide-band data were obtained in the "Low voltage mode".

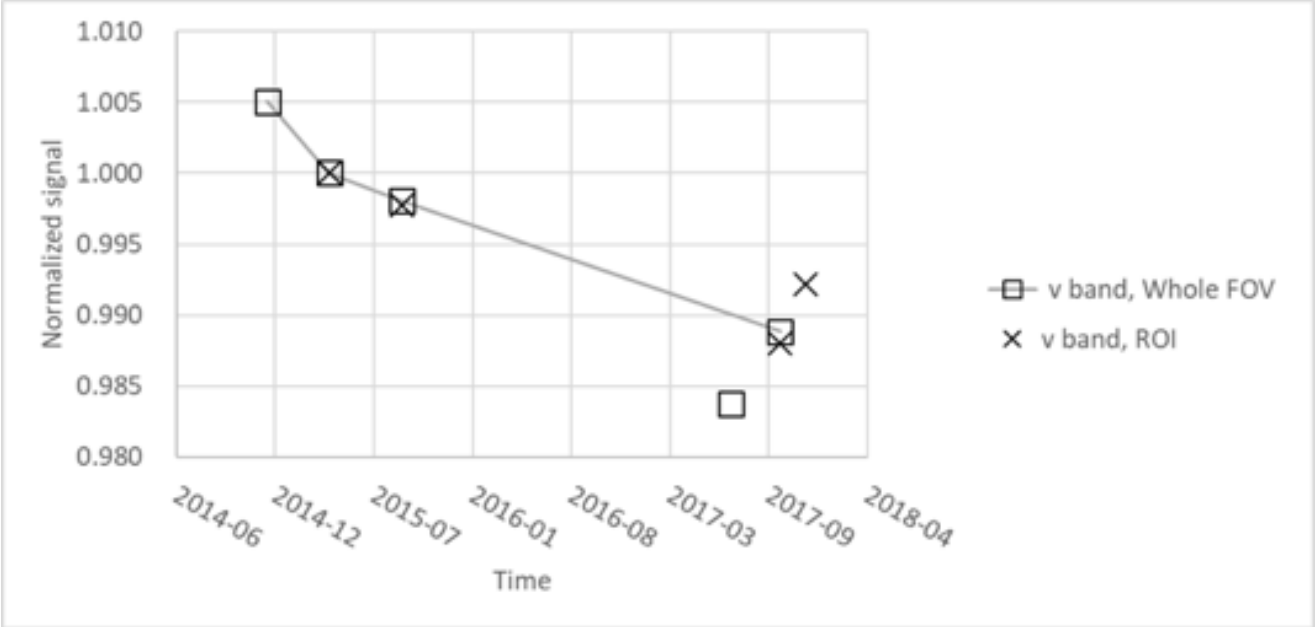

**Figure 3.31.** Time variation of the FF lamp count flux in the v-band. Square: Full FOV data (same as **Fig. 3.30**). Cross: ROI sub-frame. The data points are normalized to the 16 April 2015 observation.

### 3.9.2. Sensitivity Dependence from CCD temperature Based on Jupiter Observations

The spectral sensitivity of the ONC-T was measured using an integrating sphere at room temperature prior to launch (Kameda et al., 2017). The sensitivity at cold temperatures was estimated using the typical sensitivity dependence from CCD temperature provided by the manufacturer (E2V), and was demonstrated to be consistent with the SP/SELENE lunar reflectance model (Suzuki et al., 2018). In this section, we examine the Jupiter observations taken by the ONC-T with different CCD and ELE temperatures to provide a model of the sensitivity dependence from hardware temperatures each filter band.

We observed Jupiter using ONC-T from May 16 – 17, 2017. The CCD temperature was controlled by the heater to range between -29 °C to +25 °C and the ELE temperature from -10 °C to 0 °C (**Table 3.7**). Because of Jupiter's non-uniformity of reflectance, we measured the Jupiter's brightness over 4 rotation periods under a different temperature for each period, for a total of 4 rotations. **Figure 3.32** shows the normalized flux dependency to the CCD temperature. The effect of the CCD temperature is much larger than that of ELE temperature especially for the long wavelength filter bands. As the CCD temperature increases, the sensitivities of the ul- to w-bands decrease, while the sensitivities of the x- and p-bands increase. We assumed that the sensitivity, $S_n$, can be expressed as a function of the CCD temperature, $T_{CCD,T}$;

$$S_T = S_{0,n}\left(a_{CCD,n}\left(T_{CCD,T} + 30\right) + 1\right), \tag{3.20}$$

where $S_{0,n}$ is the sensitivity for n-band at the reference temperature $T_{CCD,T} = -30°\mathrm{C}$ and $a_n$ is the temperature dependence factor determined by $\chi^2$-fitting. The sensitivity of each band at any temperatures of $T_{CCD,T}$ (-30°C to +25 °C) can be calculated by Eq. (3.19) and the values listed in **Table 3.8**, where the 2-σ error is based on deviations of the measurements from Eq. (3.19). Similarly, the ELE temperature dependence can be expressed as (**Table 3.9**):

$$S'_T = S_{0,n}\left(a_{ELE,n}(T_{ELE} + 10) + 1\right). \tag{3.21}$$

The reference ELE temperature is -10 °C. Note that, the ELE temperature dependences for ul-to-Na bands are nearly 0 within the 2-σ errors, showing no dependence on the ELE temperature in the CCD and filter response. However, w-to-p bands show noticeable dependences from ELE temperature. The CCD temperature has the dominant effect on the CCD and filter responses. Thus, we use **Eq. (3.19)** to quantify the CCD sensitivity at CCD temperature ranging from -30°C to 25°C. Although the dependence from ELE temperature might be observed for longer wavelengths, we do not apply ELE temperature correction for first products because we do not have physical explanations of dependence difference in filters, while the dependence from CCD temperature for each band should be different due to the temperature dependence of quantum efficiency in different wavelengths. To apply the ELE temperature correction, further investigation is needed. For example, we will be able to test the correction to images obtained during close approaches. While this does not account for sensitivity variations caused by other hardware temperatures, for example, the error introduced by the ELE temperature effect is small (<0.1%/°C for the ul-to-Na bands and <0.3%/°C for the w-to-p-band).

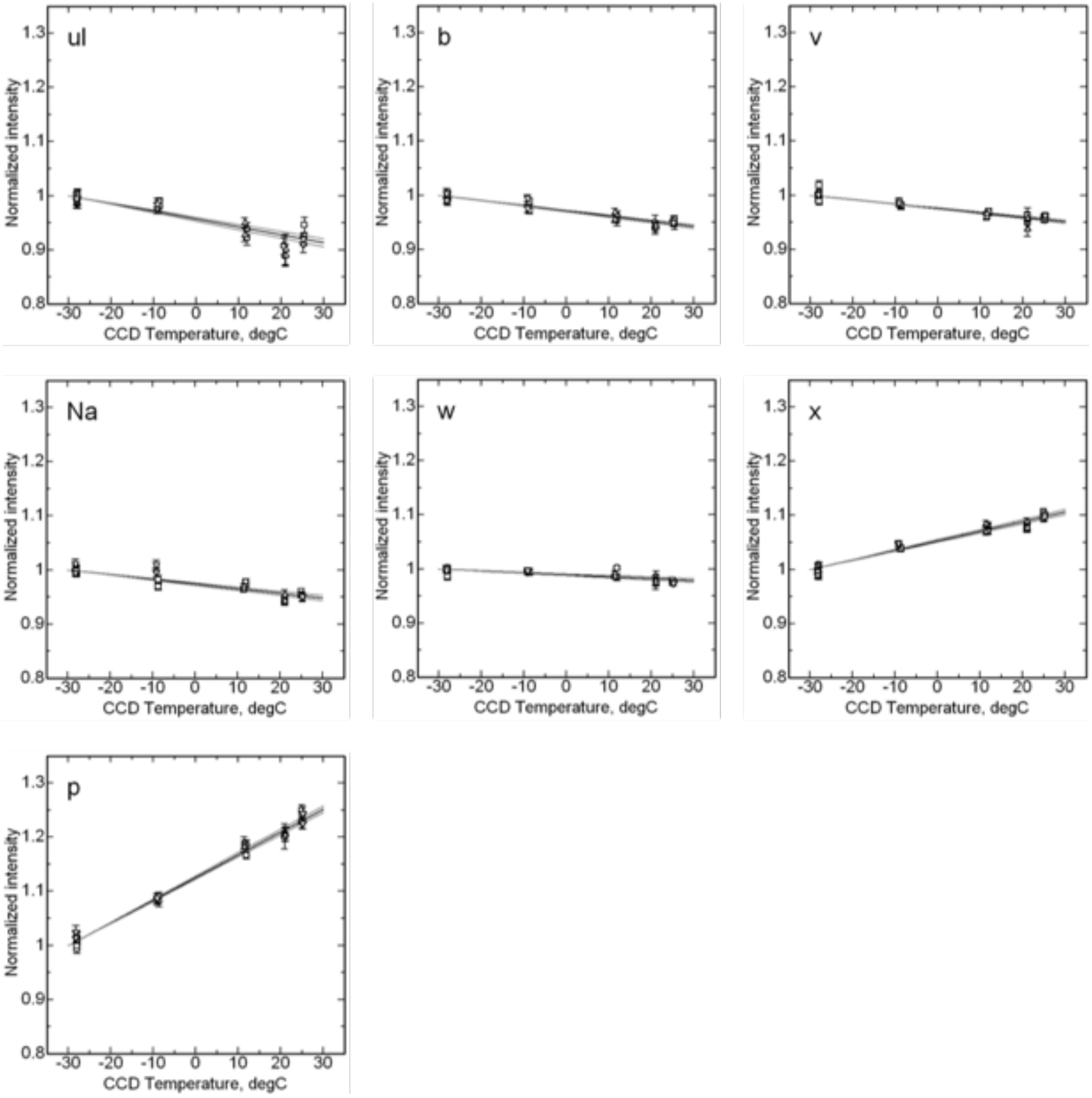

**Figure 3.32.** Observations of Jupiter using the ONC-T for each band (ul-p). The symbols, ○, △, □, and ×, indicate the normalized signal rate (DN/s) for rotational periods 1, 2, 3, and 4, respectively. The black lines show the results of $\chi^2$-fitting with the gray lines of 1- σ errors.

**Table 3.7.** CCD and ELE temperatures during the ONC-T Jupiter observations.

| Phase# | CCD Temp. [°C] | ELE Temp. [°C] | Date in May, 2017 | Start time, UT | End time, UT |
|---|---|---|---|---|---|
| 1 | 21.04 | -8.84 | 16 | 12:30 | 12:33 |
| 2 | 21.05 | -8.32 | 16 | 15:00 | 15:03 |
| 3 | -27.94 | -9.89 | 16 | 17:30 | 17:33 |
| 4 | -28.20 | -9.89 | 16 | 20:00 | 20:03 |
| 1 | -27.94 | -9.89 | 16 | 22:30 | 22:33 |
| 2 | -28.06 | -9.89 | 17 | 1:00 | 1:03 |
| 3 | -8.66 | -9.37 | 17 | 3:30 | 3:33 |
| 4 | -9.18 | -9.37 | 17 | 6:00 | 6:03 |
| 1 | 11.93 | -8.84 | 17 | 8:30 | 8:33 |
| 2 | 11.41 | -8.84 | 17 | 11:00 | 11:03 |
| 3 | 25.11 | -8.32 | 17 | 13:30 | 13:33 |
| 4 | 25.22 | -8.32 | 17 | 16:00 | 16:03 |
| 1 | -29.24 | -3.63 | 17 | 18:30 | 18:33 |
| 2 | -28.46 | -3.63 | 17 | 21:00 | 21:03 |
| 3 | -28.72 | 0.19 | 17 | 23:30 | 23:33 |
| 4 | -28.77 | -0.03 | 18 | 2:00 | 2:03 |

**Table 3.8.** Sensitivity dependence from CCD temperature for all band-pass-filters. The errors cited are the 2-σ errors.

| band | $a_{CCD,n}$[/℃] |
|---|---|
| ul (0.40 μm) | $-0.001449 \pm 0.000244$ |
| b (0.48 μm) | $-0.000968 \pm 0.000108$ |
| v (0.55 μm) | $-0.000814 \pm 0.000090$ |
| Na (0.59 μm) | $-0.000866 \pm 0.000154$ |
| w (0.70 μm) | $-0.000355 \pm 0.000112$ |
| x (0.86 μm) | $0.001771 \pm 0.000158$ |
| p (0.95 μm) | $0.004201 \pm 0.000202$ |

**Table 3.9.** Sensitivity dependence from ELE temperature for all band-pass-filters. The errors cited are the 2-σ errors.

| band | $a_{ELE,n}$[/℃] |
|---|---|
| ul (0.40 μm) | $-0.00077 \pm 0.00084$ |
| b (0.48 μm) | $-0.00055 \pm 0.0078$ |
| v (0.55 μm) | $-0.00085 \pm 0.0072$ |
| Na (0.59 μm) | $-0.00049 \pm 0.00058$ |
| w (0.70 μm) | $-0.00118 \pm 0.00034$ |
| x (0.86 μm) | $-0.00113 \pm 0.00086$ |
| p (0.95 μm) | −0.00288 ± 0.00094 |

### 3.9.3 Sensitivity Prediction Based on the Hardware Specifications

The DN derived from the ONC-T CCD can be modeled by using the hardware component specifications, such as the CCD quantum efficiency, filter and lens transmittances. Such model equations for the ONC-W1 and -W2 are described in Eqs. (2)–(4) of Suzuki et al. (2018). In this section, we quantify the ONC-T sensitivity based on the specifications of the hardware components.

We formulate that the expected signal $I_{sys,n}$ [DN/s] from ONC-T is given by:

$$I_{sys,n} = \frac{\pi}{4hcG}\left(\frac{p^2}{F_{\mathrm{opt}}{}^2}\right)\int \frac{1}{\pi} J(\lambda) RADF(\lambda, i, e, \alpha)\lambda\, \Phi_n\left(\lambda, T_{CCD,T}\right) d\lambda, \tag{3.22}$$

where $F_{\mathrm{opt}}$ is the effective F-number of the optics (= 9.05), $p$ is the CCD pixel size (=$1.33\times10^{-5}$ [m]), $J(\lambda)$ is the irradiance of the incidence light [W/m$^2$/μm] on the observation target, $RADF(\lambda, i, e, \alpha)$ is the radiance factor I/F (dimension-less) of the observation target at a specific set of illumination and viewing angle geometries ($i, e, \alpha$ ; incidence angle, emission angle, phase angle). The two $\pi$ symbols are not canceled in this equation to avoid confusion about the physical units, since the latter $\pi$ has a unit of [sr].

On the other hand, the model radiance $J_n$ for the n-band can be formulate as follows:

$$J_n = \frac{\int \frac{\lambda}{\pi} J(\lambda) RADF(\lambda, i, e, \alpha)\, \Phi_n\left(\lambda, T_{CCD,T}\right) d\lambda}{\int \lambda\, \Phi_n\left(\lambda, T_{CCD,T}\right) d\lambda}. \tag{3.23}$$

From **Eqs. (3.22) and (3.23)**, the predicted model sensitivity $S_{sys,n}$ can then be described as

$$S_{sys,n} = I_{sys,n}/J_n\ . \tag{3.24}$$

**Table 3.11** shows the values of the predicted model sensitivity derived from **Eq. (3.24)**. To derive these values, we assume the solar irradiance spectra (ASTM-E490) for $J(\lambda)$, a flat I/F spectra $RADF(\lambda, i, e, \alpha)$ equal to 1 and the CCD Temperature $T_{CCD,T}$ equal to -30 [°C].

### 3.9.4 Sensitivity Calibration Based on Star Observations

Ambiguities between the preflight CCD sensitivity (measured in the laboratory by Kameda et al. (2017)) and the inflight CCD sensitivity, due to temperature differences between preflight and inflight measurements, prompted further CCD sensitivity characterizations under inflight temperature conditions. To measure the absolute sensitivity inflight, we observed eight stars, with V-magnitude ranging from 2.02 to 4.73, during the cruise phase. The observed targets are summarized in **Table 3.10**. Those stars were selected based on spacecraft attitude constraints. The sensitivity can be obtained from the linear relationship between the photon flux and the digital count observed within each filter. For the reference star spectra, we used catalogs by Hamuy et al. (1992, 1994), Alekseeva et al. (1996), and Burnashev (1996) since we could not find a single catalog that included all our observable stars. μAqr, φSgr, and τ Sgr have not been observed in wavelengths longer than 750 nm, thus, the sensitivities of x- and p-filters are constrained with measurements from only five stars.

Although these different catalogs used the same calibration target α Lyr (Vega), they used a different reference spectrum for α Lyr, which contributes non-negligible systematic errors. Thus, we first derive the conversion factors to cancel the systematic errors caused by the reference spectrum differences. Using these conversion factors, the star energy flux by Alekseeva et al. (1996) were modified to be consistent with that of Hamuy et al. (1992, 1994) in all filters other than the p-filter. Note that we use the original flux values for the p-filter from all catalogs since the p-filter band pass partly contains the telluric water feature between 930 – 970 nm, causing particularly higher spectral errors as discussed in Hamuy et al. (1992, 1994). Burnashev (1996) used the same reference spectrum of α Lyr as Hamuy et al. (1992, 1994) and therefore the stellar spectra from this catalog did not require re-standardization to a common reference spectrum. Using these spectra, the expected flux $J_{\mathrm{star,\,n}}$ through each of the filters is obtained by using **Eq. (3.9)**. The digital counts from the star observations obtained by ONC-T are calculated as described in **Sec. 3.8**. The range of sensitivities derived from the star observations correlate to within 5%, except for ul- and p-bands (**Fig. 3.33**). Note that random scatter between the stars from different catalogs suggests that the systematic error between the catalogs has been properly removed. **Figure 3.34** displays the relationship between the star fluxes and digital counts. The sensitivities for the filters were obtained using a weighted least square linear regression fit of the 5 or 8 points listed in **Table 3.11**. The errors are derived based on the 95% confidence level. These deviations are consistent within the errors in the star observations of ~1% (average) given by Hamuy et al. (1992, 1994) and <3% (90% confidence level) for the b- to x-bands by Alekseeva et al. (1996). The large deviation in the p-filter sensitivity may be due to the lower accuracy of reference spectra, which has 5 – 10% error in the Hamuy et al. (1992, 1994) and <5% error in the Alekseeva (1996) spectra.

For comparison, the sensitivity based on the hardware specifications from **Sec. 3.9.3** is shown also in **Table 3.10**. Even though there are absolute differences between the two sensitivities, on the order of ~17%, the relative sensitivities between the different filters are consistent except for the p-filter (**Fig. 3.35**). The absolute sensitivity difference between the hardware specification and star calibration measurements can be ascribed to some flux loss in the optics, but the consistency in the relative values indicate that the transmittances for the filters were well defined preflight. This comparison also suggests another radiometric calibration is needed for the p-filter. In the next section, we update the p-filter sensitivity based on ONC lunar observations. We conclude that the CCD sensitivity derived from the star observations is consistent with the prediction from hardware specifications except for the p-filter.

**Table 3.10.** List of observed stars during the cruise phase.

| | HR No. | Observed Date | V mag[a)] | Spectral Type[a)] | Filters | Reference spectra |
|---|---|---|---|---|---|---|
| ζ Peg | 8634 | 2016-10-19 | 3.40 | B8V | ul, b, v, Na, w, x, p | Hamuy et al. (1992, 1994) |
| θ Crt | 4468 | 2017-05-23 | 4.70 | B9.5Vn | ul, b, v, Na, w, x, p | Hamuy et al. (1992, 1994) |
| ε Aqr | 7950 | 2017-10-10 | 3.77 | A1V | ul, b, v, Na, w, x, p | Hamuy et al. (1992, 1994) |
| μ Aqr | 7990 | 2017-10-10 | 4.73 | A3m | ul, b, v, Na, w | Burnashev (1996) |
| σ Sgr | 7121 | 2017-10-12 | 2.02 | B2.5V | ul, b, v, Na, w, x, p | Alekseeva et al. (1996) |
| ζ Sgr | 7194 | 2017-10-12 | 2.60 | A2III+A4IV | ul, b, v, Na, w, x, p | Alekseeva et al. (1996) |
| φ Sgr | 7039 | 2017-10-12 | 3.17 | B8III | ul, b, v, Na, w | Alekseeva et al. (1996) |
| τ Sgr | 7234 | 2017-10-12 | 3.32 | K1+IIIb | ul, b, v, Na, w | Alekseeva et al. (1996) |

a) Hoffleit and Jaschek (1991)

**Table 3.11.** Sensitivity factors at the reference temperature ($T_{CCD,T}$=-30 ℃).

| Filter name | Sensitivity at $T_{CCD,T}$=-30 ℃ (DN/s)/(W/m²/μm/sr) | |
|---|---|---|
| | star observations | hardware specifications |
| ul (0.40 μm) | 439.1±2.2 | 547.3 |
| b (0.48 μm) | 969.3±7.9 | 1104.6 |
| v (0.55 μm) | 1175.0±10.0 | 1371.9 |
| Na (0.59 μm) | 546.9±2.0 | 642.3 |
| w (0.70 μm) | 1515.0±19.3 | 1728.4 |
| x (0.86 μm) | 1499.8±24.6 | 1736.1 |
| p (0.95 μm) | 1033.0±71.1* | 1023.1 |

* for the p-filter we will use the value 961.2±28.8 (DN/s)/(W/m²/μm/sr) for the radiometric calibration pipeline based on the lunar observations (See **Sec. 3.9.5**).

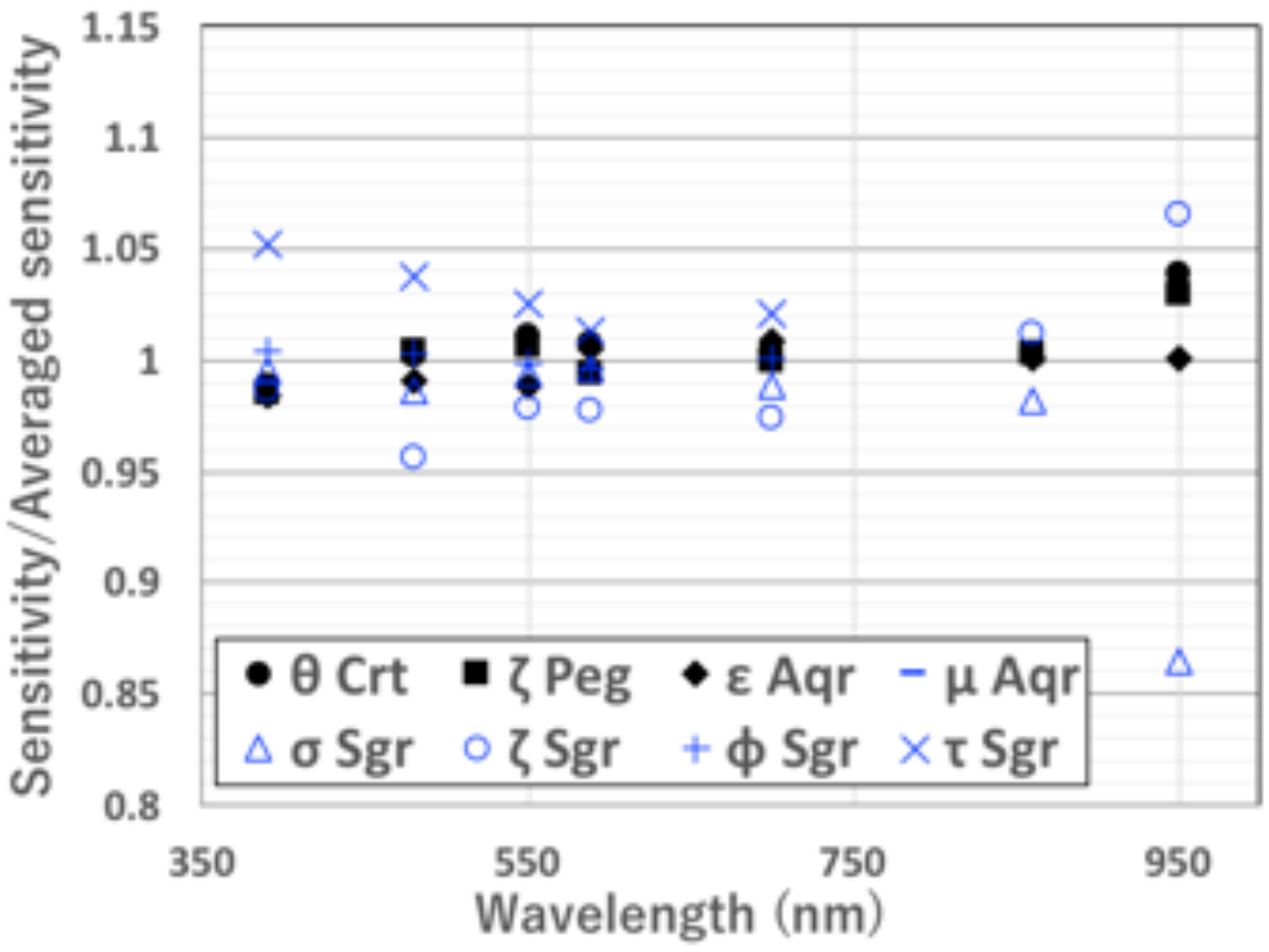

**Figure 3.33.** The sensitivity deviation measurements from the stellar observations. The sensitivities are normalized

by the average sensitivity within each band.

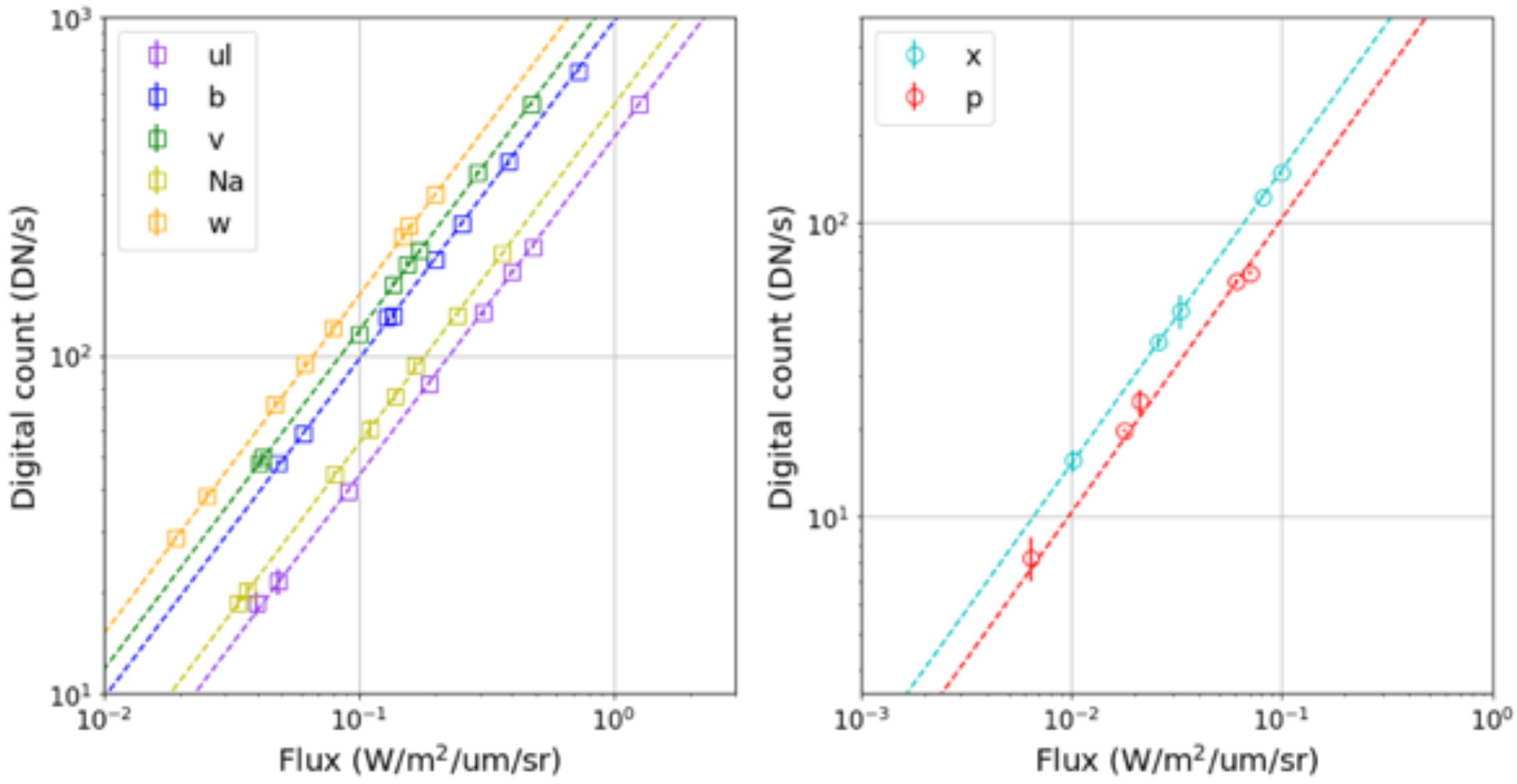

**Figure 3.34.** Star fluxes and digital counts observed by the ONC-T filters. The dotted lines indicate linear regression lines and the slopes correspond to the sensitivities of the filters.

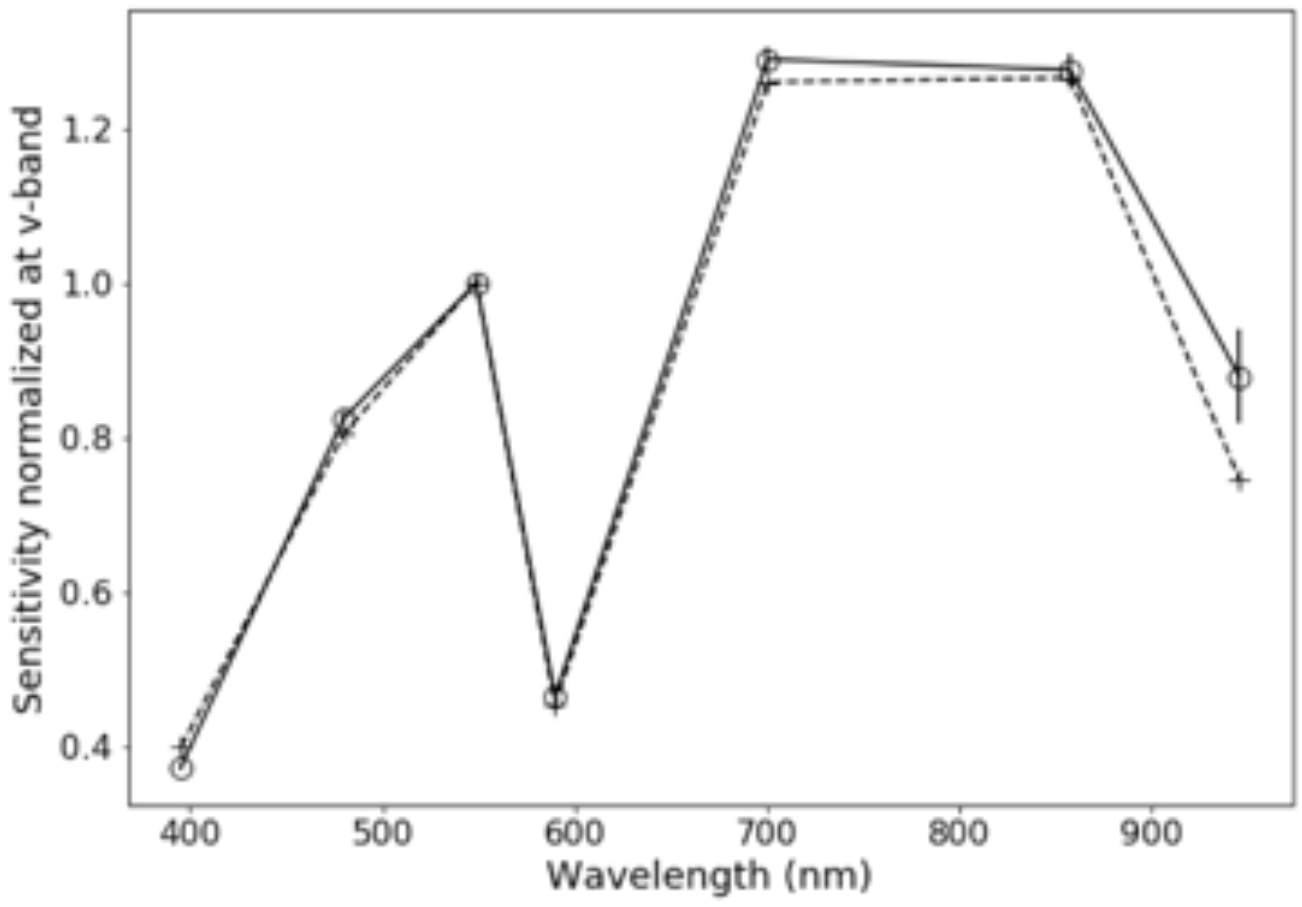

**Figure 3.35.** Sensitivity comparisons between those based on the hardware specifications (dotted line) and those based on the stellar observations (solid line).

### 3.9.5. Jupiter, Saturn, and Moon Observations

Here, we evaluate the CCD and filter sensitivities based on the ONC-T observations of Jupiter, Saturn, and the Moon. Comparisons with well-known spectra are useful for evaluating whether the derived sensitivities are accurate. Although the lunar observations have already been analyzed by Suzuki et al. (2018), the sensitivities have been updated and require re-evaluation. Thus, in order to evaluate the updated sensitivities, we revisit the ONC-T lunar observations. Comparison with lunar observations is a well-established technique for evaluating camera radiometric calibrations. The Jupiter and Saturn observations are analyzed in a similar manner as described in Sec. 3.9.2 for the characterization of temperature dependencies.

**Figure 3.36** shows the comparison between spectra obtained by the ONC-T on 5 December 2015 with SP/SELENE (Kouyama et al., 2016) and WAC/LROC (Sato et al., 2014) spectral models. Note that both models are calibrated to fit the Robotic Lunar Observatory (ROLO) spectral model (Kieffer and Stone, 2005). Comparisons of the absolute radiance comparison show the radiance based on the updated ONC-T CCD sensitivity exceeds that predicted from the lunar models. On the other hand, the relative spectral radiance is consistent (within 2.5%) between the models and the ONC-T observations The ROLO model has been shown to have a 5-10% uncertainty in its absolute value determination, whereas the relative spectral uncertainty is on the order of less than 3%. The relative uncertainty is based on the ROLO model comparison with lunar brightness measurements taken with Moderate Resolution Imaging Spectroradiometer, which is a well calibrated Earth observing sensor onboard Aqua and Terra (Stone, 2008). Thus, the updated sensitivity reproduces the lunar radiance spectrum to within the ambiguity in the models. Nevertheless, we can update the sensitivity conversion of the p-filter based on the lunar observations since the lunar reference models are more accurate than the stellar observations at this wavelength. The sensitivity derived from the lunar observations for the p-filter's normalized radiances are identical in both the SP/SELENE model and the ONC-T measurement. The sensitivity for the p-filter based on the lunar observations is 961.2±28.8 (DN/s)/(W/m$^2$/μm/sr). This value will be used for product generation during the mission's rendezvous phase.

**Figure 3.37** displays the normalized radiance spectra of Jupiter and Saturn at $T_{CCD,T}$=*-29* to *-27* °C compared to ground-based observations in 1995 by Karkoschka (1998). Karkoschka (1998) observed Jupiter and Saturn at phase angles of 6.8° and 5.7°, respectively, while the ONC-T observed these objects at 3.9° and 4.9°, respectively. The Jupiter spectrum obtained by the ONC-T shows overlap with the ground-based observation which has ~4% of error. Moreover, although the sensitivity based on the stellar observation can reproduce the radiance spectrum to within the ambiguity in the observation, the spectra reproduced by the sensitivity based on the lunar observations shows better consistency with the Jupiter ground-based observation. However, the Saturn spectra acquired by the ONC-T shows non-negligible differences between the x-band observations and the ground-based measurements by Karkoschka (1998). Because the x-band wavelength coincides with a known methane absorption, a higher reflectance from the rings compared with Saturn's disk (e.g., Irvine, 1971) is expected. Depending on the area ratio of the rings relative to the disk in the ONC-T images, the x-band could be brighter than the disk-only spectra of Karkoschka (1998). This explains the variations between the ONC-T x-band and the ground-based Saturn observations. Moreover, the bluer spectral properties of the rings may also explain the darker brightness seen in the ul-band compared to the ground-based measurements. These spectral comparisons reasonably confirm the ONC-T's spectral calibration.

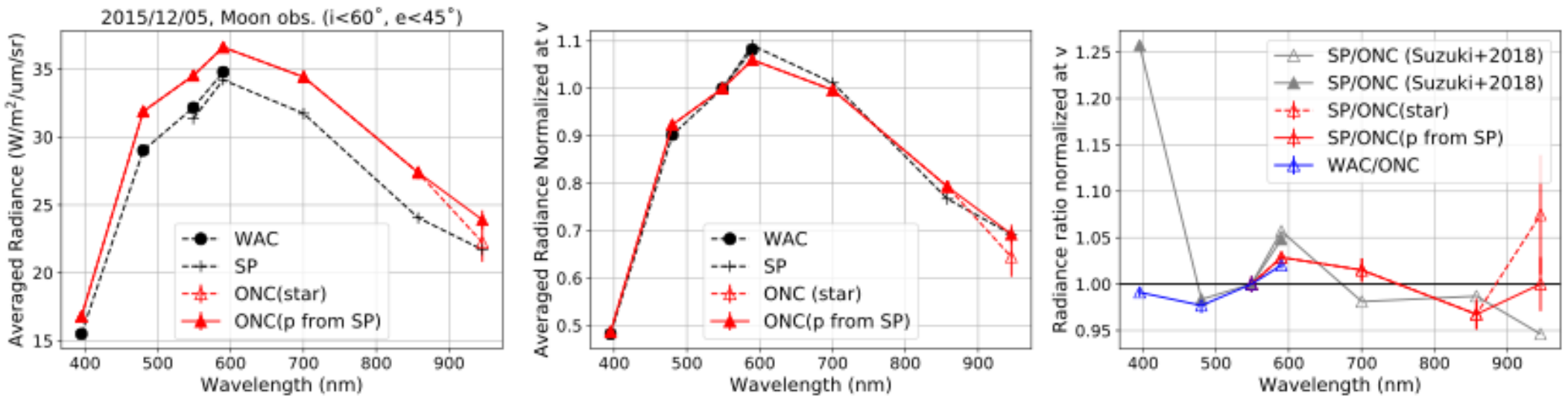

**Figure 3.36.** These graphs compare the Moon's radiance between that observed by the ONC-T and those predicted by lunar spectral models. (left) The disk-averaged radiance (i<60° and e<45°), (center) radiance normalized to the v-band, and (right) ratio of the normalized spectra (shown in the central graph) to the ONC normalized observation. Gray symbols are the results based on the sensitivity in Suzuki et al. (2018). The sensitivities for all bands derived from the stellar observations are shown by the dotted red line. The updated sensitivity for the p-band derived from comparisons between the normalized radiance of ONC-T and the SP/SELENE model radiance is shown by the solid red line.

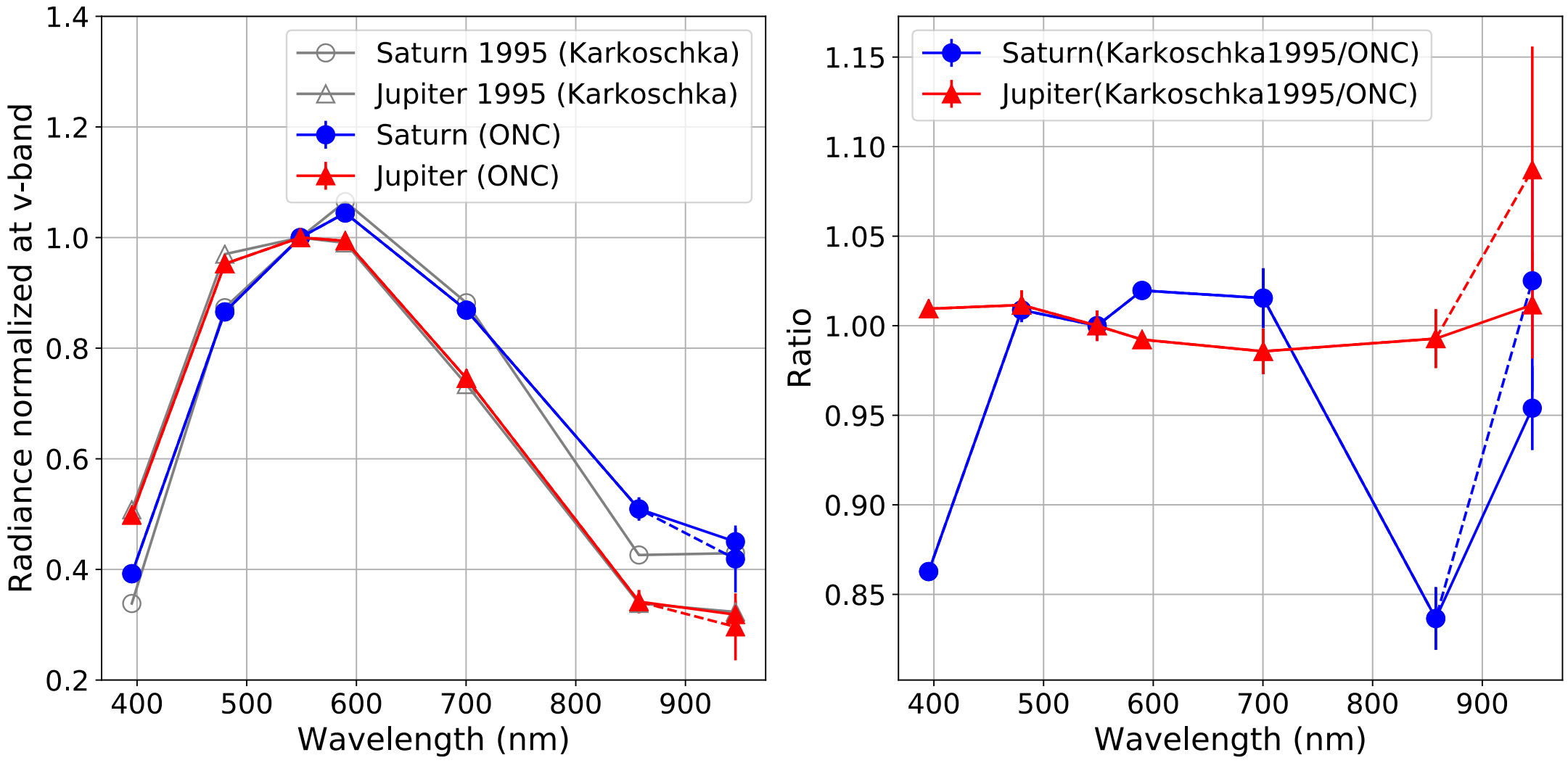

**Figure 3.37.** Normalized spectral comparisons of Jupiter and Saturn observations. (left) Comparison of normalized radiance between spectra obtained by ONC-T and Karakoschka (1998). Dotted lines indicate spectra produced based on the sensitivity from the stellar observations, and solid color lines indicate those based on the Moon observations. (right) The Jupiter and Saturn spectra of ONC-T divided by the spectra of Karakoschka (1998) are shown. The Earth-based Jupiter spectra and those from the ONC-T are in good agreement. The brightness in the x-band and the darkness in ul-band by the ONC-T in the spectra of Saturn can be explained by the differences in the spectral properties between the ring and the disk (e.g., Irvine, 1971).

## 4. Evaluation of the Upper Limit of a Detectable Sodium Atmosphere

Sodium (Na) is one of the most common elemental species surrounding solar system bodies, due to its high abundance and volatility. For example, sodium was the first metallic component detected around Mercury by ground-based telescopes (Potter and Morgan, 1985). Sodium was also detected around the Moon (~5kR) and comets (1kR~900MR depending on its distance to Sun) through remote observation using the resonant scattering of solar light at the sodium emission wavelengths of 589.594 nm (D1) and 588.997 nm (D2) (Rahe et al., 1976; Potter and Morgan, 1988; Leblanc et al., 2008). Sodium, if present around Ryugu, is expected to be observable by the ONC using the Na-band filter. The density distribution and its seasonal variation as a function of phase angle or orbital position (season) is an important variable in understanding the formation of C-type asteroids and the generation processes of planetary atmospheres. This section evaluates the upper limit for the detection of sodium by the ONC-T, based on its observations of Jupiter's sodium nebula.

### *Method outline*

The sodium emission can be detected with the Na-band filter on ONC-T. In order to eliminate the stray light effect and fluctuations in background noise, an image taken concurrently with the v-band filter should be subtracted from the Na-band image. This eliminates the stray light component, as its contribution is independent of the observing filter. Therefore, the sodium observation sequence includes the images taken alternately between the Na- and v-band filters. In addition, the background noise can be eliminated by taking the ratio of background count ($C_{\mathrm{Na/v}}$) images of "sky-data" taken in both the Na-band and v-band, acquired between the Na- ($I_{\mathrm{Na}}(\mathrm{H, V})$) and v-band ($I_{\mathrm{v}}(\mathrm{H, V})$) images of the target. Then, the reduced Na-band image of the target is calculated as follow:

$$\mathrm{Target}(\mathrm{H, V}) = I_{\mathrm{Na}}(\mathrm{H, V}) - I_{\mathrm{v}}(\mathrm{H, V}) \times C_{\mathrm{Na/v}} \quad (4.1)$$

### *Data reduction*

It is known that Jupiter is surrounded by sodium torus originating from its moon Io. The brightness of the D1 and D2 lines from this sodium torus are observed to be 20-80 R (e.g. Yoneda et al., 2015). During 18 to 20 May 2017, ONC set its target for the west side of Jupiter's sodium torus and acquired 183 sets of images (Na- and v-bands) continuously. Each image was taken with a 178 s integration time period. The geometrical configuration of the observations is shown in **Fig. 4.1**. Even though we did not detect significant counts from sodium around Jupiter, the upper limit evaluation can still be derived using this data set.

### *Upper limit for Na detection*

As a first step, we identify and remove any hot pixels using the "dark image" derived in Sec. 3.4. Any pixels with high dark count rates (>1-σ) are eliminated from the sodium-detection processing. Furthermore, pixels with higher counts (>3-σ) are also eliminated from the processing as possible cosmic ray hits. The image analysis process described in the Eq. (4.1) is then adopted for the 183 image sets.

The ratio $C_{\mathrm{Na/v}}$ is calculated to be 0.9961 using the median from the processed sky images (Na- and v-band) taken on 28 April 2017. Nine ROIs are then defined as shown in the **Fig. 4.2** for each image. In order to evaluate size effects, three types of ROIs (set A, B, and C) are defined. The standard deviation in counts for each pixel in the ROI is calculated. This value is defined as the upper limit of counts for sodium detection by the ONC (1-σ). Integrating over the multiple images statistically reduces the upper limit. As a result, an upper limit to the Na brightness detection, as a function of the number of integrated images, can be calculated and is shown in **Fig.4.3**. If more than 100 image sets are integrated, then the upper limit is for sodium detection is around 100 R. In this calculation the count-to-brightness conversion factor is assumed to be 21 R/DN/pix (from Kameda et al., 2017).

The ONC-T observations during 18 to 20 May 2017 did not detect a significant count from sodium around Jupiter. This is not surprising since the brightness of the sodium cloud is known to be less than 80 R (e.g. Yoneda et al., 2015), which is lower than the detection limit for the ONC-T.

The brightness around Ryugu is expected to be 10-10s kR, although the estimation is strongly dependent on the outgas rate assumption based on the surface and the sodium g-factor. When we assume the outgas rate at Ryugu to be same as Earth's moon or Mercury, the brightness is extremely low (10-100 R) which is difficult to detect by the ONC-T. On the other hand, if we assume the outgas rate to be the brightness of several 10s kR which even corresponds to much lower outgas rate on comets, the sodium atmosphere is detectable by the ONC-T with a single image set. It should be noted that the brightness estimation also depends on the orbit of the asteroid because of the Doppler shift in the solar light spectrum, which is the source of the sodium resonant scattering (Killen et al., 2009).

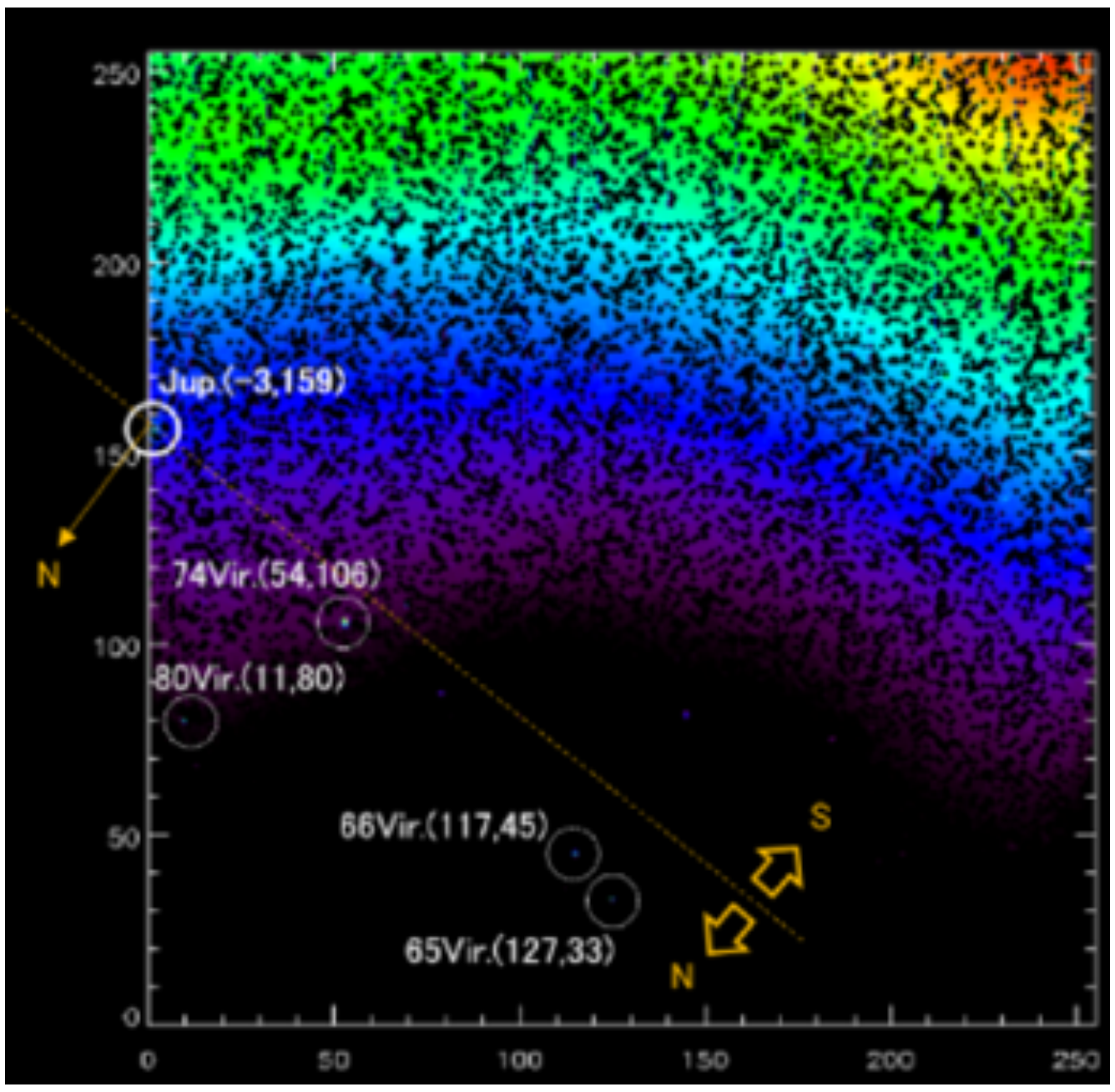

**Figure 4.1.** This image shows the geometry for the Jupiter's sodium cloud observations during 18 to 20 May 2017 (Na-band image). Jupiter and its north pole is indicated by the white

circle and yellow allow, respectively. The known stars are indicated by the white broken circles. The color gradation is caused by the radiator stray light.

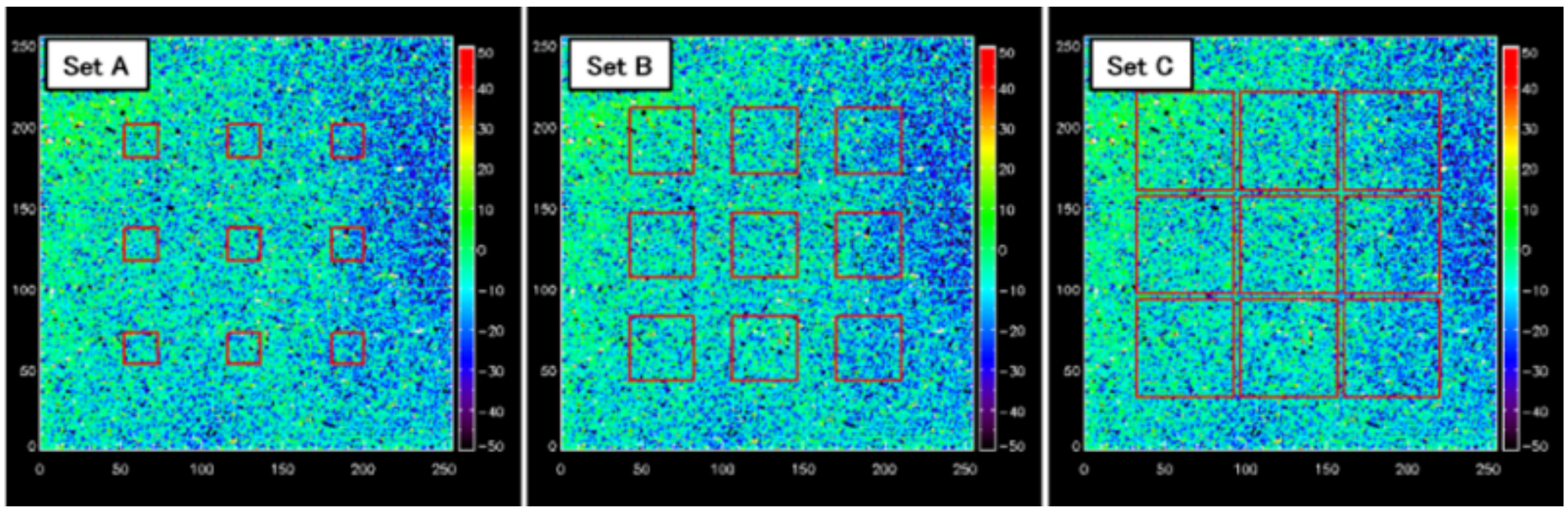

**Figure 4.2.** An example of the ROI region definitions for each set. The standard deviations for counts in each ROI were evaluated for the 183 image sets.

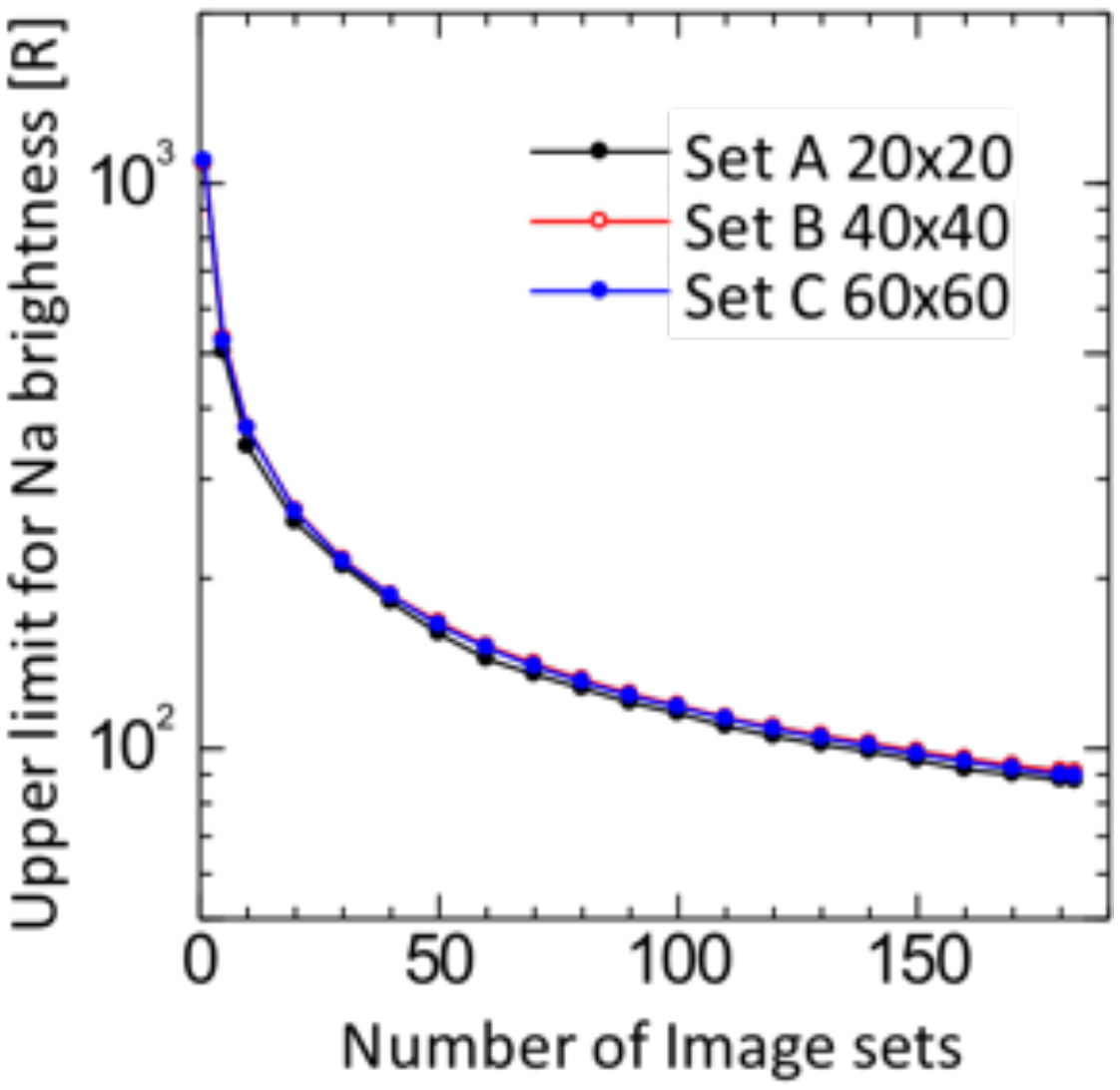

**Figure 4.3.** The upper limit of the brightness detection for sodium by the ONC as a function of number of the integrated image sets. No clear dependency on the ROI size is seen.

## 5. Inflight Calibration of the ONC-W1 and -W2

Both preflight and inflight calibrations of the flat-fields, PSFs, sensitivities, and stray light contributions are summarized in this section for the ONC-W1 and ONC-W2. Hayabusa2 proximity operations, such as touch-downs, lander-releasing, and gravity measurements, will be conducted primarily with the ONC-W1 since the long focal length of the ONC-T will be out of focus at positions closer to the surface. During touch-downs, ONC-W1 will take sequential images up to a spatial resolution of ~1.5 mm/pixel. Thus, the calibrations of the ONC-W1 and W2 are as important as the calibration of the ONC-T. The calibrations conducted both pre-flight and inflight are summarized in **Table 5.1**.

**Table 5.1.** Summary of the data used for the calibrations of the ONC-W1 and -W2.

| | Ground measurement | Inflight measurement | Inflight Observation Date |
|---|---|---|---|
| Bias | 0-sec exposures | 0-sec exposures | Continuously during the cruise phase |
| Dark current and Hot pixels | Thermal vacuum test (variation of CCD, ELE, and AE temperatures) | Deep sky observation with variation of CCD temperatures | 30 November 2017, 2 December 2017 |
| Sensitivity ($S_{cam}$) | Integrating spheres | Stars for ONC-W2 and Jupiter for ONC-W1 | 9 to 15 February 2016,<br>19 February to 21 March 2016,<br>28 June 2016, 30 June 2016,<br>2 July 2016, 17 October 2017,<br>30 November 2017 |
| Flat-Field | Flat panel | Star observation with variation of S/C attitudes | 12 March 2017<br>Same as images for Sensitivity calibration |
| Smear | 0-sec exposures | Earth for ONC-W2 | 4 December 2015 |
| Geometric distortion | Dotted pattern (Suzuki et al., 2018) | Stars (Suzuki et al., 2018) | W1: 19 February 2015<br>W2: 12 March 2017 |
| Alignment to the S/C | N/A | Stars | W1: 19 February 2015<br>W2: 11 December 2014 |
| Sharp PSFs | Pinhole | Stars, and Mars, Jupiter, and Saturn | 31 May 2016, 9 to 15 February 2016,<br>19 February to 17 March 2016,<br>28 June 2016, 30 June 2016,<br>2 July 2016, 30 June 2017, |
| | | | 17 October 2017, 30 November 2017 |
| Stray lights | N/A | Stray lights with variation of the S/C attitudes. | 11 December 2014, 19 February 2015, 26 February 2015, 23 June 2015, 12 to 17 October 2015, 9 November 2015, 14 November 2015, 22 December 2015, 9 February to 17 March 2016, 21 March 2016, 7 June 2016, 28 June to 23 July 2016, 12 March 2017, 13 to 20 June 2017, 17 to 21 October 2017 |

## 5.1 Bias Correction

We evaluated the bias level of the ONC-W1 and -W2 as functions of CCD and ONC-AE temperatures. We analyzed 0-second exposure images taken during preflight thermal vacuum tests and the cruise phase. Most of the data acquired during the cruise were taken at the nominal CCD temperature (-25°C), but at the end of 2017's, we also acquired 0-sec exposure images at a CCD temperature of +20°C to test the temperature dependency. **Figure 5.1** shows the bias level (the average for all pixels) of the ONC-W1 and W2 as functions of the CCD and ONC-AE temperatures. The bias level obtained during the inflight observations increased with CCD temperature, similar to the inflight observation results for ONC-T (**Fig. 3.3**) and preflight tests for ONC-W1 and -W2. The bias level can be empirically described as a function of the CCD temperature, $T_{CCD,(W1\ or\ W2)}$ (°C), and AE temperature, $T_{AE}$ (°C) using the relation:

$$I_{bias} = (a_0 + a_1 T_{AE} + a_2 T_{AE}^2) T_{CCD,(W1\ or\ W2)} + (b_0 + b_1 T_{AE} + b_2 T_{AE}^2). \quad (5.1)$$

Optimized values of the six coefficients in this equation are listed in **Table 5.2**. **Figure 5.2** shows comparisons between this empirical relationship and the actual bias measurements. We can predict the bias level from Eq. (5.1) within 10% error. Note that as was discussed in **Sec. 3.**2 for the ONC-T, the bias level may also depend on the temperature of the ELE of the ONC-W1 and -W2. When the ELE temperature changes drastically during proximity observations, the error might be greater and need to be corrected for by deductive methods based on the ONC-T.

**Figure 5.3** shows time-dependent change of the bias level at CCD temperatures around -25 °C using the inflight data acquired after launch. We found that the bias levels for both cameras stay nearly constant, within 1% change over 3 years, similarly to the ONC-T's bias level (**Fig. 3.5**).

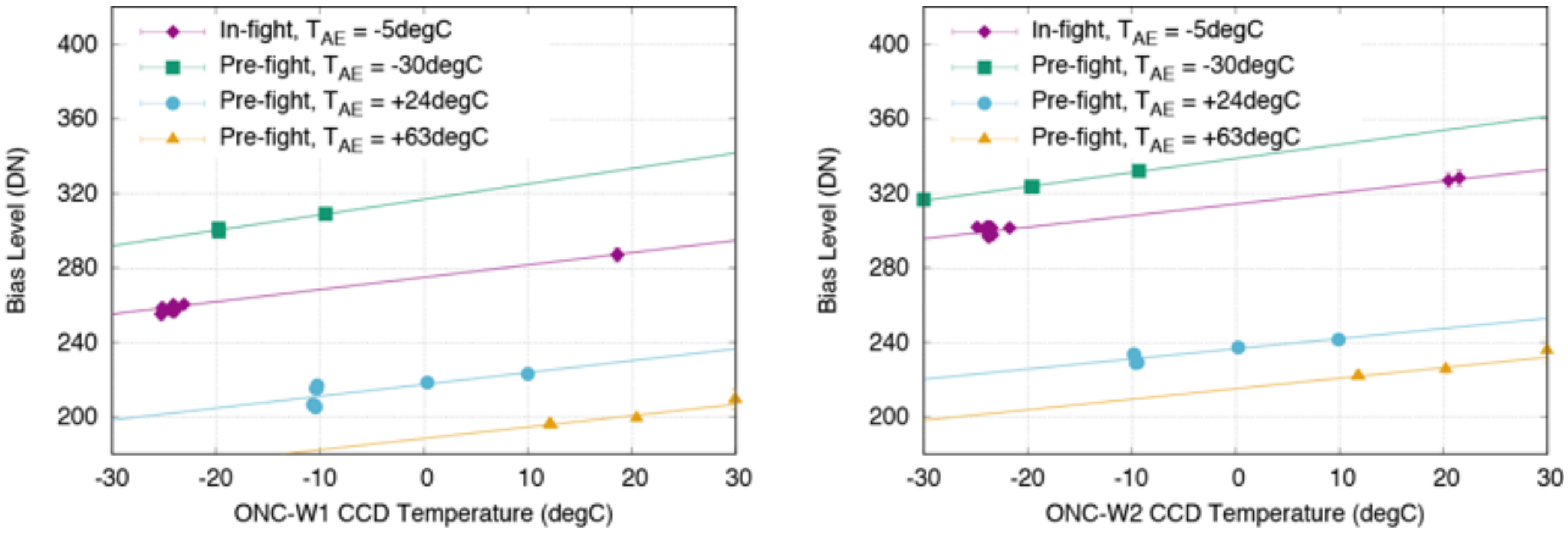

**Figure 5.1.** Bias levels for the ONC-W1 (left) and the ONC-W2 (right) as a function of the CCD temperature.

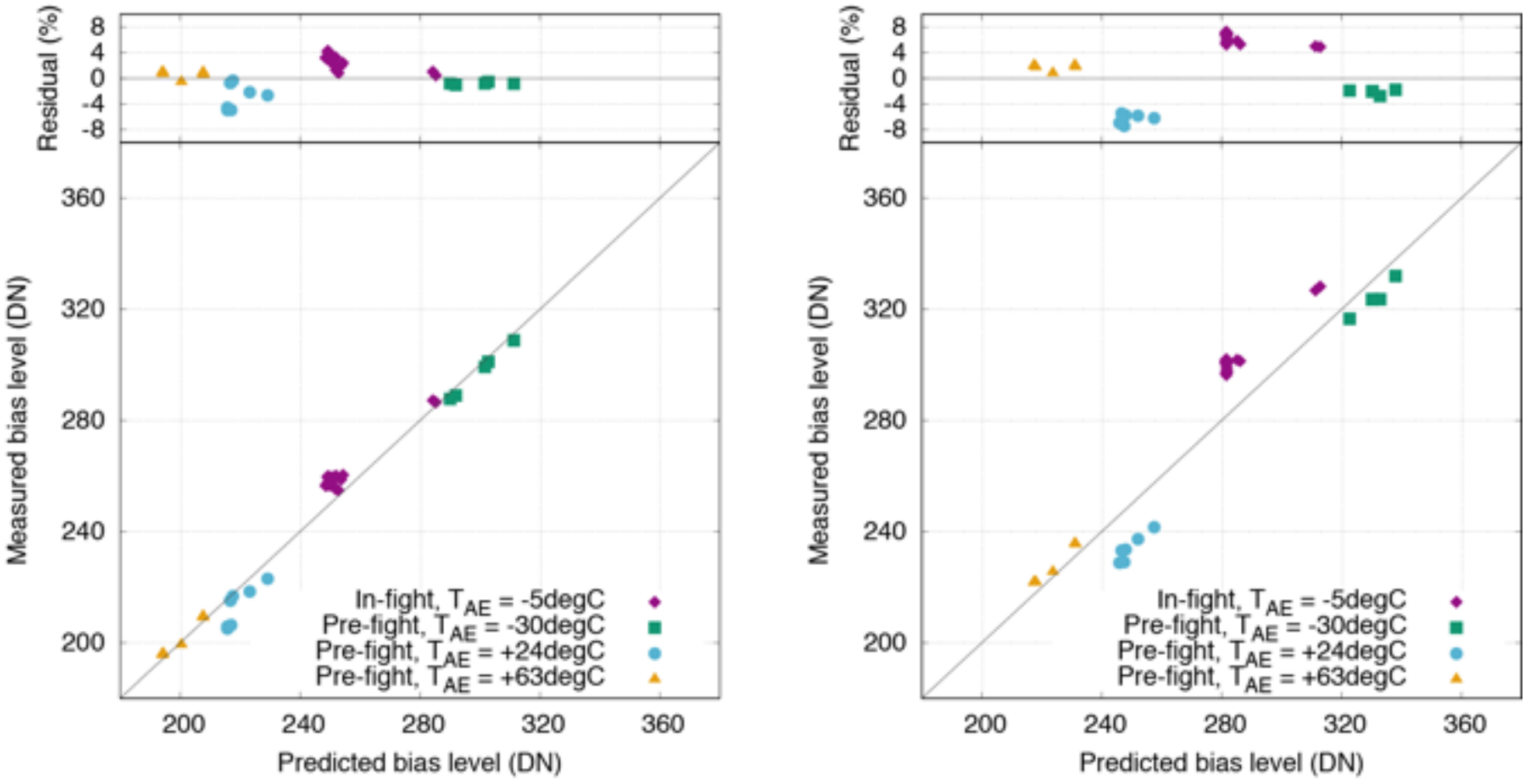

**Figure 5.2.** Comparisons between the actual bias measurements and those predicted from **Eq. (5.1)**. Left figure is for the ONC-W1and right for the ONC-W2.

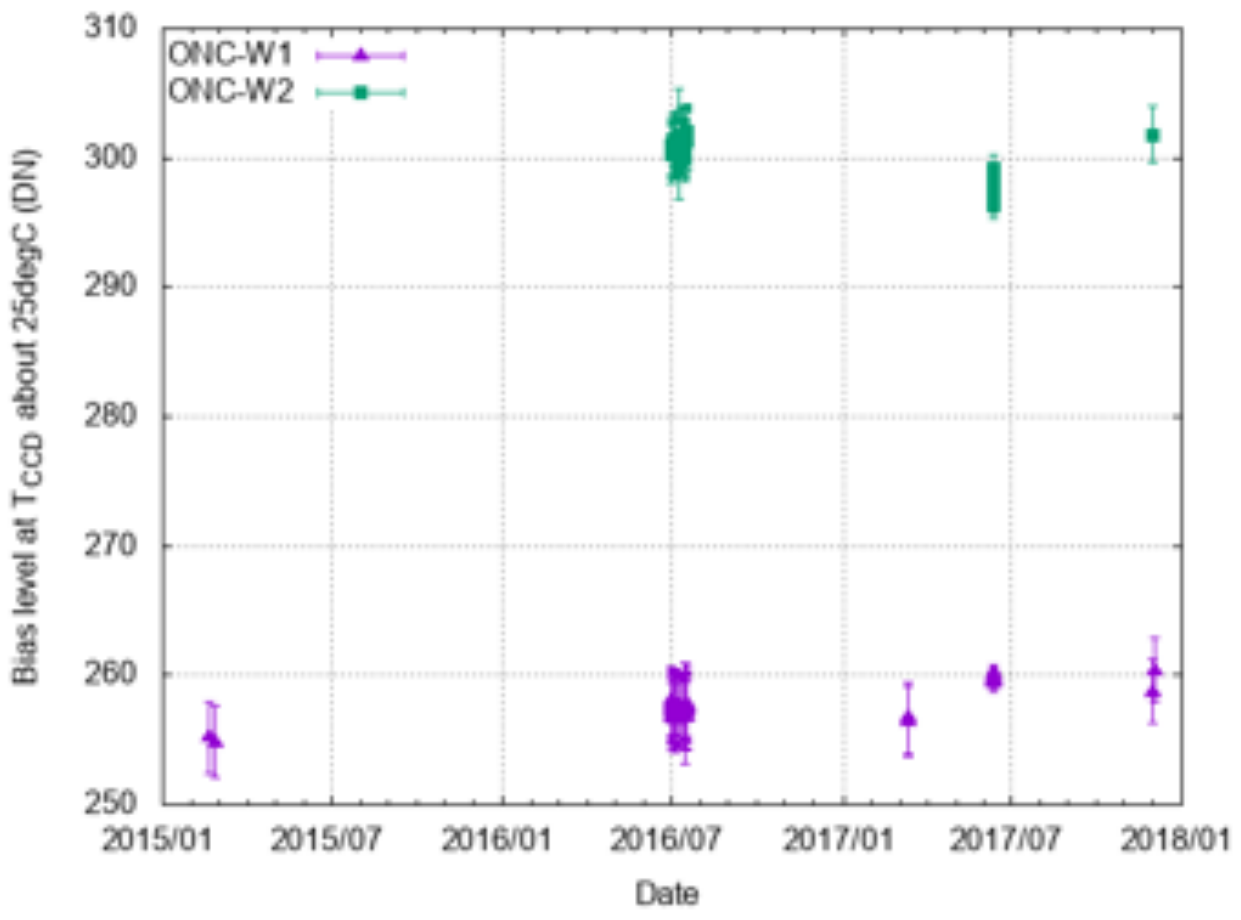

**Figure 5.3.** The time-dependent change in the bias level of the ONC-W1 and ONC-W2.

**Table 5.2.** Coefficients for **Eq. (5.1)** for the ONC-W1 and ONC-W2.

| | ONC-W1 | ONC-W2 |
|---|---|---|
| $a_0$ | 0.680 | 0.573 |
| $a_1$ | $-5.74 \times 10^{-3}$ | $-2.95 \times 10^{-3}$ |
| $a_2$ | $1.06 \times 10^{-4}$ | $8.92 \times 10^{-5}$ |
| $b_0$ | 260 | 288 |
| $b_1$ | $-1.72$ | $-1.65$ |
| $b_2$ | $8.50 \times 10^{-3}$ | $6.29 \times 10^{-3}$ |

**5.2. Dark Current and Hot-Pixel Correction**

On 30 November and 2 December 2017, we conducted deep space observations using the ONC-W1 and -W2 in order to evaluate their dark current levels. Two different CCD temperatures, $-25°\text{C}$ and $+20°\text{C}$, were used for both cameras during these observations. We took 4-image sets with exposure times of 5.57 sec at both CCD temperatures for ONC-W1, and 44.56 sec at $T_{\text{CCD,W2}} = -25°\text{C}$ and 16.8 sec at $T_{\text{CCD,W2}} = +20°\text{C}$ for ONC-W2. A few 0-sec exposure images were also obtained, which were used to analyze the bias levels (see **Sec. 5.1**). Bright spots due to cosmic rays were removed by taking a median of the 4 images, and the bias level was removed by subtracting the 0-sec exposure images. Signals due to stars were omitted by comparing two median images taken from the different two days. Unfortunately, we cannot distinguish stray light and dark current from the images taken at $T_{\text{CCD,(W1 or W2)}} = -25°\text{C}$. Thus, we estimated an upper limit for the dark current using a section of the median image, where the effects of the stray light appear to be small (rectangle area {[200, 200]; [800,600]} pixels on ONC-W1 images, and {[100, 100]; [900,900]} pixels on ONC-W2 images). At $T_{\text{CCD,(W1 or W2)}} = +20°\text{C}$, we evaluated the dark current on the entire images by subtracting the stray-light components estimated from the images at $T_{\text{CCD,(W1 or W2)}} = -25°\text{C}$.

**Figure 5.4 (a)** shows a histogram of the dark current for the ONC-W1 at $T_{\text{CCD,W1}} = -25°\text{C}$ and $+20°\text{C}$. The averages and the standard deviations are shown in **Fig. 5.4 (b)**, as functions of the CCD temperature. We define hot pixels as pixels where the dark current is larger than 30 DN/s. Figure **5.4 (c)** shows the number ratio of hot pixels to total pixels in the analyzed area of the image. We confirmed one of the pixels are greater than 30 DN/s at $T_{\text{CCD,W1}} = -25°\text{C}$. Similar data for the ONC-W2 are shown in **Fig. 5.5**. We confirmed three of the pixels are greater than 30 DN/s at $T_{\text{CCD,W2}} = -25°\text{C}$. The hot pixel ratio of the ONC-W1 at the CCD temperature of $+20°\text{C}$ seems to be less than that of the ONC-W2, and comparable with that of the ONC-T (**Fig. 3.7**).

We are planning to observe the asteroid surface closely using the ONC-W1, down to a few meters altitude during the touch-down sequence. We can understand the details of the physical properties of Ryugu's surface, such as regolith grain size and morphology, with an image resolution of a few mm/pixel at the touch-down sites. During the proximity observations, the CCD temperature is predicted to rise up to a few tens of degrees C at most, leading to high large dark noise levels, estimated to be about 10 DN/s at $T_{\mathrm{CCD,W1}} = 20$°C. Fortunately, however, the average dark current is negligible compared with the expected signal of ~50,000 DN/s from the reflection from the asteroid surface. Nevertheless, the hot pixels should be carefully examined during analyses, since 1% of the image area has dark current values higher than 100 DN/s, ~0.2% of the asteroid surface brightness (the maximum value is as high as 680 DN/s, ~1.3% of asteroid surface brightness) at the higher CCD temperature (**Fig. 5.4 (c)**). **Figure 5.6** shows a map of such hot pixels, which needs to be recognized and accounted for, especially in examinations of pixel-scale morphology.

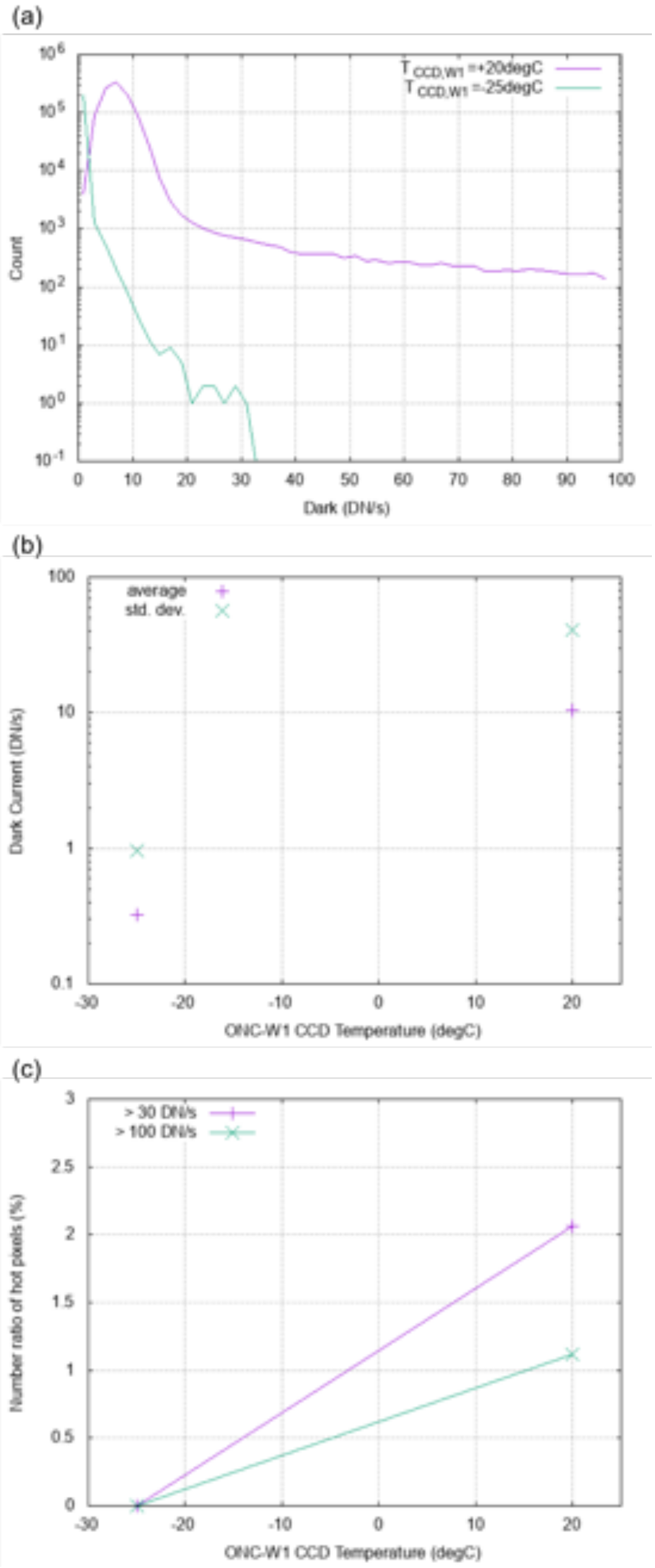

**Figure 5.4.** (a) Histogram of the dark current level, (b) the averaged values as a function of the CCD temperature, and (c) number ratio of the hot pixels for the ONC-W1.

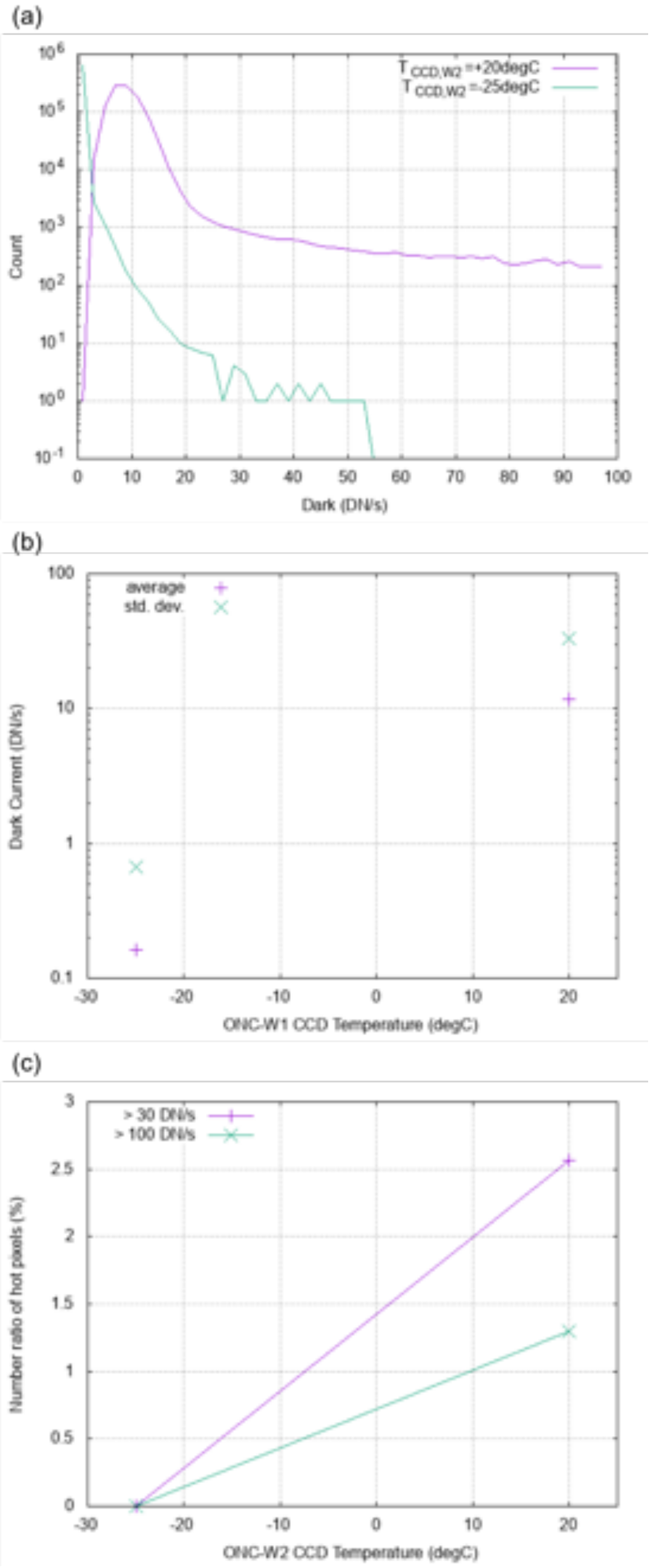

**Figure 5.5.** (a) Histogram of the dark current level, (b) the averaged values as a function of the CCD temperature, and (c) number ratio of the hot pixels for the ONC-W2.

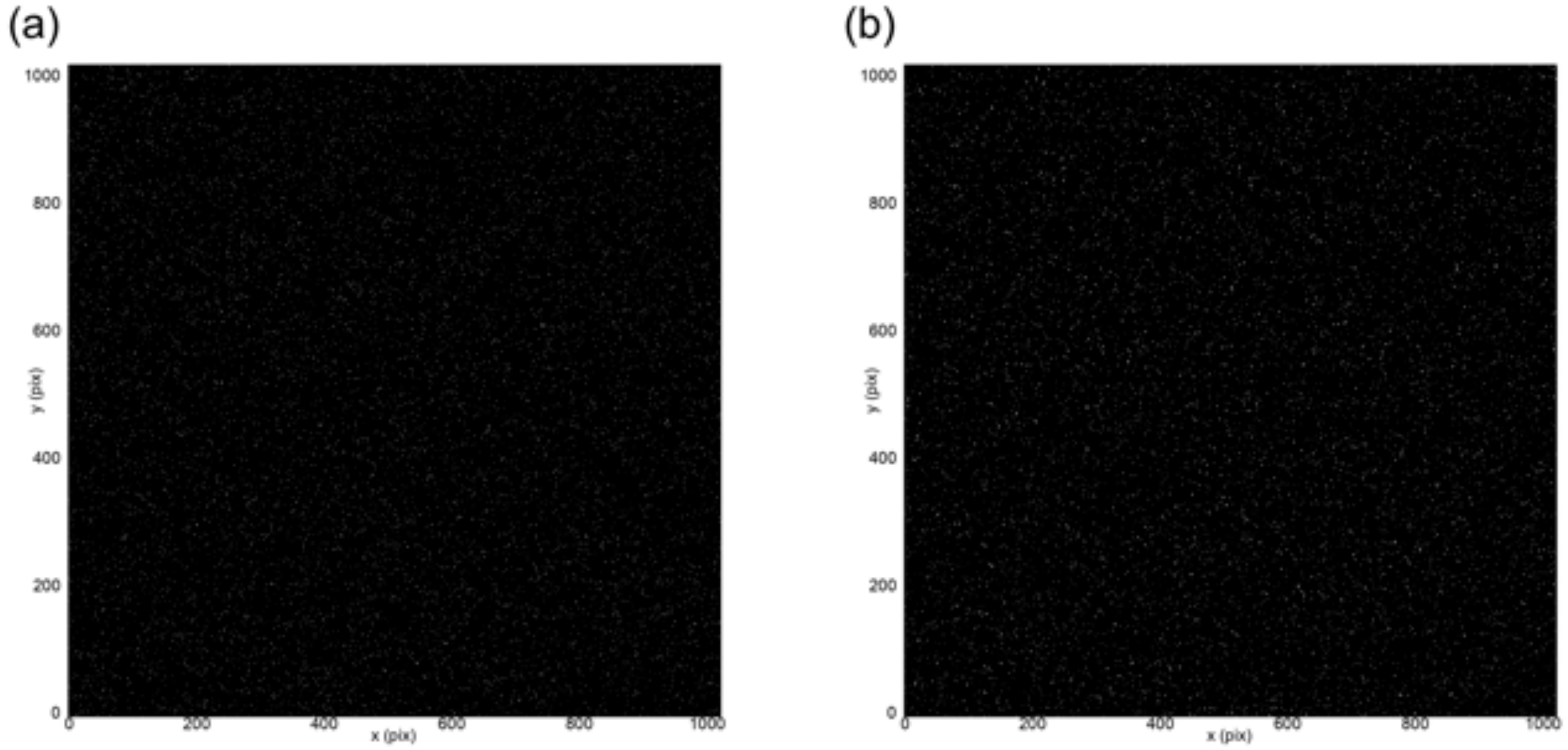

**Figure 5.6.** Hot pixel map of (**a**) the ONC-W1 and (**b**) ONC-W2 at the CCD temperature of 20°C. The hot pixels in this image are defined as the pixels where the dark current is larger than 100 DN/s.

### 5.3. Characterization of Stray Light in the ONC-W1 and -W2

Weak stray light seen in images taken by the ONC-W1 and -W2 were observed during the cruise phase. Similar to the tests with the ONC-T, we have taken images under a variety of spacecraft attitudes with respect to the Sun in order to examine the patterns and the intensity of the stray light. We found that the pattern and intensity of stray light in the ONC-W1 and -W2 depends on spacecraft attitude in a similar fashion as seen in the ONC-T.

**Figure 5.7** shows the typical patterns of stray light observed in the ONC-W1. Stray light in the ONC-W1 have a spotlight-like pattern, unlike the gradual intensity change in the FOV of the ONC-T. As the phase angle increases ($X_{PNL}$), the observed stray light also increases. The intensity of the stray light is evaluated using the median of an 11x11 pixel box in the spotlight-like zone in order to remove the effects due to cosmic rays and stars. Because optimal exposure times for the ONC-W1 during Ryugu observations will be <50 ms, stray light darker than 20 DN/s cannot be detected. The area affected by stray light is defined as the region in the FOV with greater than 20 DN/s in the image. **Figure 5.8** displays the relationship between the intensity, the area affected by stray light, and the spacecraft attitude. The spacecraft attitude is defined in the same manner as in the ONC-T discussion. Except for one attitude, the stray light was weaker than 30 DN/s, which will result in a 1.5 DN for 50-ms exposures. As seen in the ONC-T, the spot-like stray light is smaller and weaker at larger $Y_{PNL}$. Moreover, the intensity of the stray light is usually weaker for the ONC-W1 than the ONC-T at $X_{PNL} < 50°$ and is negligible when the stray light for the ONC-T are avoided by twisting the spacecraft around the $Z_{SC}$ axis.

**Figure 5.9** shows the typical patterns of stray light observed by the ONC-W2. Stray light in the ONC-W2 changes gradually in FOV. Note that bright parts are seen at the upper and lower right corners of the ONC-W2 FOV. They are the parts of the frame of the ONC-W2 reflecting the light source. We evaluated the average intensity of stray light for the ONC-W2, excluding the frame parts (red square in **Fig. 5.9**). **Figure 5.10** displays the relationship of the intensity of the stray light and the spacecraft attitude. Optimal exposure times for the ONC-W2 observations of Ryugu is estimated to be <20 ms, shorter than for the ONC-W1. Thus, the stray light <50 DN/s will not be detected in such Ryugu images. The ONC-W2 is found not to have severe stray light contributions when $X_{PNL} > 0°$, the Sun light illuminates the *+X*$_{SC}$ plane, but stray light contributions occur when the *-X*$_{SC}$ plane is illuminated. During the rendezvous-phase observations, Sun light will radiate on the *+X*$_{SC}$ panel at the nominal attitude. Thus, stray light in the ONC-W2 is expected to be negligible in most of rendezvous-phase images of Ryugu.

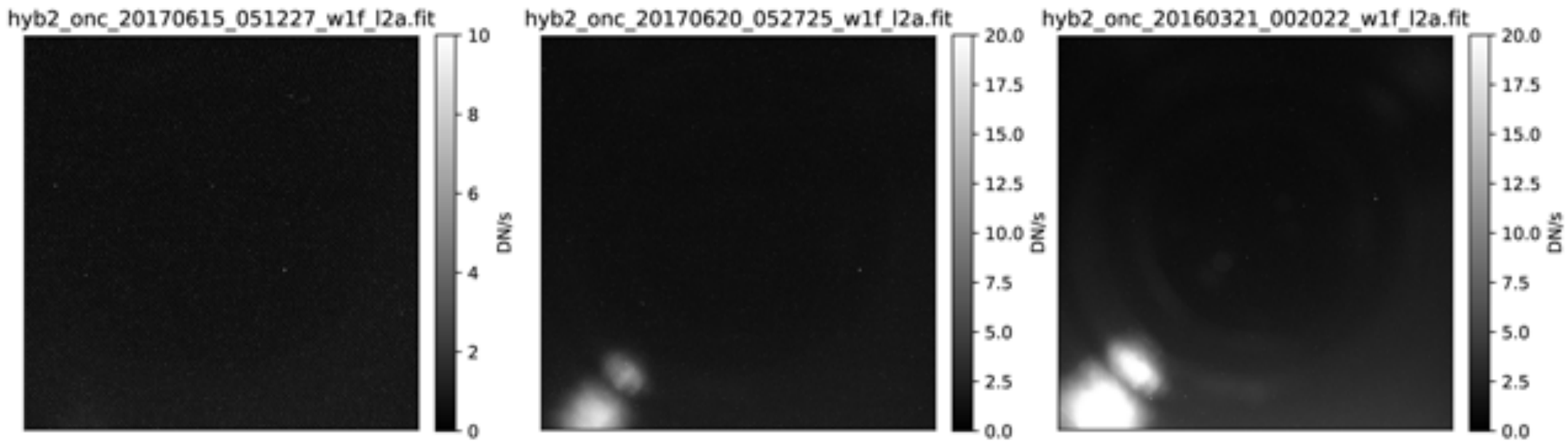

**Figure 5.7.** Stray light patterns at $Y_{PNL} = 0°$ with a changing $X_{PNL} = 15°, 30°, \text{and } 48°$ (from left to right). The stray light is observed at the lower left corners in the images with large $X_{PNL}$.

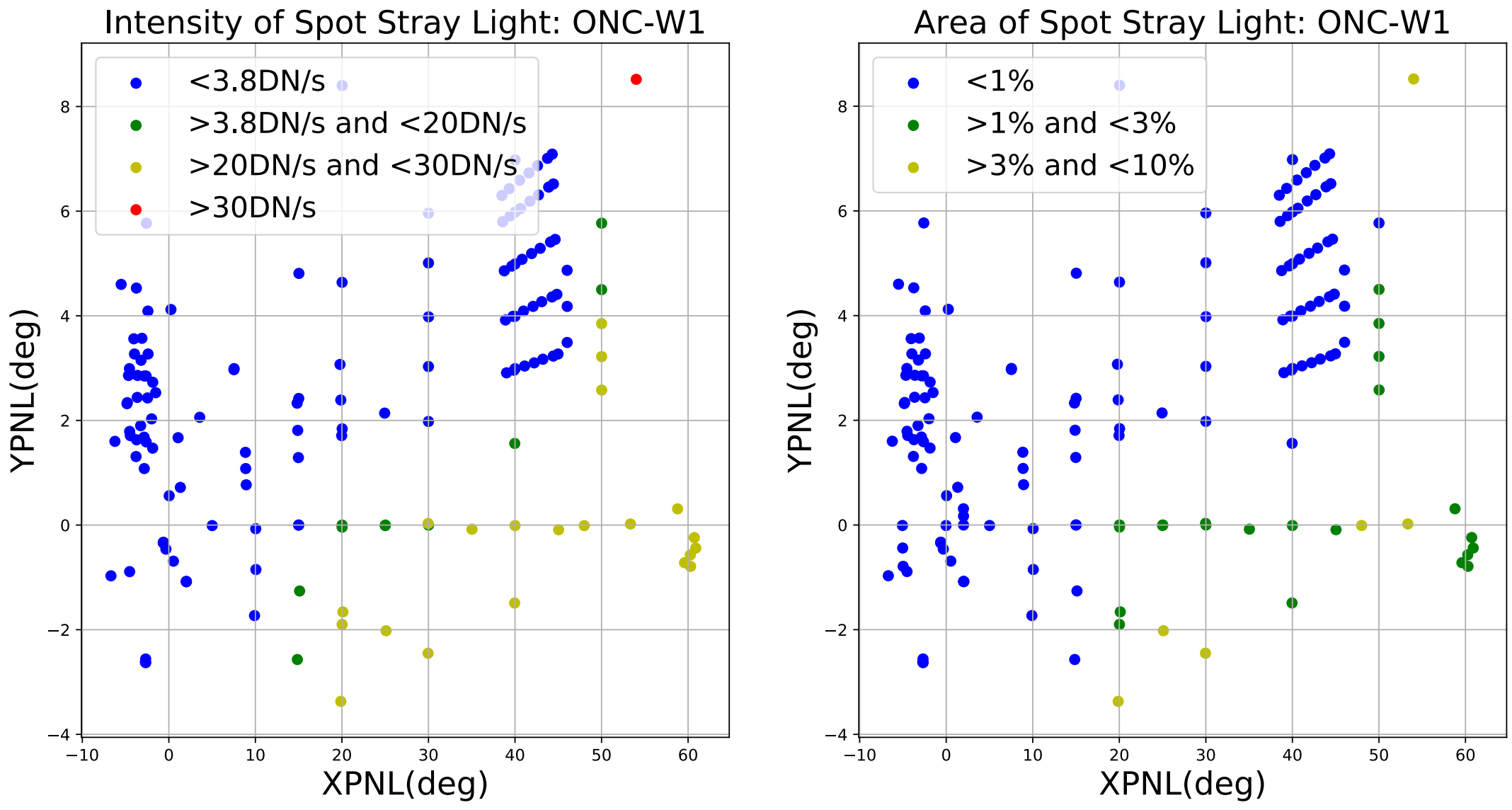

**Figure 5.8.** (left) The intensity classification of the ONC-W1 stray light. The stray light is weaker than 20 DN/s (blue and green) and will not be detected due to the short exposure times for the ONC-W1, <50 ms. (right) The area ratio classification of the ONC-W1 stray light.

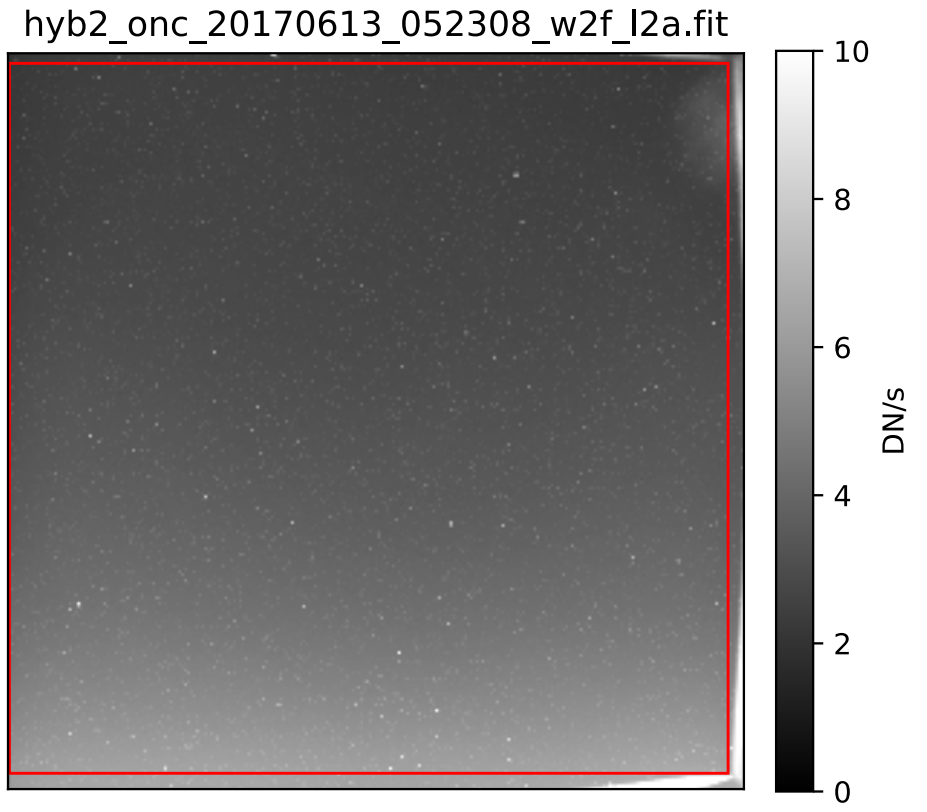

**Figure 5.9.** Stray light pattern of the ONC-W2 at $(X_{PNL}, Y_{PNL}) = (0°, 0°)$. The red square indicates the region where the stray light intensity was assessed.

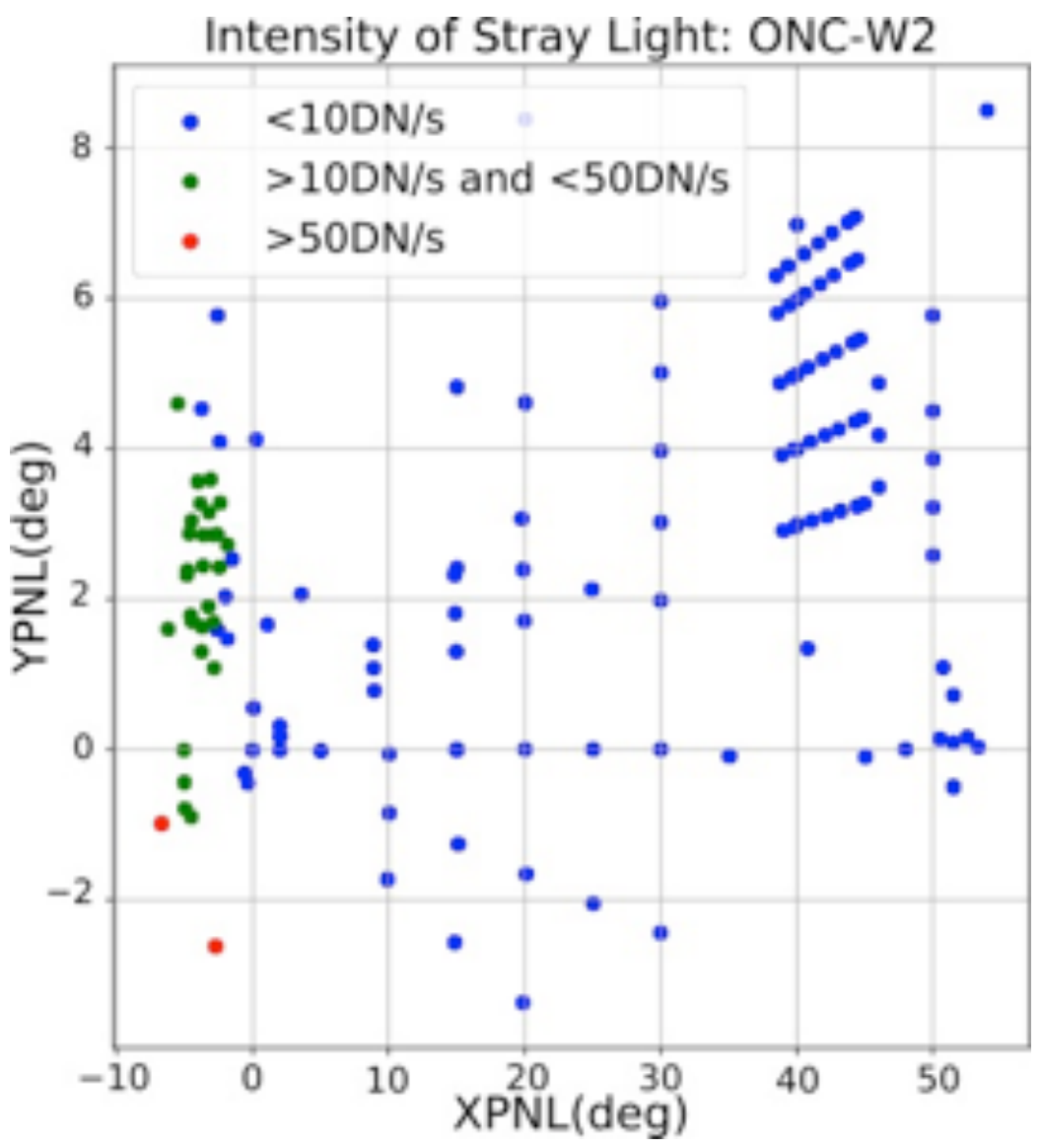

**Figure 5.10.** The intensity classification of the ONC-W2 stray light. The stray light is weaker than 50 DN/s (blue and green) and will not be detected due to short exposure time for the ONC-W2, <20 ms.

### 5.4. Flat-field Correction

The flat-fields of the ONC-W1 and -W2 were measured by using the portable self-emissive flat panel (CABIN CL-5300L), which was also used to measure the flat-field of ONC-T prior to launch in Tanegashima (launch site) (see Suzuki et al., 2018). In this section, first, we evaluate the uniformity in reflectance of the flat panel used for the preflight measurement. Then, the effect of emission angle (*e*) from the flat panel is discussed and the emission angle effect-corrected flat fields of the ONC-W1 and -W2 are shown.

#### 5.4.1 Flat-Fields Based on the Preflight Calibration Data

**Figure 5.11** shows the radiance distribution of the panel taken by the calibration camera (Orion StarShoot G3 (Monochrome)) when the emission angle is set to be zero to show the reflectance uniformity of the flat panel calibration source. An edge of luminous area with a dimension of 350 x 280 mm is clearly seen in the left figure. The ONC-W1 and -W2 took images of this panel from a 30 cm distance during the preflight calibration measurements. The approximate FOV of the ONC-W1 and -W2, during the calibration, are shown by the area enclosed by white rectangles in the figure. Since the reflectance uniformity of the panel within this area is essential for a flat field evaluation of the ONC, the color scale in both figures shows values normalized by the mean radiance within this area. **Figure 5.11(b)** is the same as **Fig. 5.11(a)** but the color scale is expanded to enhance the reflectance uniformity of the panel within FOV. The standard deviation of the radiance within the FOV is measured by the calibration camera to be 1.7%. Unlike the ONC-T, an emission angle dependence of the irradiance from the panel needs to be taken into account since the ONC-W1 and -W2 have wider FOVs. Since the flat plane is a self-emitting luminous plane, only the emission angle should be of concerned in this experiment. **Figure 5.12** shows the configuration for the measurements of the emission angle dependence of the radiance properties of the panel. In this measurement, the radiances from the panel were measured by a calibration camera at $e$ = 0°, 10°, 20°, 30°, and 35°. **Figure 5.13** shows the emission angle dependence of the radiance measured by the calibration camera. The radiance is defined as the mean counts within the area of FOV (enclosed by white rectangle in **Fig. 5.11**) normalized by the equivalent value at $e$=0°. This shows that the radiance of the panel gradually decreases as emission angle increases. For example, the radiance is 14% darker at $e$ = 35° which corresponds to the edge of the FOV of the ONC-W1. Due to this range in emission angle across the ONC-W1 and –W2 FOV, a single raw image of the flat panel taken at $e$=0° by the ONC-W1 or -W2 cannot be used as a flat field image, unlike the case of the narrow angle FOV ONC-T. Thus, a correction to account for this effect is necessary to acquire the actual flat field data for the ONC-W1 and -W2 from a single flat panel image. The count distributions in the flat panel images taken at $e$=0° by the ONC-W1 and -W2 contain both the actual flat field specification of the camera and the emission angle dependency of radiance across the FOV. From the pre-flight measurement, we calculated the emission angle dependent radiance model by fitting the data in **Fig. 5.13** (red line). An actual emission angle from the center of the FOV to any pixel located in an image can be calculated by using the distortion coefficients for both cameras (Suzuki et al., 2018). The raw counts in the image of the flat panel are then divided by the normalized radiance model at each angle shown in **Fig. 5.13**. By applying this correction to the raw counts for all the pixels, the actual sensitivity for the flat field images for the ONC-W1 and -W2 are obtained. **Figure 5.14** shows the sensitivity flat fields of the ONC-W1 and -W2 calculated using this methodology. The values in these images are normalized by a mean value within 50 pixels from the center of the image (the center area). The precision of these flat fields is defined by the spatial inhomogeneity of the flat panel (~1.7%) since a sensitivity of the calibration camera was confirmed to be much more stable than this value. The absolute sensitivity of the center area measured by using the integrating sphere is presented for both wide angle cameras in **Sec. 5.5**.

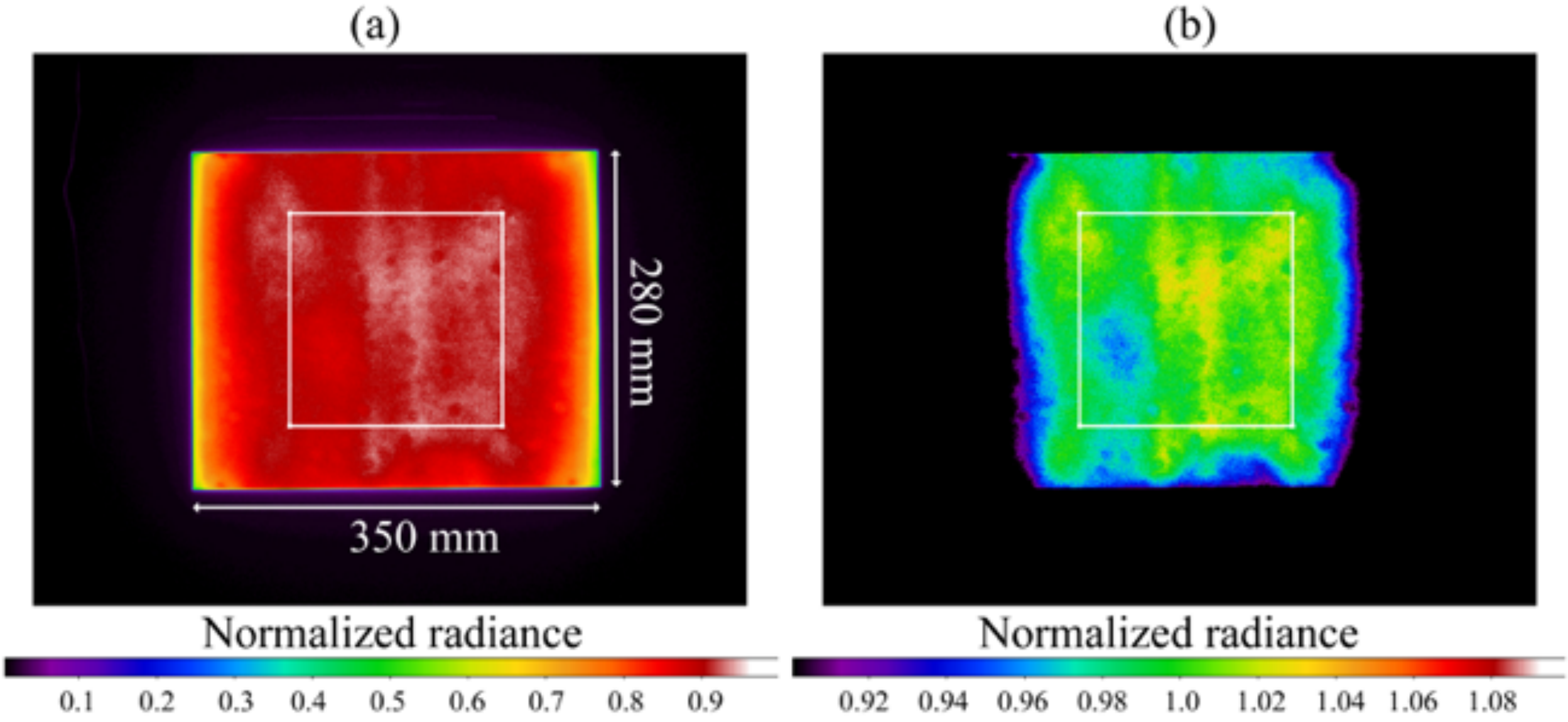

**Figure 5.11.** Radiance distribution of the self-emissive flat panel taken by the calibration camera when the emission angle is 0°. The two images are same except for the color scale in order to enhance the inhomogeneity in image (b). The white square in image (a) indicates the approximate FOV of the ONC-W1 and -W2 in the pre-flight calibration.

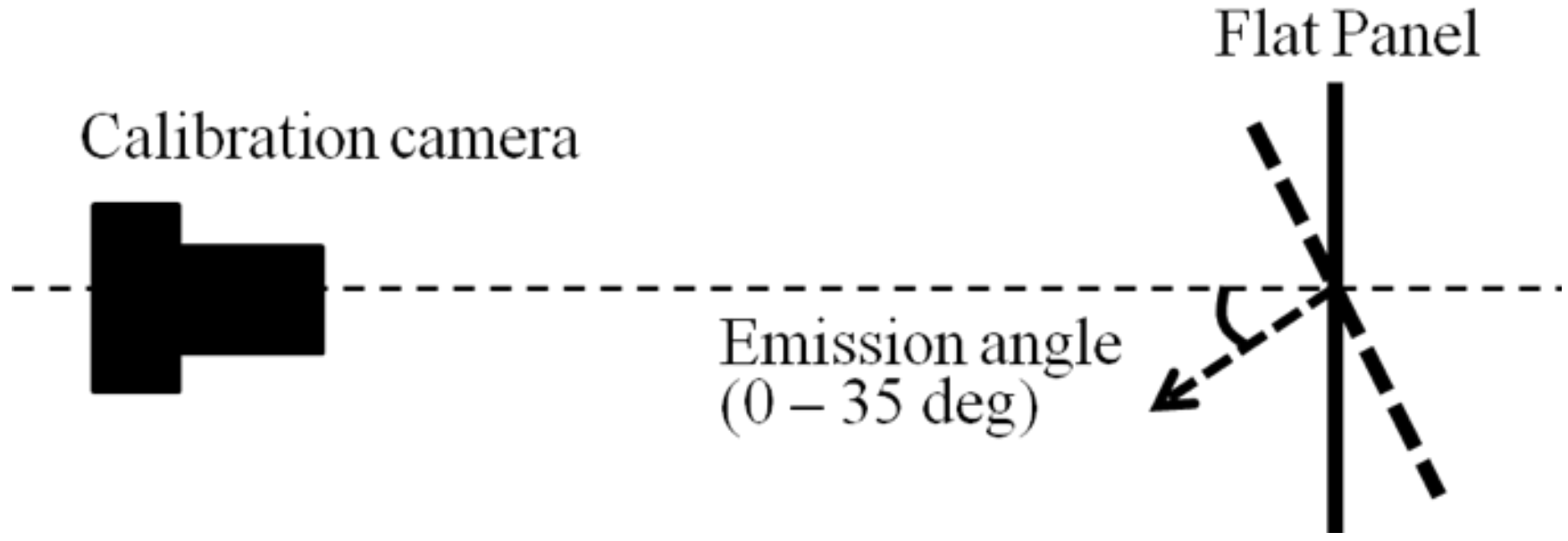

**Figure 5.12.** The configuration for the measurement of the emission angle dependence of the radiance across the wide-angle camera FOVs.

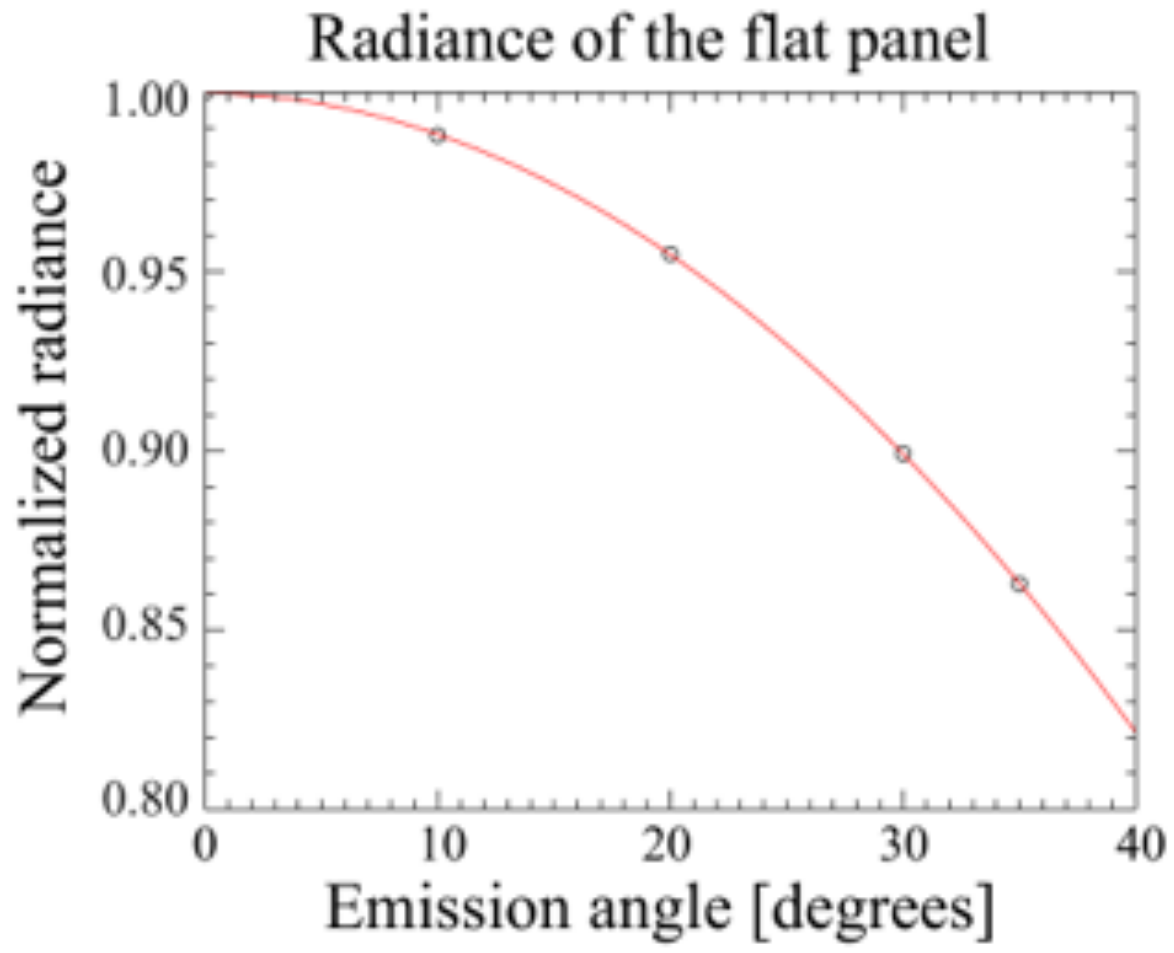

**Figure 5.13.** The relationship of normalized radiance and the emission angle taken by the calibration camera.

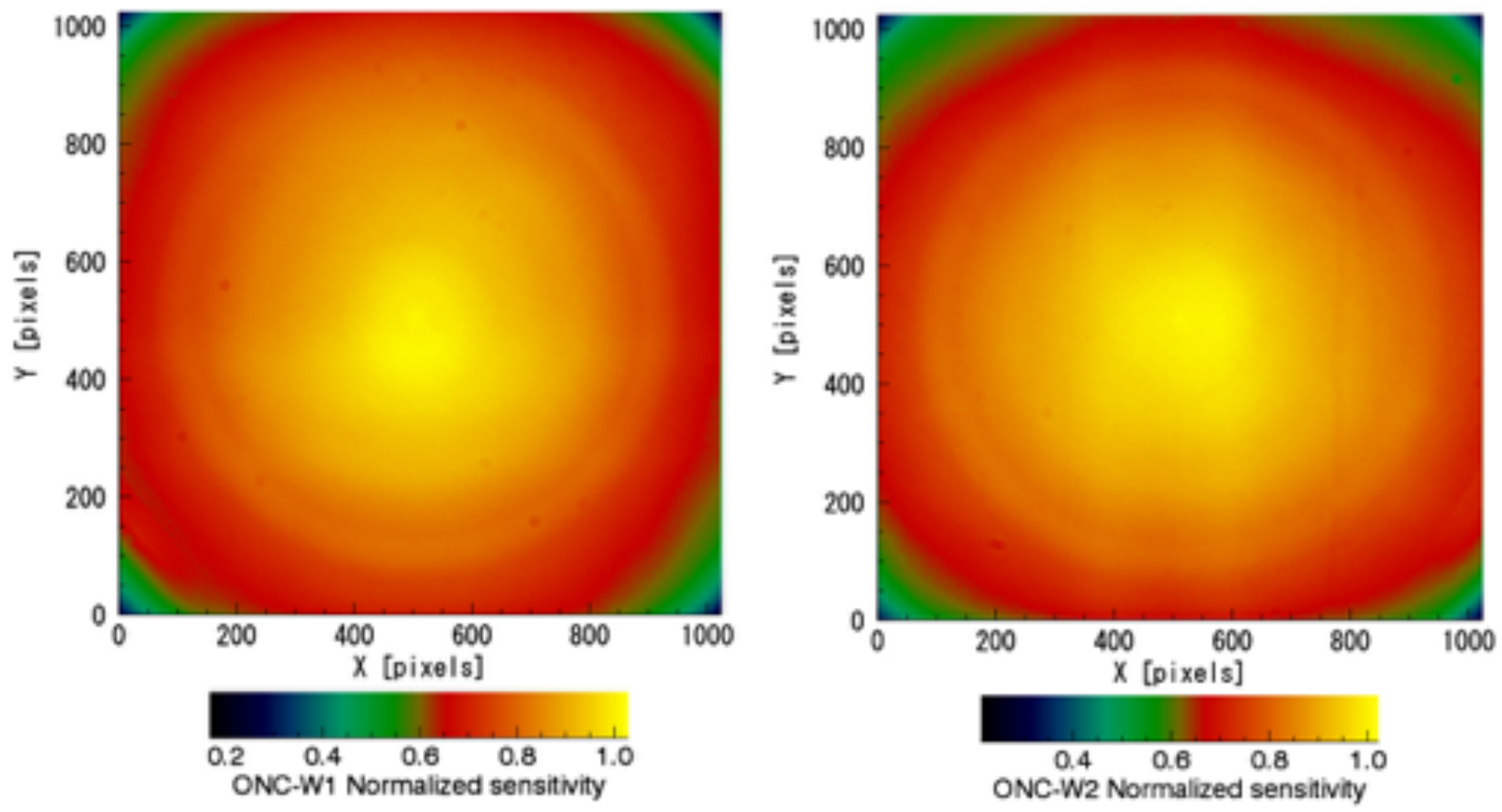

**Figure 5.14.** The sensitivity flat field for the ONC-W1 (left) and ONC-W2 (right).

### 5.4.2. Flat-Field Evaluation Based on Inflight Star and Planet Observations

The validity of the sensitivity across the FOV of the ONC-W1 is examined using 35 Jupiter images obtained from February to March 2016, approximately two years after Hayabusa2's launch. In this observation period, Jupiter is located at various positions within theONC-W1's FOV. Same as described in Sec. 3.6, we measure Jupiter's brightness by integrating the digital counts around Jupiter (to include brightness distributed outside the physical disk by the PSF). A dark image for this observation period is created by averaging the 35 images where bright spots, such as Jupiter, bright stars, and counts from cosmic rays hits are excluded. Flat field correction based on the pre-flight measurements for the ONC-W1 was not performed for the Jupiter images due to difference in the brightness dependence on the incident angle in the optics between a point light source and a surface light source, which is significant at a large incident angles. Though the Jupiter observations may not be used for validating the flat field, they can be used to monitor the consistency in the ONC-W1's sensitivity across its FOV.

Since Jupiter's brightness varies with phase angle, we perform a phase angle correction based on the Jupiter brightness measurements with the Cassini/Imaging Science Subsystem (ISS) (Mayorga, et al., 2016). Here, we use the coefficients for the green-band of the ISS, which has a similar wavelength coverage as the ONC-W1. The possible error from the phase angle correction is within ~1% (Mayorga, et al., 2016). The phase angle value between the Sun, Jupiter and the Hayabusa2 spacecraft decreased from 4° to 2°, and then again increased to 4° during this observation period. Finally, we corrected the brightness variation due to the distance between the Sun and Jupiter and between Jupiter and the Hayabusa2 spacecraft. After these corrections, the measured brightness of Jupiter by the ONC-W1 is stable within a standard deviation of 1.8% (**Fig. 5.15(a)**) with a weak incident angle dependence with $\cos^{0.5} i$, where $i$ is the incident angle. An ideal pinhole optics has a $\cos i$ dependence for a point light source.-Note that the CCD temperature was -25.1±0.3°C in this observation period, and the sensitivity variation due to CCD temperature was not expected to be significant.

The sensitivity validation of the ONC-W2 was examined with a series of 60 star images taken between February 2016 and November 2017. Only stars brighter than 0.5 V-magnitude (Achernar, Canopus, Procyon, Rigel, Sirius, and Alpha Centauri A and B) were used since they provided larger DN signals than many of the hot pixels and cosmic ray hits due to long exposure time of the ONC-W2 observations. **Figure 5.15(b)** shows the distribution in the sensitivity deviations deduced from the star observations, in which each sensitivity deviation is measured with the same approach described in **Sec. 3.8**. The standard deviation of the sensitivity in the FOV of ONC-W2 is 4.4 %, which is somewhat larger than those of the ONC-T and ONC-W1. The large standard deviation could be from the larger number of hot pixels in ONC-W2, which may occasionally overlap star locations. If the largest (15%) and the second largest (9%) deviations are excluded from the calculations, then the standard deviation reduces to 3.7 %.

In summary, it has been confirmed that both the ONC-W1 and -W2 do not have any clear sensitivity defects over their FOVs. The deviations in sensitivity, after flat-field correction, are within a few percent in large parts of their FOVs. In addition, we have not observed any clear sensitivity degradation within the ONC-W2 detector during the 1.5-year star observation period. The validity of the flat field correction for a surface light source will be re-evaluated with close-up observation of Ryugu.

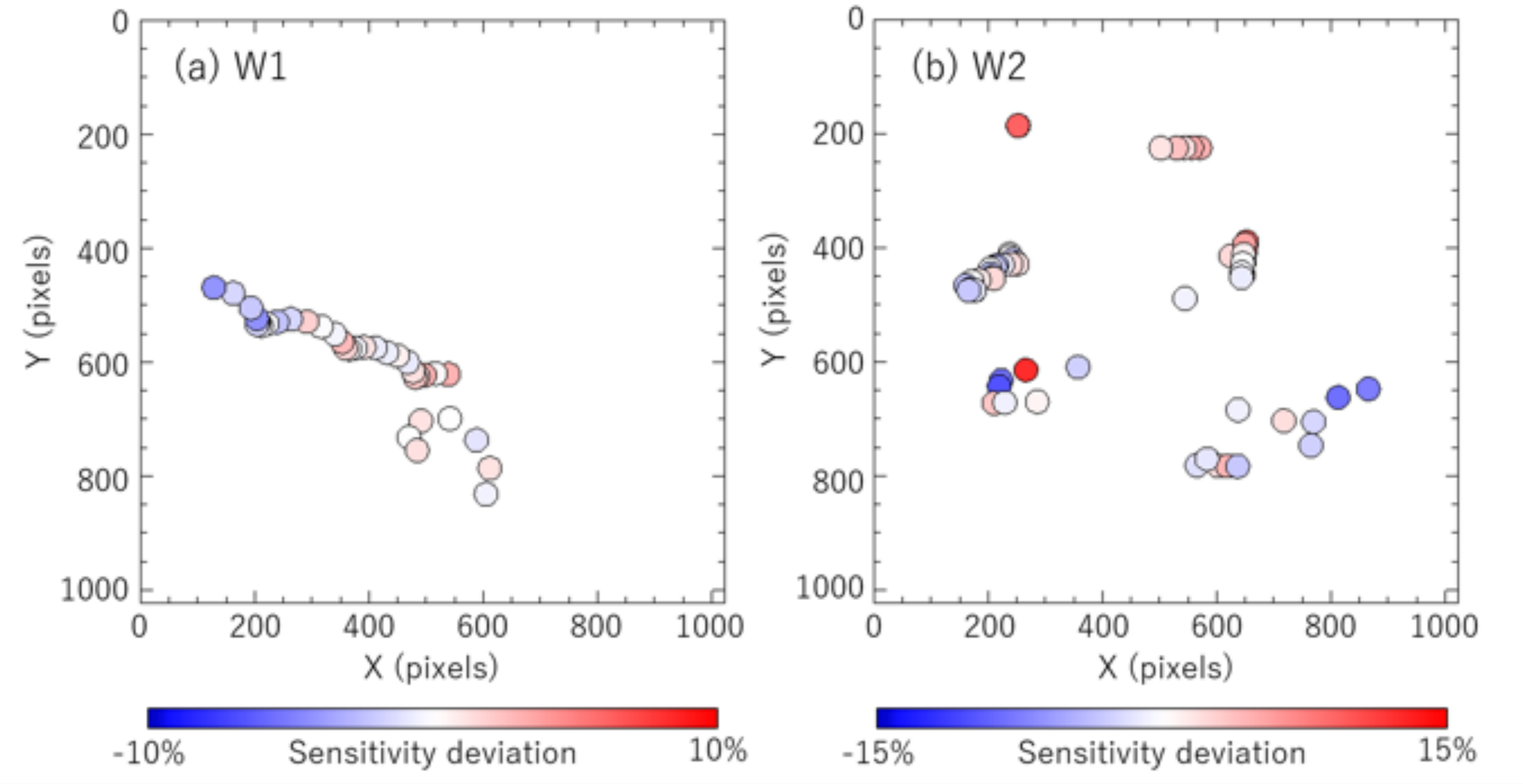

**Figure 5.15.** Sensitivity deviations in (a) ONC-W1 FOV and (b) ONC-W2 FOV as measured from Jupiter and star observations, respectively.

### 5.5 Radiometric Calibrations

#### 5.5.1 Sensitivity Calibration Based on the Preflight Data

The radiometric sensitivities of the ONC-W1 and -W2 were measured by using an integrating sphere prior to launch. The details of the integrating sphere and the procedures to quantify the sensitivity for the ONC-T are described in Kameda et al. (2017). Thus, the differences between the procedure used to quantify the sensitivity of the ONC-T and those used to quantify the sensitivity of the ONC-W1 and -W2 are described here. The major differences come from the wider FOVs of the ONC-W1 and -W2 (~70 deg) relative to narrow FOV of the ONC-T (~6.5 deg). Due to this FOV difference, the luminous plane of the integrating sphere could not fully fill the FOVs of the ONC-W1 and -W2. **Figure 5.16** displays the image of the integrating sphere taken by the ONC-W1, which shows a bright circular area (the luminous plane) surrounded by a dark area. Although it is difficult to derive the sensitivity for all pixels by using this single image, the averaged sensitivity of the pixels within a 50-pixel radius from the center of the image can be measured. By combining these sensitivity values and the normalized sensitivity flat-fields shown in **Fig. 5.14**, the sensitivity of all pixels across the FOV can be characterized. Since the ONC-W1 and -W2 are panchromatic cameras with a broad band pass, the mean radiance of the integrating sphere, $\overline{F_{IS,\mathrm{cam}}}$ [W/m²/sr/μm] is calculated by

$$\overline{F_{IS,\mathrm{cam}}} = \frac{\int_{0.45\ \mu\mathrm{m}}^{0.65\ \mu\mathrm{m}} F_{IS}(\lambda)\ \Phi_{\mathrm{cam}}(\lambda)\ \mathrm{d}\lambda}{\int_{0.45\ \mu\mathrm{m}}^{0.65\ \mu\mathrm{m}} \Phi_{\mathrm{cam}}\ (\lambda)\ \mathrm{d}\lambda}, \qquad (\mathrm{cam=W1\ \ or\ \ W2}) \tag{5.2}$$

where $F_{IS}(\lambda)$ is the known spectral radiance profile of the integrating sphere (see **Fig. 7** of Kameda et al., 2017) and $\Phi_{\mathrm{cam}}(\lambda)$ is the system efficiency function of the ONC-W1 or -W2 (See **Fig. 2** of Suzuki et al. 2018). The current of the integrating sphere was set to 2.75 A for these measurements. The radiometric sensitivities, $\overline{S_{\mathrm{cam}}}$ [DN/s]/[W/m²/sr/μm], for the center area of the FOVs are obtained by

$$\overline{S_{\mathrm{cam}}} = \frac{\bar{I}}{\overline{F_{IS,\mathrm{cam}}}}\ , \tag{5.3}$$

where $\bar{I}$ is the signal [DN/sec] averaged over pixels within a 50-pixel radius from the center of the image. The $\overline{S_{\mathrm{cam}}}$ values obtained for the ONC-W1 and -W2 are $1.38 \times 10^3$ and $3.84 \times 10^3$ [DN/s]/[W/m²/sr/μm], respectively. It is noted that these coefficients are reliable only when the spectral radiation of an object has a comparable color temperature with the integrating sphere. Thus, the mean radiance obtained by using these coefficients is reliable only within an order of magnitude estimation, in general. The results of this radiometric calibration are summarized in **Table 5.3**.

### 5.5.2 Sensitivity Calibration Based on the Inflight Jupiter and Star Observations

The sensitivity of the ONC-W1 was measured based on the Jupiter observations, using the same images described in **Sec. 5.4.2**. The sensitivity was evaluated by comparing observed Jupiter digital counts and expected Jupiter counts based on the ONC-W1's specification (Suzuki et al., 2018), which was determined prior to launch. We used the Jupiter images since even bright stars (0 V-magnitude) provide only less than 200 DN in total signal in an ONC-W1 image. These stellar observations are comparable to counts from hot-pixels and typical cosmic ray hits, due to the short exposure times of the ONC-W1 (up to 5.57 sec), while Jupiter provides more than 2000 DN in total signal.

Same as with star calibrations, the digital counts expected from Jupiter can be calculated by using **Eqs. (3.18) and (3.19)**. Unlike stars, however, the disk-integrated Jupiter brightness observed by the ONC-W1 has a geometrical dependence on the observation geometry. The expected irradiance (W/m²/μm) from Jupiter, $F_J$, can be expressed as

$$F_J(\lambda) = A_J(\lambda, \alpha)\Omega_J \frac{J_S(\lambda)}{\pi}\left(\frac{1AU}{D_{S-J}}\right)^2, \tag{5.4}$$

where $\alpha$ is the phase angle at the time of observation, $A_J(\lambda, \alpha)$ is the disk-equivalent reflectance of Jupiter, $\Omega_J$ is the solid angle of Jupiter as seen from the Hayabusa2 spacecraft, $J_S(\lambda)$ is the modeled solar irradiance (W/m$^2$/μm) at 1 AU, and $D_{S\text{-}J}$ is the distance between the Sun and Jupiter in AU. The disk-equivalent albedo is calculated from

$$A_J(\lambda,\alpha) = A_J(\lambda,\alpha_0)\frac{\Psi(\lambda,\alpha)}{\Psi(\lambda,\alpha_0)} , \tag{5.5}$$

where $\Psi(\lambda,\alpha)$ represents the phase angle dependence of $A_J(\lambda, \alpha)$, and $\alpha_0$ is the reference phase angle. In this study, we used a solar irradiance model from Gueymard (2004), the Jupiter reflectance data from Karkoschka (1998) taken in 1995 at a phase angle of 6.8° (i.e., $\alpha_0$ = 6.8°), and the phase angle dependence for the ISS green-band from Mayorga et al (2016), as described in **Sec. 5.4**. Since the apparent size of Jupiter was much smaller than 1-pixel in the ONC-W1 FOV, the solid angle of Jupiter in steradian is approximately

$$\Omega_J = \pi\frac{r_{Je}r_{Jp}}{{D_{J-H}}^2} , \tag{5.6}$$

where $r_{Je}$ and $r_{Jp}$ are the equatorial and polar radii of Jupiter, respectively, and $D_{J\text{-}H}$ is the distance between Jupiter and the Hayabusa2 spacecraft at the time of observation. Both $D_{S\text{-}J}$ and $D_{J\text{-}H}$ were more than 5 AU during the observation period.

From the 16 Jupiter images in which Jupiter was located near the center of the FOV, where variations in the sensitivity are not significant (**Sec. 5.4.2**), the sensitivity of the ONC-W1 was measured to be $(1.44 \pm 0.02) \times 10^3$ [DN/s]/[W/m$^2$/sr/μm], which is highly consistent with the sensitivity measured prior to launch. The error range is evaluated from the standard deviation of measured sensitivities from 35 Jupiter observations. It should be noted that the sensitivity value from the Jupiter observations has errors from possible temporal variations in Jupiter's reflectance (for example, a couple of percent (Karalidi et al., 2015)), uncertainties in measuring Jupiter's brightness in the ONC-W1 images (2% based on standard deviation), errors in the Jupiter reflectance (4% in absolute value by Karkoschka (1998)), and errors due to the phase angle dependence (up to 1% by Mayoruga, et al. (2016)). In addition, Jupiter and the integrating sphere we used in the pre-flight measurements have different spectral profiles. These could provide differences between the sensitivities derived from pre-flight and inflight measurements.

The sensitivity of the ONC-W2 was measured based on star observations conducted from February 2016 to November 2017. Stars images used for this measurement are same for the evaluation of the ONC-W2 flat-field described in **Sec. 5.4**. Comparing the observed DN to the expected DN based on star irradiances, the sensitivity of the ONC-W2 in 6 star observations was evaluated as $(4.03 \pm 0.07) \times 10^3$ [DN/s]/[W/m$^2$/sr/μm] at the center of FOV, which is consistent with the sensitivity from the preflight measurement. In addition, this value corresponds to a 47% lower sensitivity than expected from the ONC-W2 specification, which was commensurate with the results from the Earth observations, where the ONC-W2 measured Earth's reflectance to be 40% darker than the reference value (Suzuki et al., 2018).

It can be concluded that both the ONC-W1 and -W2 basically retained their preflight sensitivities during the cruise phase. Continuous sensitivity monitoring with bright targets will provide a continuing measure of the stability of the sensitivity inflight for these camera detectors.

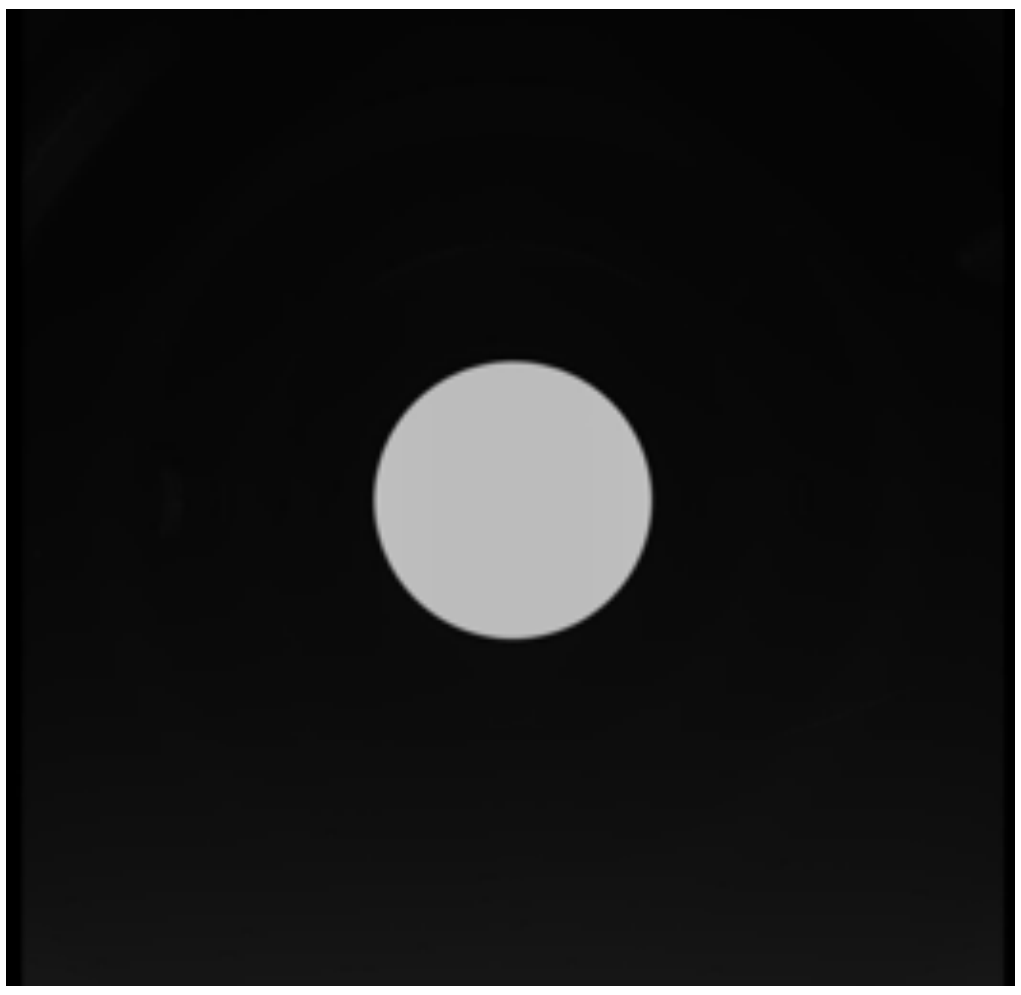

**Figure 5.16.** An image of the integrating sphere taken by the ONC-W1.

**Table 5.3.** Summary of the radiometric calibration of the ONC-W1 and -W2 based on preflight integrating sphere measurements .

| Camera | Date | Current setting of the integrating sphere. | $S_{\mathrm{cam}}$ [DN/s]/[W/m$^2$/μm/sr] |
|---|---|---|---|
| ONC-W1 | 27 March 2014 | 2.75 A | $1.38 \times 10^3$ |
| ONC-W2 | 27 March 2014 | 2.75 A | $3.84 \times 10^3$ |

### 5.5.3. Point Spread Functions (PSFs)

The PSF of the ONC-W1 is evaluated from observations of bright targets, i.e. Mars, Jupiter, and stars brighter than 1 V-magnitude (Betelgeuse, Capella, and Procyon). Mars images were taken on 31 May 2015, and Jupiter images are the same images used for the flat-field and sensitivity evaluation. The star images were taken on 12 March 2017. The PSF of the ONC-W2 was also evaluated from stars brighter than 1 V-magnitude, taken between February 2016 to November 2017. **Figure 5.17** shows a summary of the FWHM of each bright target for the ONC-W1 and -W2. The FWHM values are evaluated from a two-dimensional Gaussian fitting method, same as that used for measuring the PSF of the ONC-T (Suzuki et al., 2018). Due to the short exposure time of the ONC-W1 (up to 5.57 sec), only a limited number of star images were used, while 142 star samples were available for measuring the PSF of the ONC-W2, whose exposure time was 44.56 sec.

Around the center of the FOV, the FWHM was ~1.8 pixels, similar to the ONC-T for both ONC-W1 and -W2, and they are almost same as (slightly better than) the FWHM determined from the pre-flight measurements. The larger FWHM at the image corner in the FOV of the ONC-W2 could be caused by a vignetting effect from a baffle, same as for the ONC-T (Suzuki et al., 2018). While the FWHM of the ONC-W1 shows an increasing trend with distance from the center of the FOV, the FWHM values in large parts of both the ONC-W1 and -W2 FOVs are smaller than each FHWM value at the image corner from the pre-flight measurements, indicating no significant degradation of the PSF after launch. It should be noted that since the stars are 20-50 times darker than either Jupiter or Mars in the ONC-W1, due to the short exposure time of the ONC-W1 stellar observations (5.57 sec), star observations provide rather large errors in the FWHM of the ONC-W1.

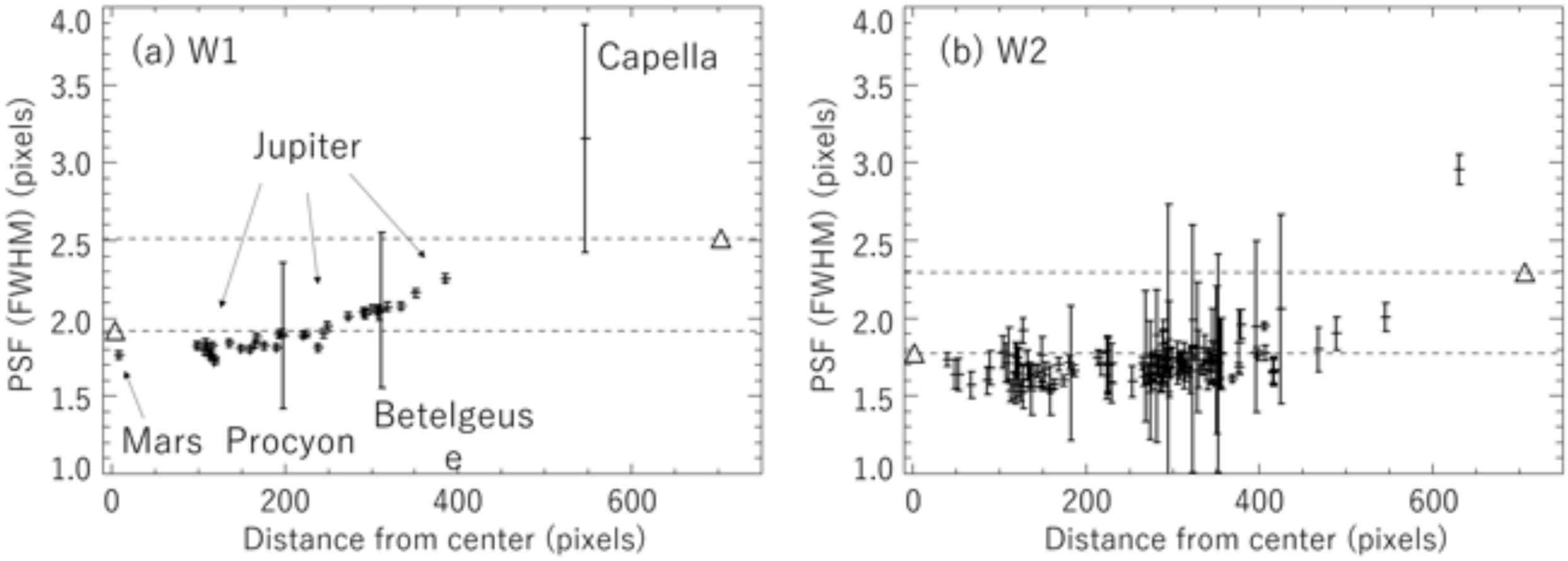

**Figure 5.17.** The measured FWHM of (a) the ONC-W1 and (b) the ONC-W2 PSF as a function of the pixel distance from the center of FOV. Triangles and dotted lines indicate PSF values at the center and corner of the FOV measured in pre-flight experiments.

## 6. ONC System Alignment with the Spacecraft

Understanding the alignment of the ONC system with respect to the spacecraft is essential for comparing the measurements within the different FOVs or matching footprints with other instruments, such as the TIR and NIRS3, or even when we compare images taken from the ONC-T and ONC-W1. Moreover, the alignment information is necessary to calculate the accurate attitude and position of the spacecraft from images during operation. In this section, we present the results of measurements deriving the viewing direction of the ONCs relative to the spacecraft attitude. As mentioned in Suzuki et al. (2018), the FOVs of the ONC-W1 and ONC-T were designed to align in the $-Z_{SC}$ direction within the spacecraft coordinate system, while that of the ONC-W2 was slanted by approximately 30° from the $-Z_{SC}$ direction. However, actual alignments may have small offset from these designed values due to an ineluctable error in assembly procedures prior to launch or offsets created by the strong vibrations during launch of the Hayabusa2 spacecraft. These offsets are measured precisely by using star filed images taken during the cruise phase. Star field images used in the determination and verification of the distortion coefficients (see Suzuki et al., 2018) of the three cameras are also validated in this measurement. In **Secs. 3 and 6** of Suzuki et al. (2018), celestial coordinates were fitted to the star field images to verify the estimated distortion coefficients (See **Fig. 5** for the ONC-W1 and -W2, and **Fig. 12** for the ONC-T). From these results, right ascension and declination angles corresponding to the center of the FOV at the observation time for each camera can be determined. On the other hand, information of the spacecraft attitude at these observation periods are also available from the telemetry data. This allows us to predict pointing directions and the centers of the FOVs of three cameras during the observation periods by assuming that the cameras were perfectly assembled to the spacecraft as designed. Thus, the offset values between the predicted direction of an optical axis of each camera and that estimated from the actual star image can be defined as alignment information. **Table 6.1** summarizes the observation dates and times of the star field images, the observed center of the FOV in the celestial coordinate system, the center of the FOV predicted by the attitude information, and the offset between them (i.e., the alignment information) in terms of angles and in image coordinate (in pixels). The last column of **Table 6.1** gives the alignment information in the image coordinate (for definitions see **Fig. 1** of Suzuki et al., 2018) defined as

$$\Delta H = H_{obs} - 511.5 \text{ [pixels]} \quad (6.1)$$

where, the $H_{obs}$ is observed location of the predicted center (e.g., the direction of $-Z_{SC}$ for the ONC-W1 and ONC-T) and the value 511.5 pixels is the location of image center. We also can express alignment by Euler angles with respect to the spacecraft coordinate system. The matrix transforming from the spacecraft coordinate system is

$$M = \begin{pmatrix} \cos\Theta_Z & \sin\Theta_Z & 0 \\ -\sin\Theta_Z & \cos\Theta_Z & 0 \\ 0 & 0 & 1 \end{pmatrix} \begin{pmatrix} \cos\Theta_Y & 0 & -\sin\Theta_Y \\ 0 & 1 & 0 \\ \sin\Theta_Y & 0 & \cos\Theta_Y \end{pmatrix} \begin{pmatrix} 1 & 0 & 0 \\ 0 & \cos\Theta_X & -\sin\Theta_X \\ 0 & \sin\Theta_X & \cos\Theta_X \end{pmatrix}, \quad (6.3)$$

where $\Theta_X, \Theta_Y,$ and $\Theta_X$ are Euler angles from the spacecraft coordinate system to the instrument coordinate system. The Euler angels for ONC are also listed in **Table 6.1**.

| Camera | Observation time for the star images. YYYY-MM-DDTHH:mm:ss | Observed center of the FOV in (RA, DEC) [°]. | Predicted center of the FOV in (RA, DEC). [°] | Offset in angle. [°] | Offset of the center of FOV in the image coordinate (alignment information) (ΔH, ΔV) [pixels]. | Euler angles ( $\Theta_Z, \Theta_Y, \Theta_X$ ) [Degree] | Distortion parameters* $(\epsilon_1, \epsilon_2)$, $r' = r + \epsilon_1 r^3 + \epsilon_2 r^5$. | Focal length* (mm) |
|---|---|---|---|---|---|---|---|---|
| W1 | 2015-02-19T09:50:10 | (138.880,6.150) | (138.789,6.475) | 0.313 | (-3.5, -2.5) | (0.0, -180.2728, 0.1948) | (3.134×10⁻⁷, -1.716×10⁻¹³) | 10.22 |
| W2 | 2014-12-11T13:38:12 | (136.500,6.020) | (136.484,6.051) | 0.051 | (0.5, -0.5) | ( -270, -121, 0 ) | (2.893×10⁻⁷, -1.365×10⁻¹³) | 10.38 |
| T | 2014-12-11T14:52:53 | (77.388,21.961) | (77.532, 21.996) | 0.1510 | (9.5, -22.5) | (0.0, -179.9415, 0.1386) | (-9.28×10⁻⁹, 0) | 120.50±0.01 |

*Suzuki et al. (2018)

**Table 6.1.** Alignment information for the three cameras estimated by using star field images.

### 7. NIRS3

Hayabusa2 payload includes the near infrared spectrometer NIRS3, whose purpose is to detect and measure the absorption around 3 μm, associating with the hydration of mineral species (Iwata et al., 2017). NIRS3 is designed to observe wavelengths from 1.8 to 3.2 μm. It is important to connect the spectra from the ONC-T and NIRS3 instruments to measure the spectral characteristics over a broad range of wavelengths, from the ultraviolet (UV) through the visible (vis) to the near infrared (NIR). This will allow us to examine the mineralogy of Ryugu using the integrated spectra from both instruments. In this section, we show the ONC-T and NIRS3 spectra of the Moon observed almost simultaneously on 5 December 2015 (Kitazato et al., 2016). The consistency of the spectra is discussed based on the comparisons with the SP/SELENE model (Yokota et al., 2011; Kouyama et al., 2016) for vis-NIR spectra, since there is no overlap in the wavelength range between the two instruments. In this comparison, we used a modified SP/SELENE model which covers longer wavelengths (~2 μm) based on the updated reflectance data by Yokota et al. (2012) than the model in Yokota et al. (2011) and Kouyama et al. (2016). The reflectance spectrum from the SP model was corrected with the ROLO reflectance model since the SP model shows somewhat darker and brightener tendencies in shorter and longer wavelength ranges, respectively (Ohtake et al., 2013).

First, the alignment of the ONC-T and NIRS3 are calculated based on the SPICE Instrument Kernel. The NIRS3 alignment has been determined through scan observations of the Earth combined with slews of the spacecraft attitude. The error in alignment is within +/-0.005°, corresponding to ~5% of the size of field of view. **Figure 7.1** shows the alignment of the NIRS3 in the FOV of the ONC-T, showing little offset from the center of the FOV. The footprint of the NIRS3 is {[ 476.5, 454.3]; [492.7, 472.1]} pixels in the ONC-T image coordinate system. NIRS3 only obtained spectra of part of the Moon's surface due to its narrow footprint (**Fig. 7.2**). The comparisons of the lunar spectra from different parts of the surface are shown in **Fig. 7.3**. We also display the model spectra from SP/SELENE to reasonably extrapolate the ONC-T spectrum to the NIRS3 spectrum. There seems to be little offset between the spectra of the ONC-T, NIRS3, and model spectra. Note that the radiometric calibration of the NIRS3 was primarily based on preflight calibration test data. The SP model spectra may underestimate the radiance for large incidence angle regions as shown in Kouyama et al. (2016), which is of importance since the footprints of the NIRS3 are located at higher latitude regions where the incident angle is large. Thus, we shift the SP-model spectra to fit the ONC-T v-band radiance and the NIRS3 spectra to fit the SP-model in the range between 1819 to 2038 nm. We calculated the offset factor between the ONC-T and NIRS3 as 0.87 (**Fig. 7.4**). Using this value, we can evaluate slope of the intermediate wavelength from 0.95 to 1.8 μm. This wavelength range was observed by ground-based telescopes (Moskovitz et al.,2013; Le Corre et al., 2017) and is reflective of the mineralogical composition of the surface. Moreover, there is a significant difference in the slopes in this wavelength range, which was observed by Moskovitz et al. (2013) and Le Corre et al. (2017). Thus, we may observe some heterogeneity in the slope in this wavelength range over the surface of Ryugu.

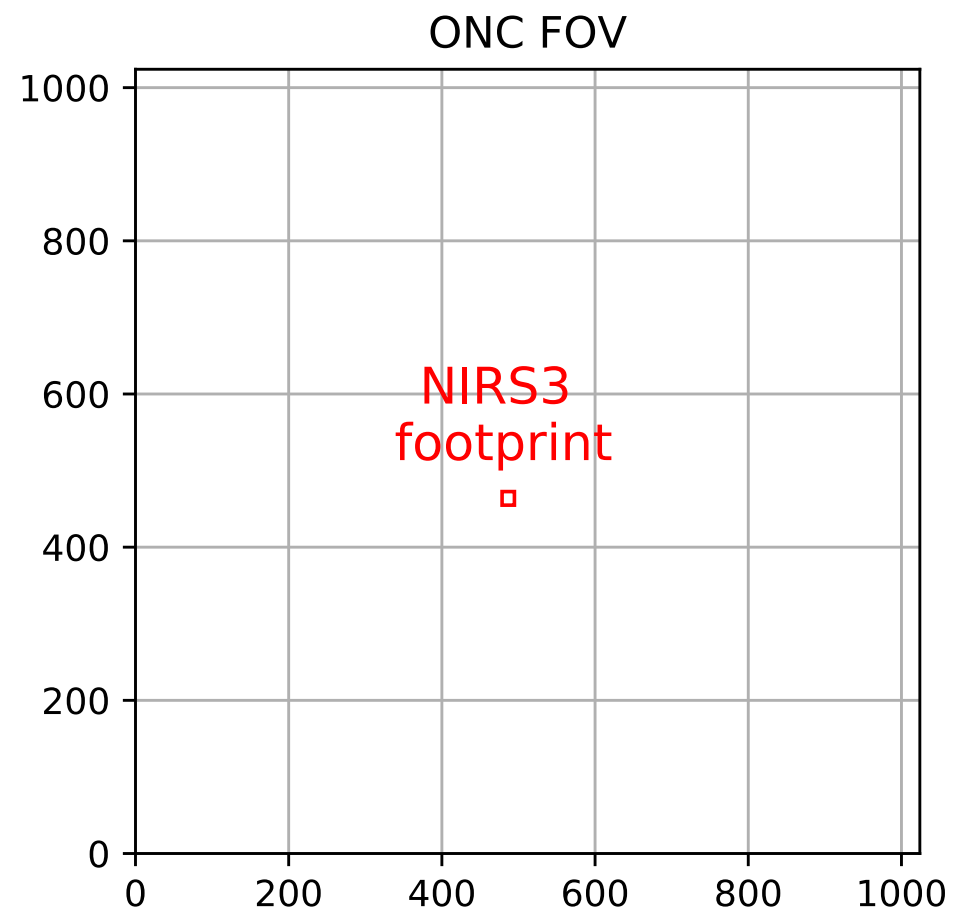

**Figure 7.1.** The footprint of the NIRS3 in the FOV of the ONC-T.

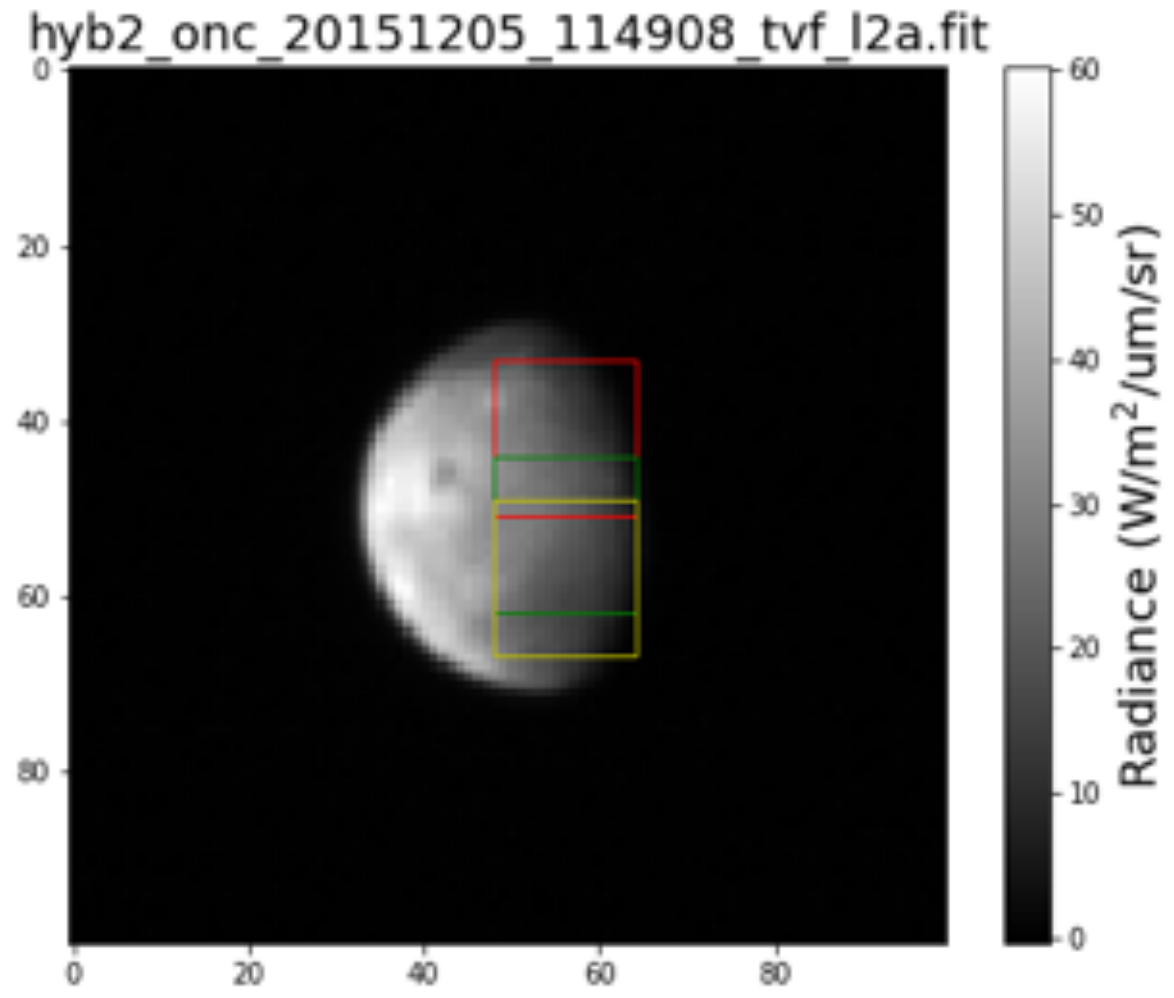

**Figure 7.2.** Squares indicate the ROIs of the NIRS3 observations in an ONC-T image (red (ROI1): 11:35:14 on 5 December 2015, green (ROI2): 11:42:27 on 5 December 2015, yellow (ROI3): 11:45:29 on 5 December 2015).

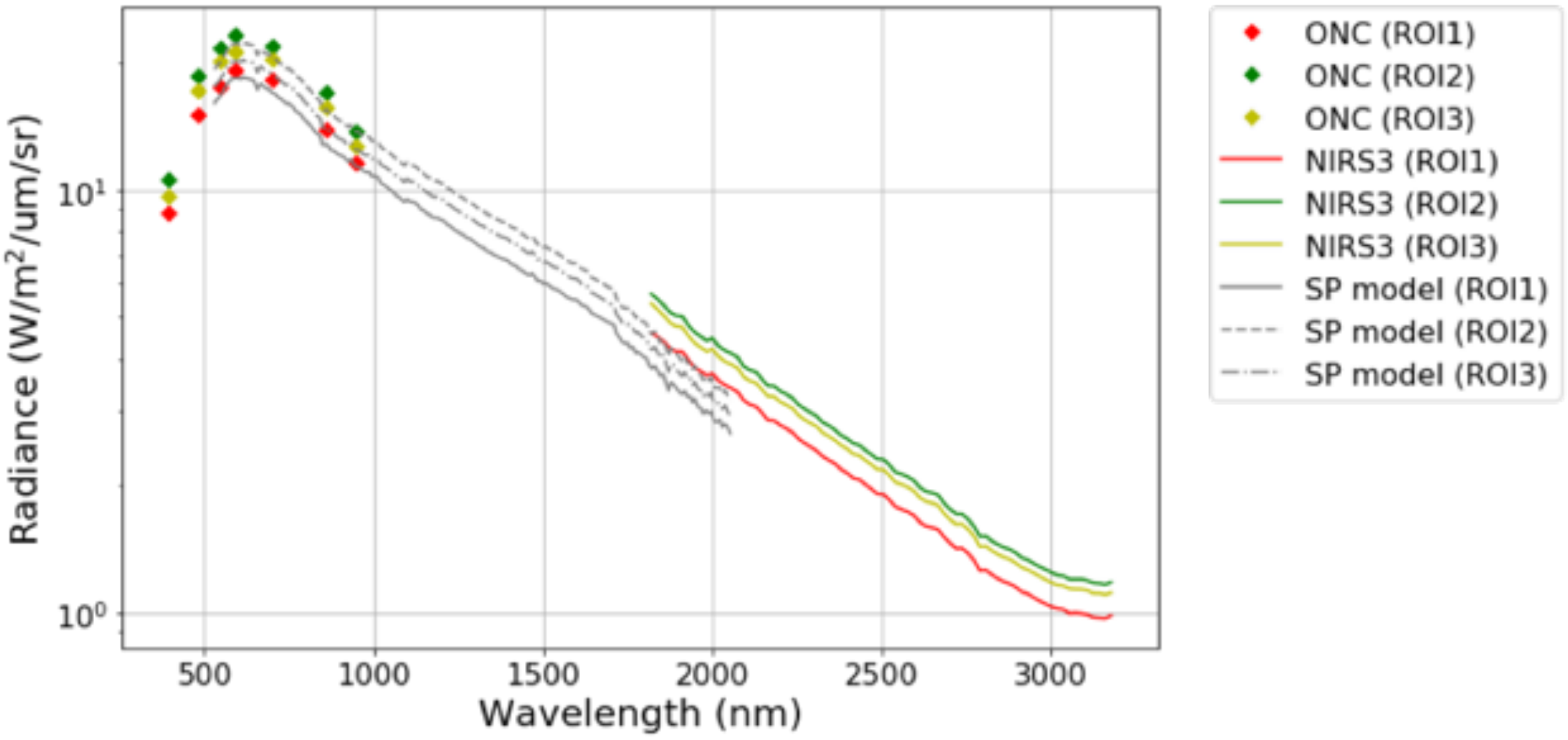

**Figure 7.3.** Radiance comparison of the lunar spectrum obtained by the ONC-T and NIRS3. The gray lines indicate the SP model spectra by Kouyama et al. (2016) and the gray symbols indicate the WAC model spectra by Sato et al. (2014). Note that radiometric calibrations of the ONC-T and NIRS3 were conducted independently.

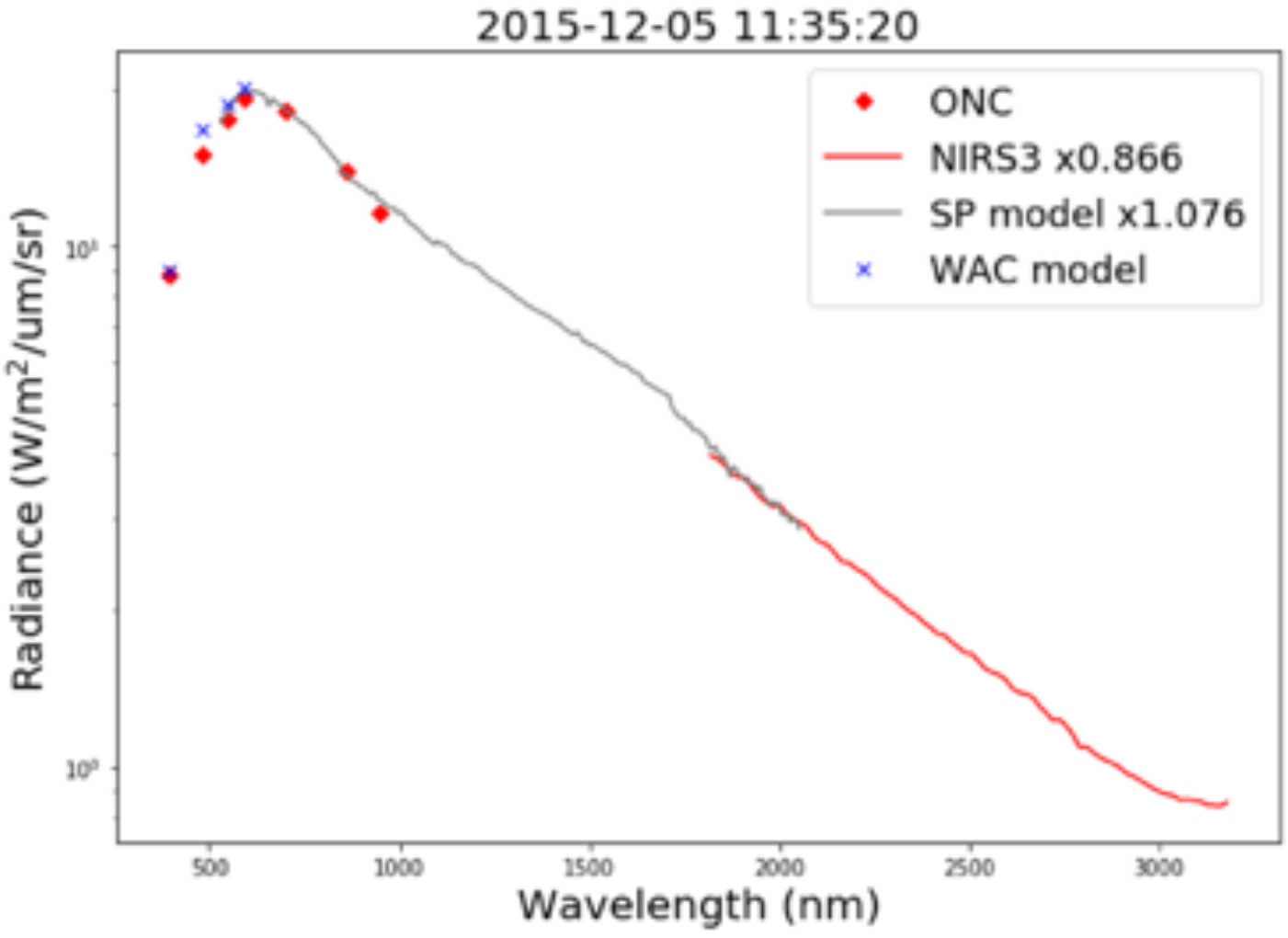

**Figure 7.4.** Corrected radiance of the Moon observed at 11:35:20 on 5 December 2015. The NIRS3 and SP/SELENE model spectra are shifted by the coefficients in the legend to fit the ONC-T radiance.

## 8. Applications to Scientific Analyses on Ryugu

We have evaluated the bias level, EMI noise, dark level, hot pixels, linearity, flat-fields, stray light, and sensitivity of the detectors and filters for the ONC system. This section addresses to the evaluation of accuracy the key scientific analyses to answer important nature of the asteroid. More specifically, one is to detect the presence of hydrated minerals, which is an indicator of the presence of water (past or present) on the asteroid, and the another is to characterize the real radiance variations on its surface, even when temperature conditions are changing. We evaluate the accuracy that is expected based on the calibration steps which are described in previous sections.

### 8.1. Signal-to-Noise Ratio

The signal-to-noise ratios (SNRs) of the ONC-T, -W1, and -W2 are important for establishing the detectability of radiance variation over the asteroid surface. The scientific quality of the images is changed by temperature conditions. Unlike orbiting missions, the Hayabusa2 spacecraft will experience a large range of temperatures, because the change in the distance between the spacecraft and the asteroid includes radiative heat change when at low altitudes. Especially, during touchdowns the CCD temperatures can be as high as 20 °C. Thereby, understanding the effects of the temperature on the detector performance is necessary in order to analyze the close-approach images. Our calibration results show the temperature causes large effects on dark noise. Here, we discuss the SNRs dependence on the CCD temperature.

The total noise $\Delta I$ can be defined as the combination of read-out noise $\Delta I_{ro}$, EMI noise $\Delta I_{EMI}$, shot noise $\Delta I_{shot} = \sqrt{I[\mathrm{DN}] \cdot g[\mathrm{e}^-/\mathrm{DN}]}/g[\mathrm{e}^-/\mathrm{DN}]$ where $g$=20.95 [e⁻/DN] is the gain factor, and dark noise $\Delta I_{dark}$:

$$\Delta I = \sqrt{\Delta I_{ro}^2 + \Delta I_{EMI}^2 + \Delta I_{shot}^2 + \Delta I_{dark}^2} \ \ [\mathrm{DN}], \tag{8.1}$$

where, only the dark noise is a function of time, but has a negligible contribution at the reference temperature (-30 °C). Because the three cameras within the ONC use the same manufactured CCDs with the almost same characteristics, the ONC cameras all image with a SNR~200 at the reference temperature. However, once the CCD temperature gets high as 20 °C, the dark noise contribution will dominant in the total noise and yields a SNR~150 for the ONC-T v-band, while the ONC-W1 and W2 will be less affected due to the shorter exposure times during operation. These value suggest that the variation across the asteroid surface over a single image can be measured within errors of 0.7% at the worst case. Moreover, there may be unavoidable effect of hot pixels (>100 DN/s), so that ~1% of the FOV area will not be useable.

### 8.2. Detectability of 0.7-μm Absorption Band

One of the most important inquiries for the spectral mapping on Ryugu is the determination of the presence of hydrated minerals, which are indicators of aqueous alteration processes on primordial asteroids. Asteroids are considered to be a possible source of water to the inner Solar System, including our own Earth (e.g., O'Brien et al, 2014; Altwegg et al., 2014). Moreover, the degree of aqueous alteration also provides insight into the heating history of asteroids. Here, we discuss the total error revealed by the calibrations, especially sensitivity, and their impact on the possible detection of hydrated minerals from ONC-T multiband images.

Our target asteroid Ryugu is categorized as a C-complex asteroid, which are theorized to have formed in the volatile-rich region in the Solar System, in the Bus and Binzel taxonomy (Bus and Binzel, 2002). The presence of hydrated minerals is usually discussed by the most prominent and unambiguous 3-micron absorption (e.g., Lebofsky, 1980; Takir and Emery, 2012). However, the 3-μm band is difficult to observe on small asteroids from the Earth. Thereby, the attributed iron charge-transfer 0.7-μm absorption, which is easier to observe, is used as the reliable proxy for hydrated minerals (e.g., Vilas et al., 1994). Ground-based observations have established the correlation between the 0.7- and 3-μm absorptions among C-complex asteroids (e.g., Howell et al., 2011). Many observations of Ryugu have been made to date to obtain mineralogy information prior to Hayabusa2's encounter (Binzel et al., 2001; Vilas 2008; Lazzaro et al., 2013; Moskovitz et al., 2013; Sugita et al., 2013; Perna et al., 2017). Most of these spectra display very flat spectral shapes, while only one of them shows evidence of a 0.7-μm absorption (Vilas et al., 2008). This suggests there may be variations in the composition across Ryugu's surface that are characterized by the presence or absence of the0.7-μm absorption feature. Thus, it is important to map out the 0.7-μm absorption by the ONC-T based on v-, w-, and x-band observations. The depth of 0.7-μm absorption can be measured by

$$d_{0.7} = 1 - \frac{3.1R_w}{1.6R_v + 1.5R_x}. \quad (8.2)$$

Kameda et al. (2015) reported that the 0.7-μm absorption detection on CM2 chondrites, such as Murchison and Nagoya, by the flight model of ONC-T. From our updated sensitivity, the ambiguity in the sensitivities for the v-, w-, and x-bands are 0.85%, 1.3%, and 1.6%, respectively. The error of 0.7-μm absorption depth calculated from error propagation of Eq. (8.2) are 1.6%. Thereby, the typical absorption of serpentine (3-4%) can be detected by a SNR~2. This value can be improved by decreasing the statistical errors based on more star observations.

### 8.3. UV-Slope Evaluation

As is noted above, most of the ground-based observations of Ryugu do not report an absorption around 0.7-μm. Rivkin (2012) reported that one third of C-complex asteroids in Sloan Digital Sky Survey, which includes large number of asteroids down to small objects, show the 0.7-μm feature, whereas two thirds do not. Moreover, the 0.7-μm feature seen in CM chondrite spectra easily disappears when dehydrated by heating (Hiroi et al., 1996) and space weathering (Matsuoka et al., 2015). Thus, there is possibility we will not find 0.7-μm feature on Ryugu. However, this does not indicate that there are no hydrated minerals on the surface. Other than 0.7-μm absorption feature, the strength of UV-absorption shortward of 0.55 μm, which is caused by $Fe^{2+}$ and $Fe^{3+}$ in silicates, also correlates positively with the 3-μm feature (e.g., Freierberg et al., 1985). Vilas et al. (1993) investigated the relationship of 0.7- and 0.43-μm absorptions and suggested that asteroids with 0.7-μm absorption also display the 0.43-μm absorption. Moreover, asteroid spectra that did not display the 0.43-μm feature also did not show the 0.7-μm feature. Thus, the UV absorption can be also a proxy for aqueous alteration in the ONC-T wavelength, such as a spectral turnover near the v-band. Thus, we can use the UV to blue spectral slope index instead of the 0.7-μm absorption. The UV slope can be calculated by $\ln(R_{ul}/R_v)$. According to Hiroi et al. (1996), there are large discrepancies of this value from -1 to 0 among C-, G-, B-, and F-type asteroids in the Tholen taxonomy (Tholen, 1984), and this value correlates nearly linearly with the 3-μm band absorption. Although they used 0.34 μm for the shorter wavelength, ul-band wavelength range, 0.4-μm, can serve as the same proxy. Based on our sensitivity ambiguity, the error for this value will be ~1% in the center of the FOV. However, it should be noted that the spatial error could be large because of the deviation in ul-band flat-field caused by the round-about stray light on the preflight flat-field measurement. For that reason, the spatial uncertainty of measuring the UV slope will be dominated by flat-field uncertainty of ~4%. This error estimation still suggests that we can distinguish unheated CMs of ln($R_{ul}$/$R_v$)~-0.6 to -0.3 from highly dehydrated (700-1000 °C) CMs of ~0 quite easily, assuming that $R_{\mathrm{ul}}$ and $R_{0.34\mu\mathrm{m}}$ are linearly correlated.

### 8.4. First Inflight Observation of Ryugu

The first inflight observations of Ryugu were conducted on 26 February 2018 when the angle between the Sun-Ryugu-Hayabusa2 was ~1.1-1.6°. Taking advantage of the illuminate condition close to opposition, Ryugu was apparently bright and we succeeded in observation with the ONC-T by longest exposure time (178 sec). We conducted two different observations for different purposes; 1) light curve observation and 2) Ryugu multi-band imaging. On the lightcurve observation, the wide-band was used in order to stack the signal as much as possible. On the multi-band observation, we observed different rotational phases. We took four sequential image sets for each band at one rotational phase. The summary of observations are listed in **Table 8.1**. After dark and flat corrections, invalid images, which include cosmic ray hits close to Ryugu due to long exposures were omitted from the analyses. The background noise, such as dark current and hot pixels, were main cause of errors in measurements. The lightcurve is obtained by taking three moving averages over time (**Fig. 8.1**).

On the multi-band observations, the images were acquired every ¼ rotational phase for 7 sets. Due to the very small signals, we employ two analysis methods to obtain robust results. One method is the aperture analysis, which is usually applied to star flux analyses as discussed in **Sec. 3**. Another method is the rough estimation from the peak signal of Ryugu. More specifically, we obtained the relationship between the peak signal and total signal from star observations, in which signals are much larger than the background signals. However, the signal ratio of peak to total may change from 0.15 to 0.33 with the Ryugu detected area in one CCD pixel. Because the signals of the ul- and p-bands are smaller than the dark level, <0.05 DN/s, the spectral shape is discussed without these two filters. **Figure 8.2** displays normalized reflectance spectra of Ryugu from the two methods from b- to x-bands. We found that this case with very low signal rate, the rough estimate gives smaller errors. Although, the shapes of reflectance spectra are not perfectly reproducible at two similar rotational phase, such as observation set of 2 and 6 in **Fig. 8.2**, the slope of reflectance spectra for all rotational phases varies to include 0 (flat) considering errors in measurements. Such flat spectra are consistent with ground-based observations (Binzel et al., 2001; Vilas 2008;Lazzaro et al., 2013; Moskovitz et al., 2013; Sugita et al., 2013; Perna et al., 2017). There may be darkening in the b-band compared with the v-band for most rotational phases. Furthermore, we compare the magnitude of the v-band observed with the ONC-T and the corresponding V-magnitude of the Earth-based observations from Ishiguro et al. (2014). Because we stack the signal from all 28 v-band images to reduce the error, the v-band magnitude here corresponds to a global average, or disk-integrated value. **Figure 8.3** shows the compiled phase curve of Ryugu, displaying consistency between ONC-T and Earth-based measurements, although the error bar is still large ~10%. This result also supports the robustness of the calibrations.

**Table 8.1.** Summary of Ryugu first inflight observations.

| | Observation Time (UTC) | Number of Images |
|---|---|---|
| Lightcurve observation (wide-band) | 26 February 2018 03:01 – 10:28 | 100 |
| 7-band observation (1) | 26 February 2018 10:39 – 12:01 | 28 |
| 7-band observation (2) | 26 February 2018 12:33 – 13:56 | 28 |
| 7-band observation (3) | 26 February 2018 14:28 – 15:50 | 28 |
| 7-band observation (4) | 26 February 2018 16:22 – 17:45 | 28 |
| 7-band observation (5) | 26 February 2018 18:15 – 19:39 | 28 |
| 7-band observation (6) | 26 February 2018 20:11 – 21:34 | 28 |
| 7-band observation (7) | 26 February 2018 22:04 – 23:28 | 28 |

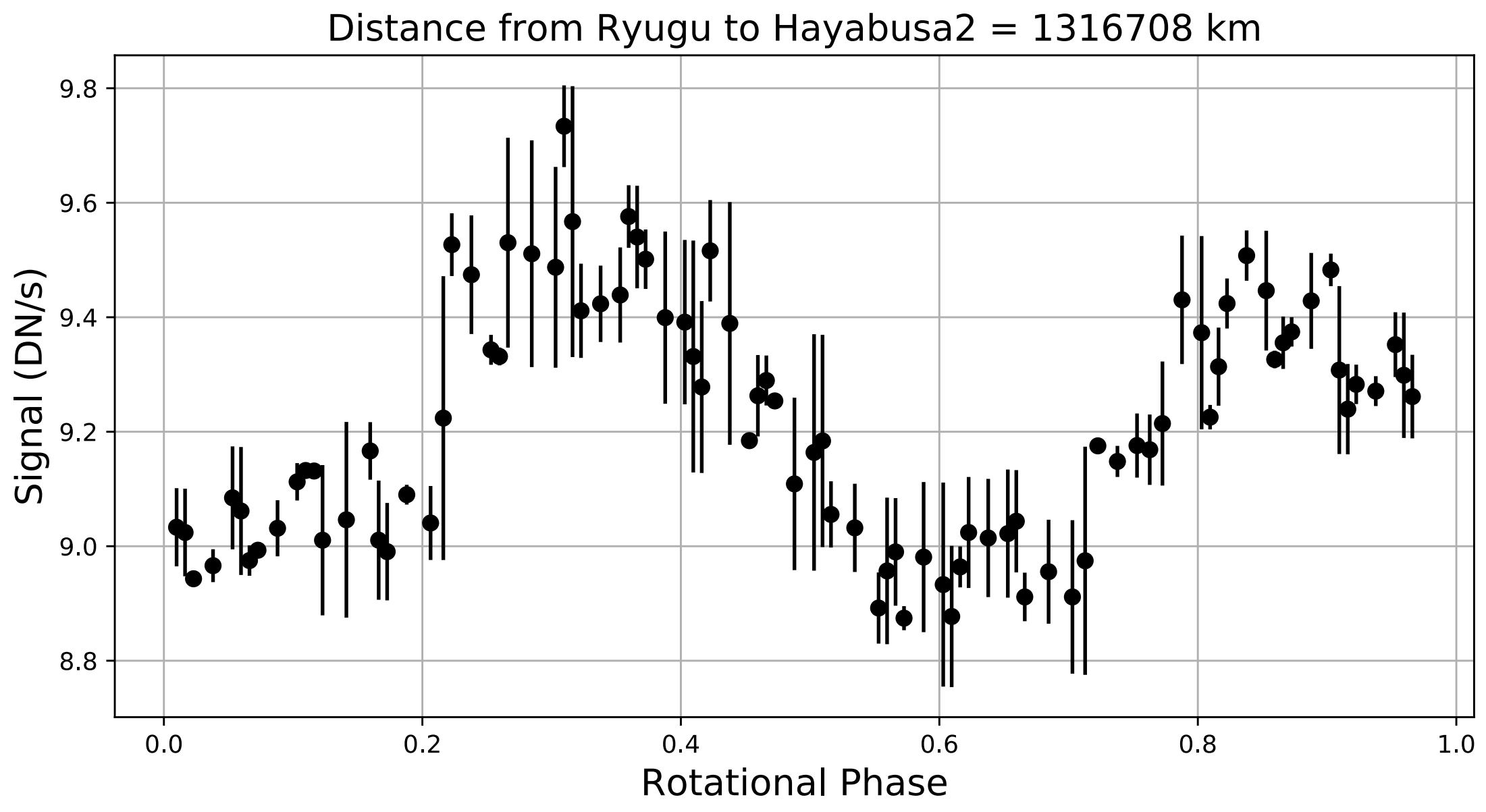

**Figure 8.1.** First inflight lightcurve observation of Ryugu taken on 26 February 2018. Every point shows a three points moving average and errors are evaluated by the deviation between three points. Due to the change in distance between the spacecraft and Ryugu, the signals are normalized by the square of the distance.

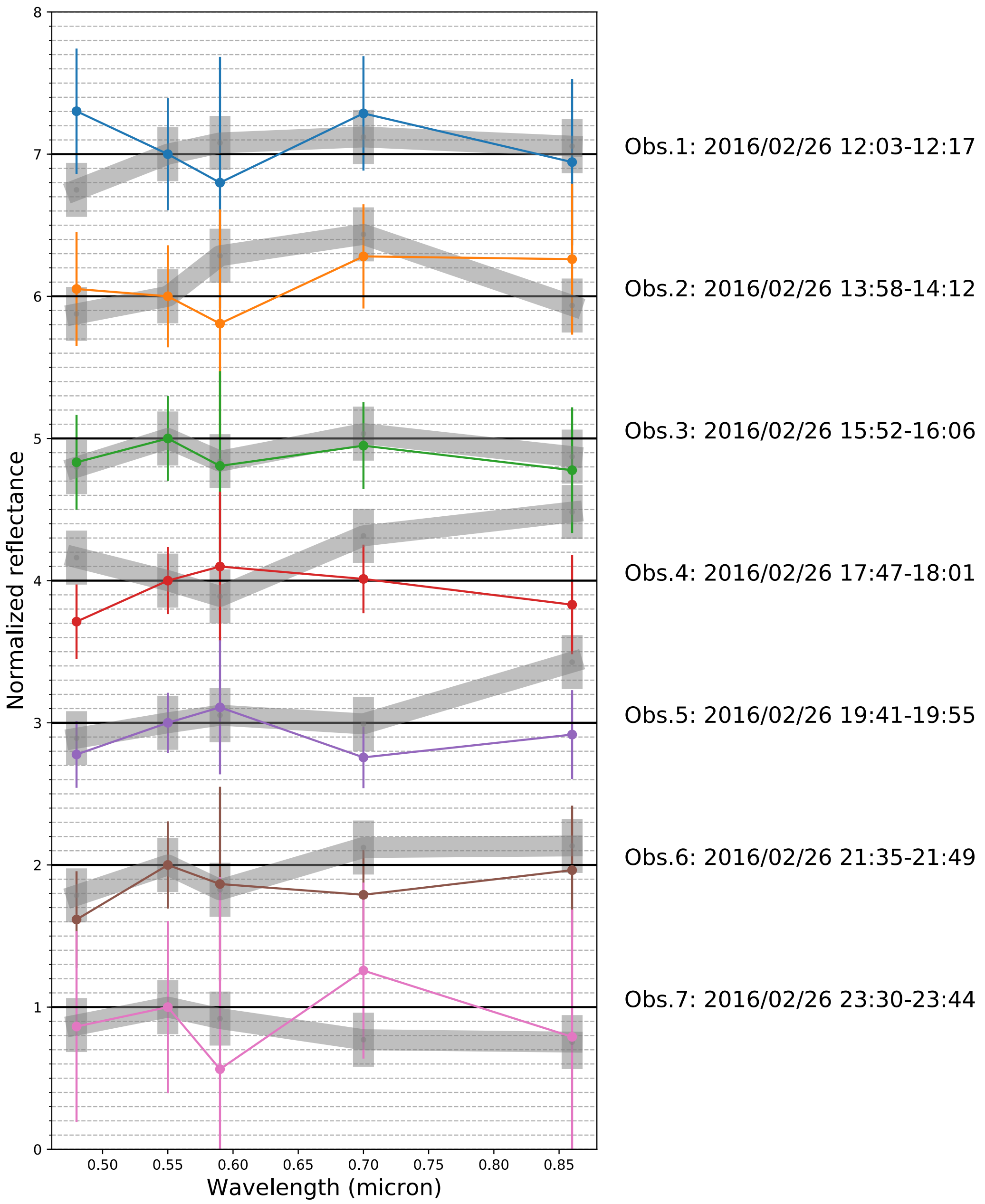

**Figure 8.2.** 5-band observations of Ryugu with the ONC-T on 26 February 2018. The color lines indicate the reflectance obtained by aperture analyses and gray lines indicate rough estimates based on peak signal rate.

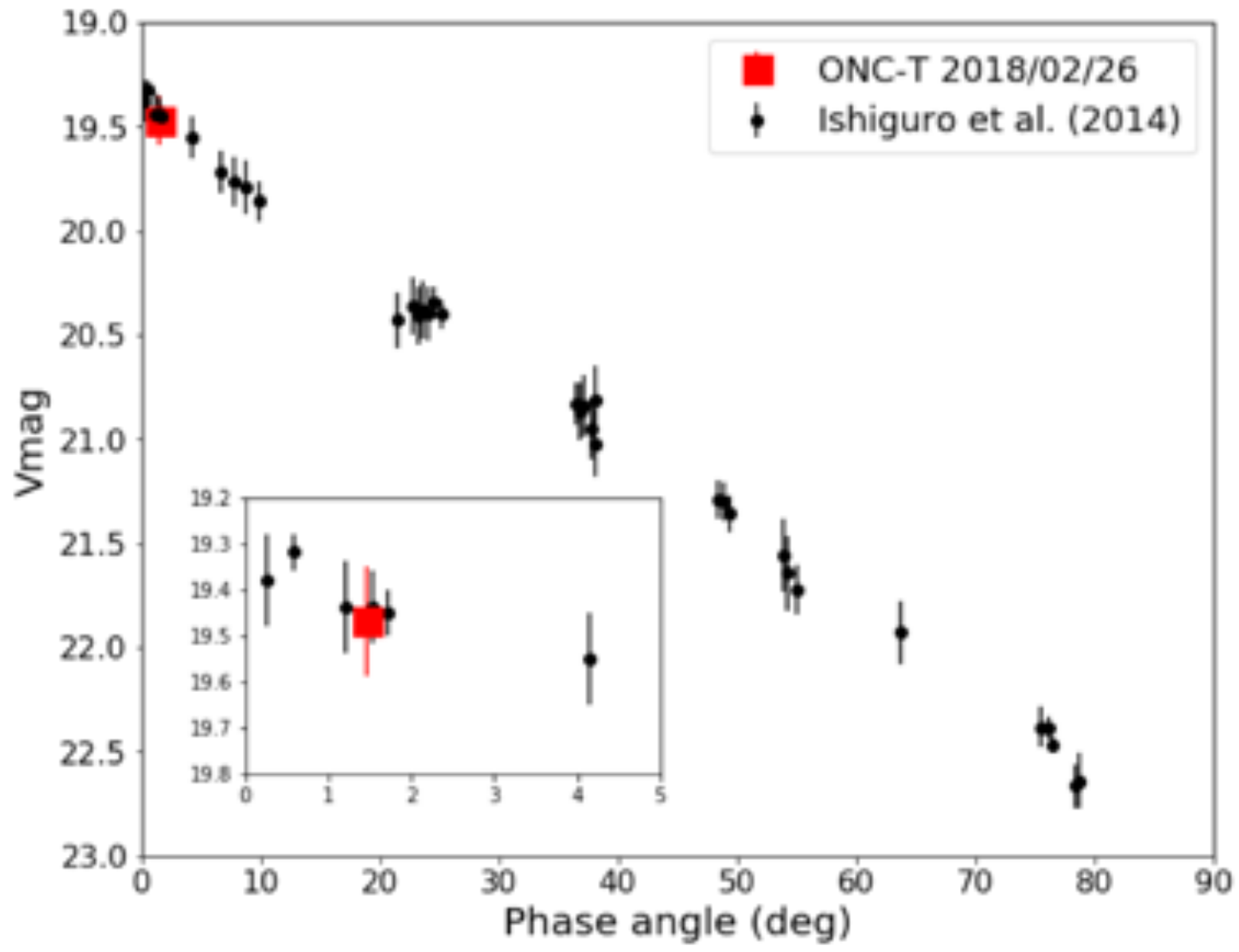

**Figure 8.3.** The phase curve of Ryugu, compiling the Earth-based data from Ishiguro et al. (2014) (black) and the inflight measurements from this study (red). The inflight ONC-T measurement overlaps with Earth-based measurements.

## 9. Summary

In this study, we have presented the calibrations of the three ONC cameras during the cruise phase over the past three years; including corrections for read-out smear noise, bias, EMI noise, dark, hot pixels, linearity, flat-field, stray light, and CCD sensitivity. Note that the CCDs of ONC-T, -W1, and -W2 have same manufacturer product number. Thus, the basic behavior is expected to be similar for all three cameras.

***ONC-T***

- Read-out smear can be omitted using inflight 0-sec exposure image subtraction, or by modeling using the empirical relationship in **Eq. (3.4)**, similar to the method described by Ishiguro et al. (2010). We showed the validity of this smear correction using an ONC-W2 image since the vertical charge transfer time has not been measured for the ONC-T. We will measure the vertical charge transfer time of the CCD for the ONC-T during the approach phase to Ryugu.
- Bias level depends on the temperature and can be described by an empirical bias model. Based on the model the bias level was shown to be stable within 1% deviations over the past three years.
- EMI noise causes wave-like patterns on images. The amplitudes have been < 5 DN for the cruise phase.
- Dark current and hot pixels are greatly changed by the CCD temperature. At the reference temperature, the dark current is smaller than 0.09DN/s, which is consistent with preflight measurements and less than 10 hot pixels >30 DN/s are found. At the CCD temperature of 20 °C, the dark current is 20 DN/s and hot pixels >100DN/s occupy ~1% of the FOV.
- The linearity of the CCD was found to be slightly degraded from the 0.6% error preflight to 1% inflight.

The empirical linearity correction model corrects the linearity to <0.1%.

- Flatness of the flat-fields for v-band and ul-band is evaluated based on star observations with slightly different attitudes. The sensitivity variations in the v-band flat-field and the ul-band flat-field are about 2% and 4%, respectively. The ul-band flat-field discrepancy is larger than the v-band because the round-about stray light might contaminate and make little gradation in the FOV during preflight tests.
- Scattered light in the optics are characterized and the PSF models are derived for all color filters. The effect of a broad PSF in the ONC-T is smaller than that on the AMICA/Hayabusa. The effect can be corrected in a way similar to that described by Ishiguro (2014).
- The ghost effect was observed and is most apparent in the p-filter, with a signal as much as 0.65% the intensity of the object observed.
- Suzuki et al. (2018) reported the radiator stray light dependence on spacecraft attitude with respect to the Sun. The radiator stray light has been intensively observed during the cruise phase. The attitudes with negligible stray light were found and reported in Suzuki et al. (2018). In this study we present the stray light model based on PCA. This model reduces the radiator stray light contamination to 25DN/s/pix, 0.25% the expected intensity of the asteroid disk.
- CCD sensitivity was carefully examined based on star, Jupiter, Saturn, and Moon observations in addition to preflight measurements. We characterized the sensitivity in two ways; 1) based on the hardware specifications of the ONC-T, and 2) based on star observations. Moreover, the sensitivity of the p-band is more accurately determined based on the lunar observations than stellar observations due to the inherent ambiguity in the stellar reference spectra. Since the sensitivity based on stellar observations for the ul- to x-bands and the lunar observations for the p-band reproduced the spectra of the Moon, Jupiter, and Saturn, current calibrations use this characterization of the sensitivity, summarized in Table. 3.10. Moreover, we measured the temperature dependency of the sensitivity using Jupiter observations acquired with different CCD and ELE temperatures. Using the relationship derived from these observations, we obtained a method for determining the sensitivity for different temperature circumstances, such as during touchdowns, and rover and lander release operations.

***ONC-W1 and -W2***

- Empirical bias models of the ONC-W1 and -W2 were derived based on the preflight measurements. The bias level during the cruise phase was found to be stable over the past three years.
- Dark level and the number of hot pixels are similar to the ONC-T.
- The flat-fields for the ONC-W1 and -W2 were derived based on the pre-flight measurements. However, because a surface light source is needed to assess the accuracy of these flat-fields, we are planning to evaluate the flatness using whole disk images of Ryugu during the approach phase.
- CCD sensitivities for the ONC-W1 and -W2 are newly derived based on the preflight integrating sphere

measurements and inflight star and Jupiter observations. The inflight sensitivities are consistent within a 10% difference. The difference in spectral shapes between the sensitivities derived from the integrating sphere and the inflight reference objects are suspected to be the major source of the differences.

- Stray light in the ONC-W1 and -W2 was characterized. The stray light contamination is basically weaker than that seen in the ONC-T. Moreover, shorter exposure times for ONC-W1 and -W2 yields quite small stray light effects.

In addition to the global spectral mapping, we are planning to conduct highlighted observations, such as sodium atmosphere observations, hydrated mineral observations, and high-resolution imaging associated with touchdowns.

A sodium atmosphere on Ryugu is highly possible, as Ryugu is expected have been formed in a volatile-rich region in the early Solar System. We evaluated the detectability of a sodium atmosphere by testing observations of Jupiter's sodium torus. Mainly due to the dark noise, the sodium atmosphere on Jupiter was not detected, but the upper limit for detection was evaluated. We expect to detect Na atmosphere of several 10s kR, similar to comets (Leblanc et al., 2008), by a single image set (v and Na) and of several 100 R by using 100 of sets.

Both 0.7-μm absorption and steep UV slope are possible indicators of Hydrated minerals on asteroids. Our calibration results suggested that the error in measurement of 0.7-μm absorption is 1.6% and the in measurement in UV slope is <4% using ONC-T. These calibration accuracies imply that we can detect 0.7-um absorption band with SNR~2, assuming the typical serpentine absorption of 3-4% and also we may distinguish unheated and highly heated CM-like materials based on UV slope.

During the touchdowns up to a few meters of altitude, the radiative heat from Ryugu is expected to raise the CCD temperature as much as 20 °C. This high temperature may degrade the SNRs of the cameras to 150 at most. However, this may still be sufficient to detect the surface radiance variations of 3% caused by illumination conditions or albedo variation. Thus, quantitative imaging with very high resolution up to a few mm/pixel is possible.

Our first inflight observation of Ryugu on 26 February 2018 shows very good agreement with the Earth-based observation by Ishiguro et al. (2014). This confirms that our radiometric calibration is robust. Finally, drastic degradation in the ONC-T system has not been observed during the 3.5 year cruise phase. However, some observations such as charge transfer time measurement, flat-field evaluation, and round-about stray light investigation based on Ryugu disk observations, and updating sensitivity by more star observations are desired to improve and monitor our calibration. And also time dependent change should be recorded at regular intervals to monitor the health of the ONC system for one and a half years of the rendezvous period.

**Acknowledgement**

We would like to show our greatest appreciation to Hayabusa2 team members who gave enormous contribution to achieve observations. We would also like to express our gratitude to Dr. D. Domingue who gives insightful and constructive comments. We also wish to acknowledge the constructive reviews from Dr. C. Güttler and an anonymous reviewer. This research is supported by Japan Society for the Promotion of Science (JSPS) Core-to-Core program "International Network of Planetary Sciences".

## Appendix A. Specifications of ONC

Detailed specifications of ONC three cameras are summarized here. Most of the information are already published in Kameda et al. (2017) and Suzuki et al. (2018).

**Table A1.** Performance of CCD (E2V CCD47-20 (AIMO)). Note that three cameras of ONC use same CCD products.

| | ONC-T | ONC-W1 | ONC-W2 |
|---|---|---|---|
| Format | 1056 (H) pixels × 1024 (V) pixels (16 × 1024 pixels on both sides are Optical Black pixels) | | |
| CCD pixel size | 13 μm | | |
| Gain factor (measured value) | 20.95 $e^-$/DN | 20.86 $e^-$/DN | 20.11 $e^-$/DN |
| Read-out noise (measured value) | 38.5 $e^-$ | 36.3 $e^-$ | 37.0 $e^-$ |
| A/D conversion | 12 bit | | |
| Full-well (measured value) | 91,000 $e^-$ | 84,000 $e^-$ | 96,000 $e^-$ |
| Pixel sampling rate | 3 MHz | | |

**Table A2.** Weight, size, and electric consumption of ONC hardware components. The basic design including CCD, electronics, and optics of W1 and W2 are the same. Thus, their electric consumption is the same, but their dimension and weight are slightly different because of difference in radiator and hood design.

| | Weight [kg] | Size [mm] | Electric consumption [W] |
|---|---|---|---|
| ONC-T | 2.1 | 430 × 155 × 134 | 8 |
| ONC-W1 | 1.355 | 249 × 155 × 95 | 8 |
| ONC-W2 | 1.295 | 222 × 155 × 95 | 8 |
| ONC-AE | 1.285 | 200 × 220 × 50 | 14 |
| DE | 2.67 | 95 × 221 × 170 | 15 |

**Table A3.** Focal length and distortion coefficient for wide-, v-, and Na-bands of ONC-T.

| Band | Focal length [mm]* | Distortion coefficient $\varepsilon_1$ [1/pix$^2$]* | Note |
|---|---|---|---|
| wide | 120.50 | $-9.28 \times 10^{-9}$ (RMS error 0.1 pixels) | Suzuki et al. (2018) |
| v | 120.50 | $-6.766 \times 10^{-9}$ (RMS error 0.3 pixels) | Newly derived from the star image hyb2_onc_20151105_065349_tvf_l2a.fit. |
| Na | 120.49 | $-1.024 \times 10^{-8}$ (RMS error 0.3 pixels) | Newly derived from the star image hyb2_onc_20171015_000153_tnf_l2a.fit. |

*Details for the distortion model are described in Suzuki et al. (2018).

It is noted that the focal lengths listed in this table are valid only when combined with each distortion coefficient.

## Appendix B. Components of ONC-T Photometric System

The quantum efficiency of CCD, originally measured in room temperature (20°C), is shown in **Fig. B1**. Moreover, the CCD temperature can change this function shape. An example of functions of quantum efficiency with two different temperatures based on the manufacturer (E2V) derived values. The transmissions of band-pass filters (same as Kameda et al., 2017), the ND filter, lenses and the CCD cover glass are displayed in **Fig. B2**.

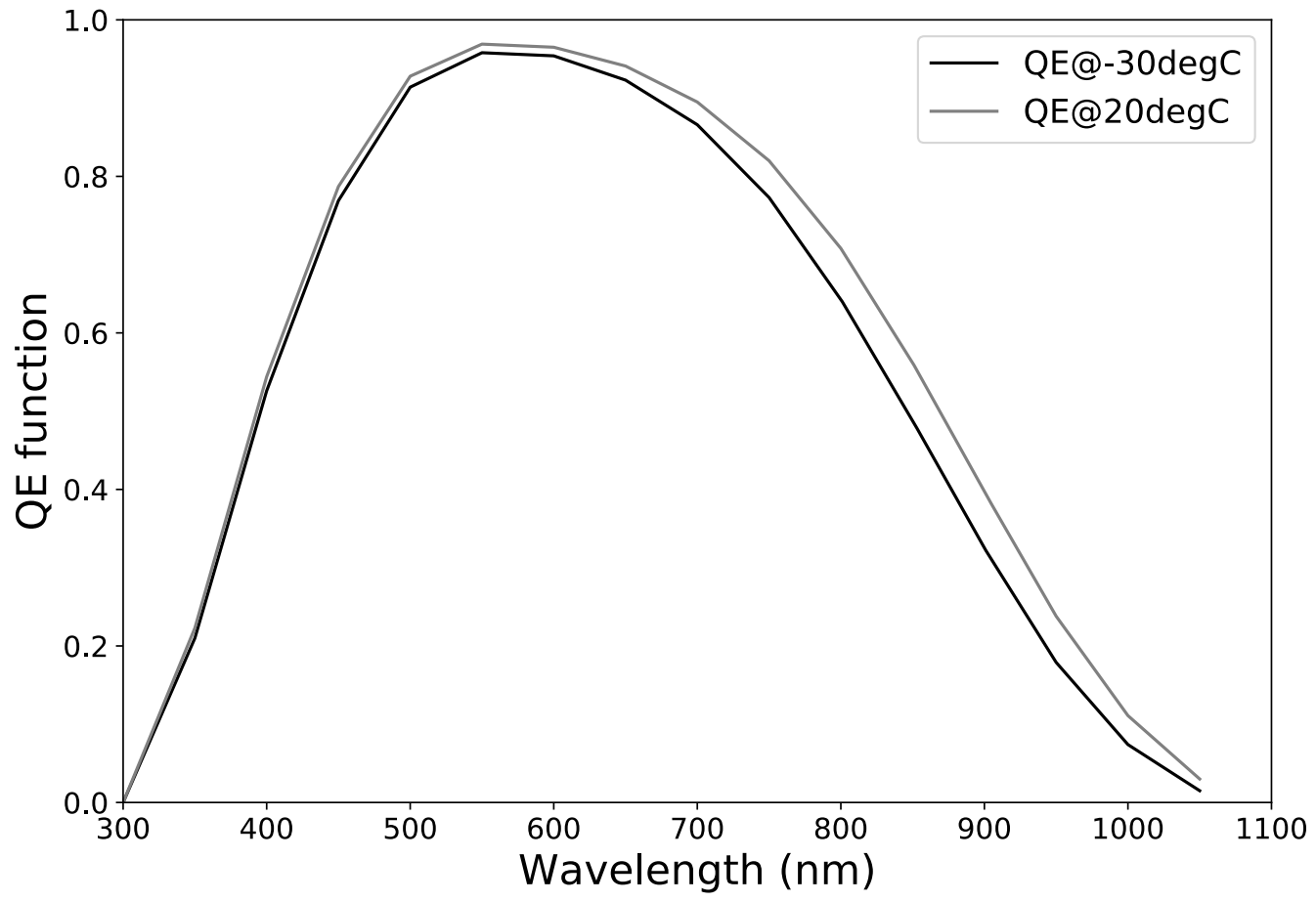

**Figure B1.** Functions of quantum efficiency in different CCD temperature. High CCD temperature shows the higher efficiency in longer wavelength range.

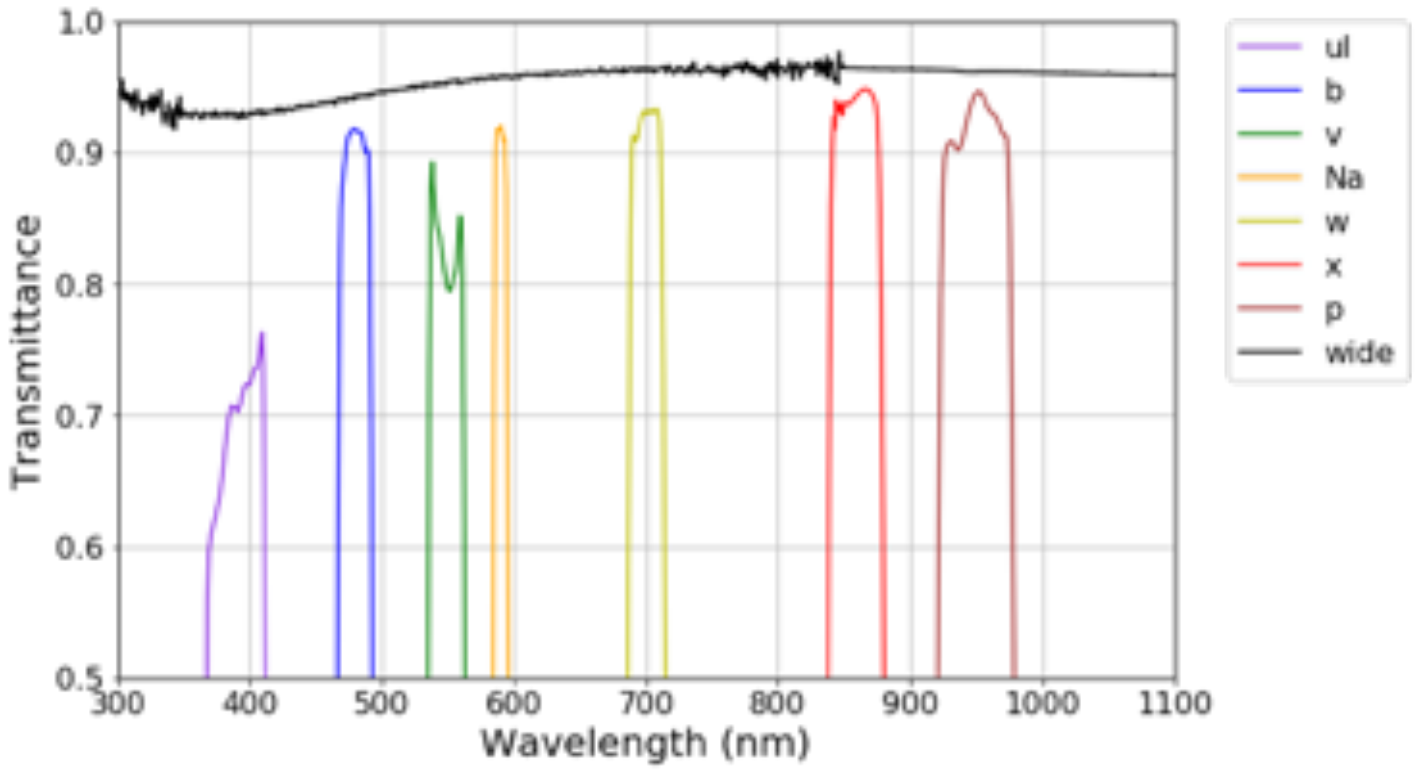

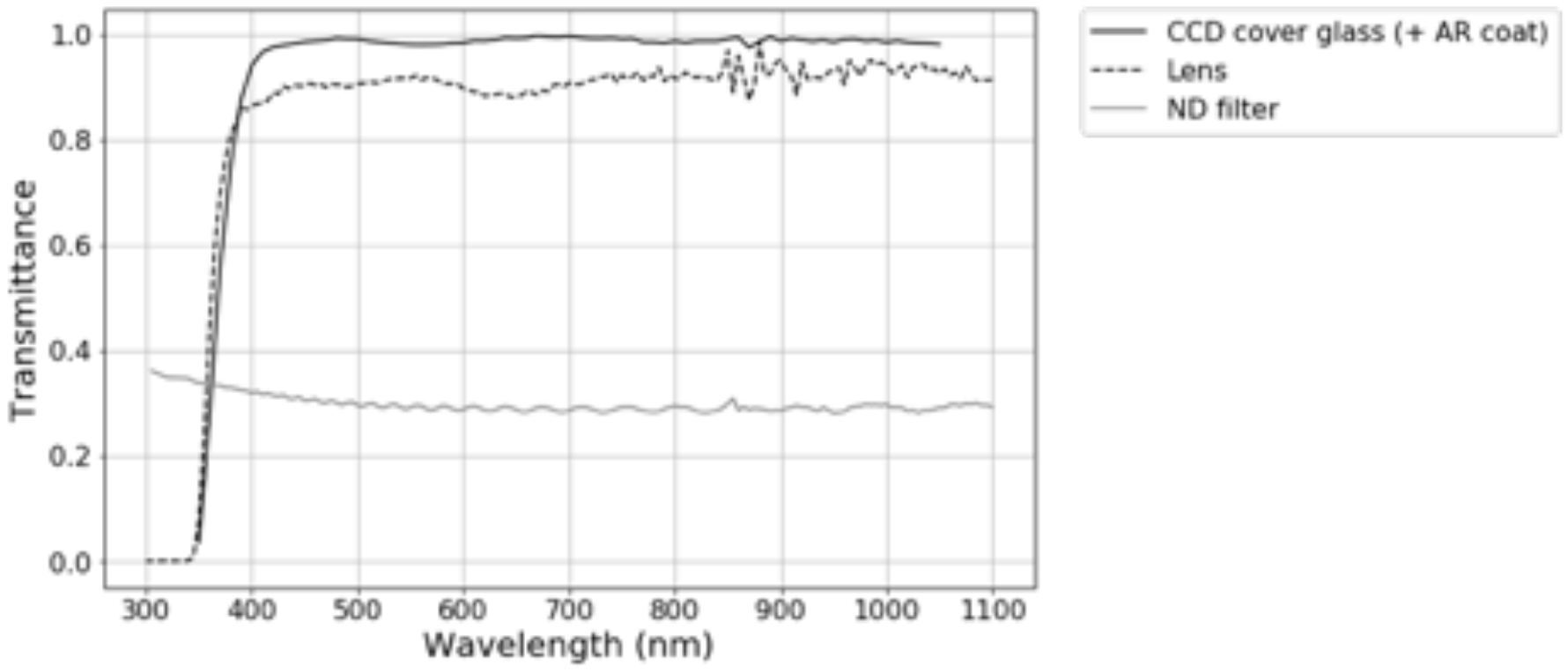

**Figure B2.** (Top) Transmittances of the band-pass filters. (bottom) Transmittances of the ND filter, lenses, and the CCD cover glass.

## Appendix C. Coordinate System and Definition of Attitude of the Spacecraft

**Figure C1** displays the spacecraft coordinate system. The spacecraft attitude with respect to the Sun could be defined by $X_{PNL}$ and $Y_{PNL}$, which correspond to the angles of the Sun to $+X_{SC}$ and $+Y_{SC}$ plane of the spacecraft.

$$\boldsymbol{s} \cdot \boldsymbol{X}_{SC} = \sin X_{PNL}, \boldsymbol{s} \cdot \boldsymbol{Y}_{SC} = \sin Y_{PNL},$$

where $\boldsymbol{s}$ is a unit vector from the spacecraft to the Sun. The angles $\phi$ and $\gamma$ are the spacecraft twisting angle described in Suzuki et al. (2018). The relationship between $X_{PNL}$ and $Y_{PNL}$ are derived as

$$-\sin\phi\cos\gamma = \sin X_{PNL}, \sin\phi\sin\gamma = \sin Y_{PNL}.$$

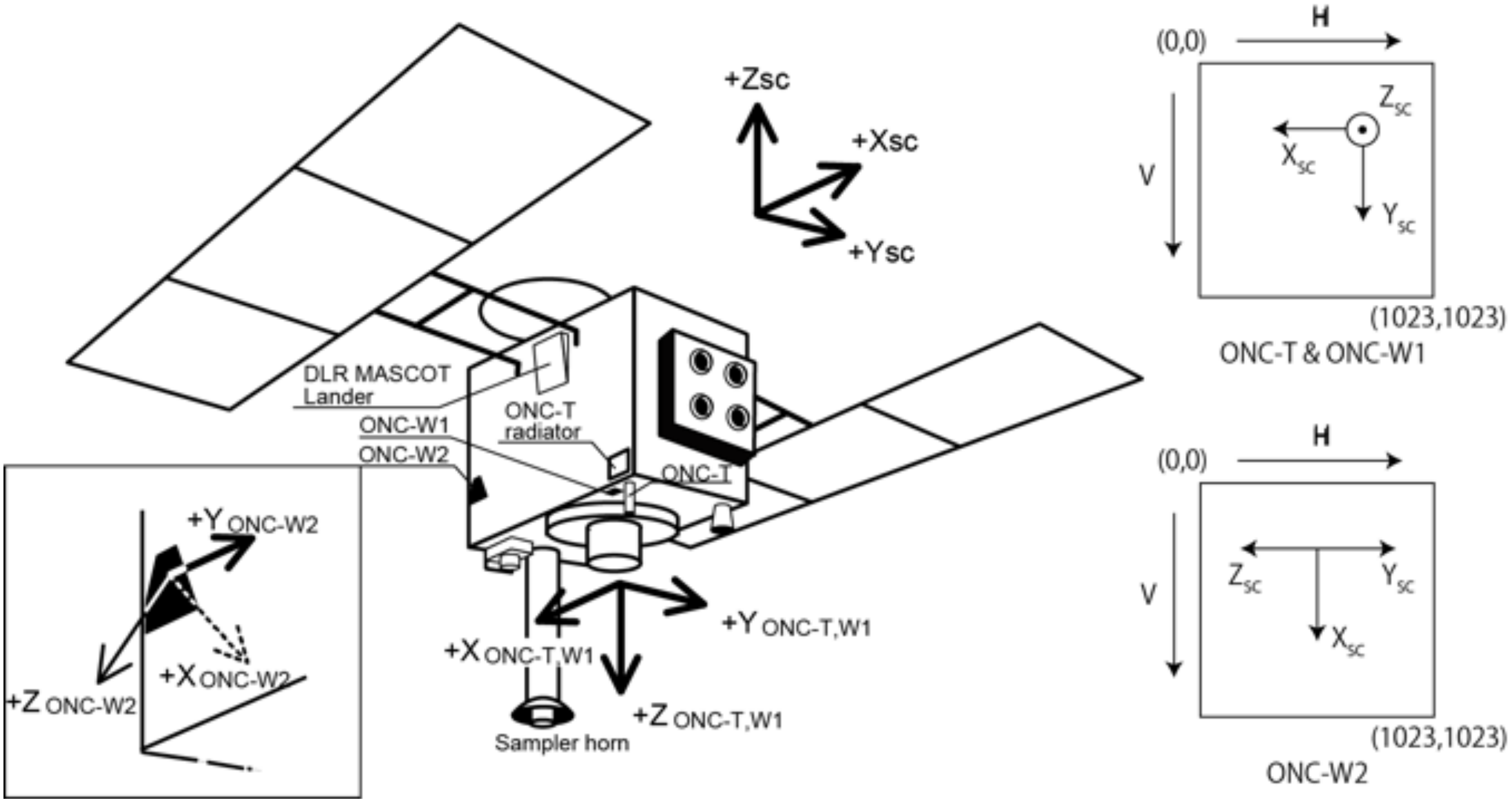

**Figure C1.** The spacecraft coordinate system and the image coordinate system. (after Suzuki et al. (2018))